\documentclass[manuscript,nonacm]{acmart}
\AtBeginDocument{%
  }

\renewcommand\footnotetextcopyrightpermission[1]{}
\setcopyright{none}

\usepackage[utf8]{inputenc} % allow utf-8 input
\usepackage{multirow}
\usepackage{colortbl}
\usepackage{xspace}
\usepackage{enumitem} % to reduce the item indentation in lists
\usepackage{pifont} % used for \ding
\usepackage{url} 
\usepackage{tabularx} % tabularx package for the task distribution table
\usepackage{makecell} % for longer toprule, midrule, bottomrle of tables
\usepackage{subcaption} % for subfigures
\usepackage{xcolor}
\usepackage{framed}  % alternate for tcolorbox
\definecolor{shadecolor}{gray}{0.9}
\newenvironment{FramedColorBox}[4]{%
  \def\FrameCommand{%
    \setlength{\fboxrule}{#3}%
    \setlength{\fboxsep}{#4}%
    \fcolorbox{#2}{#1}%
  }%
  \MakeFramed{\advance\hsize-\width \FrameRestore}%
  \noindent
}{%
  \endMakeFramed%
}

\newenvironment{WhiteContainer}{%
  \begin{FramedColorBox}{white}{black!30}{0.8pt}{6pt}%
}{%
  \end{FramedColorBox}%
}

\newenvironment{AppendixDocumentBox}[2][blue!10!white]{%
  \begin{FramedColorBox}{#1}{blue!50!black}{1.5pt}{6pt}%
  {\bfseries\large #2}\par\medskip
}{%
  \end{FramedColorBox}%
}

\newenvironment{ArtifactBox}[1]{%
  \begin{FramedColorBox}{gray!5}{gray!60}{0.8pt}{5pt}%
  {\bfseries #1}\par\smallskip
}{%
  \end{FramedColorBox}%
}

\newcounter{plancardctr}

\newenvironment{numberedPlanCard}{%
  \refstepcounter{plancardctr}%
  \begin{FramedColorBox}{gray!5}{black!20}{0.6pt}{6pt}%
  \noindent\textbf{\theplancardctr.}\hspace{0.6em}%
}{%
  \end{FramedColorBox}%
}
\newenvironment{quizCard}{%
  \begin{FramedColorBox}{gray!5}{black!20}{0.6pt}{8pt}%
}{%
  \end{FramedColorBox}%
}

\newcommand{\notick}{\ding{55}}
\newcommand{\yestick}{\ding{51}}
\definecolor{urlcolor}{rgb}{0.0, 0.53, 0.74}
\definecolor{customgreen1}{RGB}{119,221,119}
\definecolor{customred1}{RGB}{221,119,119}
\definecolor{customgreen2}{RGB}{0,128,128}
\definecolor{customred2}{RGB}{128,0,0}

\newcommand{\nocff}{\textsc{None}}
\newcommand{\assumptioncff}{\textsc{Assumptions}}
\newcommand{\whatifcff}{\textsc{WhatIf}}
\newcommand{\bothcff}{\textsc{Both}}

\begin{document}

\title{An Experimental Comparison of Cognitive Forcing Functions for Execution Plans in AI-Assisted Writing: Effects On Trust, Overreliance, and Perceived Critical Thinking}
\thanks{This manuscript is a preprint.}

\author{Ahana Ghosh}
\email{gahana@mpi-sws.org}
\authornote{This work was conducted during an internship at Microsoft Research Cambridge.}
\affiliation{%
  \institution{Max Planck Institute for Software Systems}
  \city{Saarbrücken}
  \country{Germany}
}

\author{Advait Sarkar}
\email{advait@microsoft.com}
\affiliation{%
  \institution{Microsoft Research}
  \city{Cambridge}
  \country{UK}
}

\author{Siân Lindley}
\email{sianl@microsoft.com}
\affiliation{%
  \institution{Microsoft Research}
  \city{Cambridge}
  \country{UK}
}

\author{Christian Poelitz}
\email{cpoelitz@microsoft.com}
\affiliation{%
  \institution{Microsoft Research}
  \city{Cambridge}
  \country{UK}
}

\renewcommand{\shortauthors}{Ghosh et al.}

% !TEX root =  main.tex
%%%%%%%%%%%%%%%%%%%%%%%%%%%%%%%%%%%%%%%%%%%%%%%%%%%%%%%%%%
%%%%%%%%%%%%%%%%%%%%%%%%%%%%%%%%%%%%%%%%%%%%%%%%%%%%%%%%%%

%%%%%%%%%%%%%%%%%%%% 357 words
\begin{abstract}
%%%%% the risk of GenAI + CFF to mitigate the risk
Generative AI (GenAI) tools offer productivity benefits across knowledge workflows such as writing and programming, but also introduce risks of overreliance and reduced critical thinking. To address these risks, prior work has explored cognitive forcing functions (CFFs), design interventions that compel the user to cognitively engage with the task in order to proceed. 
%%%%% CFFs situated within AI plans not explored
As GenAI workflows become longer and more complex, an increasingly common design idiom is that of the ``plan'', a description of the steps that the system proposes to execute in order to fulfil the user request. The plan, in principle, allows the user to verify that the system has interpreted their intent correctly and will follow the correct procedure. Yet the plan, in turn, suffers from the same risks of overreliance and reduced critical thinking as any other AI-generated output. Crucially, until now, the effectiveness of CFFs when applied to AI plans has not been studied.
%%%%%%%%% our study
We conduct a controlled experiment ($n=214$) where participants completed AI-assisted writing tasks which included reviewing AI plans for the writing request. We compared four conditions: \assumptioncff{} (a CFF targeting the critical thinking skill of argument analysis), \whatifcff{} (a CFF targeting the critical thinking skill of hypothesis testing), \bothcff{} (\assumptioncff{} followed by \whatifcff{}), and a no-CFF control (\nocff{}). In a separate think-aloud and interview study ($n=12$), participants completed tasks with both CFFs as well as under the no-CFF control and qualitatively compared them.
%%%%%%% our findings
Our study finds that the \assumptioncff{} CFF most effectively reduced overreliance without increasing cognitive load. This suggests that for evaluating AI execution plans for writing tasks, it is better to target argument analysis than hypothesis testing, and that combining multiple CFFs may be ineffective or even detrimental. Interestingly, this did not align with user perception: users \emph{perceived} the \whatifcff{} CFF to be more helpful. Participant traits such as cognitive disposition and GenAI familiarity shaped their likelihood of revising judgments about the AI response, but CFF design primarily influenced their overreliance behavior.
%%%%%%%% final takeaway
Our findings provide design implications and opportunities for balancing critical reflection with usability in GenAI workflows, such as selecting plan-focused CFFs that target argument analysis during AI output review to reduce overreliance without increasing cognitive load.
\end{abstract}

\begin{CCSXML}
<ccs2012>
<concept>
<concept_id>10003120.10003123</concept_id>
<concept_desc>Human-centered computing~Interaction design</concept_desc>
<concept_significance>500</concept_significance>
</concept>
</ccs2012>
\end{CCSXML}

\ccsdesc[500]{Human-centered computing~Interaction design}

\keywords{AI assisted knowledge work, cognitive forcing functions, argument analysis, hypothesis testing, overreliance, critical thinking}

\maketitle
% !TEX root =  main.tex
%%%%%%%%%%%%%%%%%%%%%%%%%%%%%%%%%%%%%%%%%%%%%%%%%%%%%%%%%%
%%%%%%%%%%%%%%%%%%%%%%%%%%%%%%%%%%%%%%%%%%%%%%%%%%%%%%%%%%

\section{Introduction}\label{sec:intro}

%%% GenAI in the workplace, its effect on human cognition
Generative AI (GenAI) systems, particularly large language models (LLMs), are now increasingly integrated into professional workflows, including drafting content, planning, summarizing documents, generating code, and synthesizing information across sources~\citep{DBLP:conf/chi/WoodruffSKRSW24,wang2025ai,davila2024industry}. While these systems offer significant productivity benefits, they also raise critical questions about individual cognition, skill development, and trust calibration in high-stakes or open-ended tasks~\citep{DBLP:conf/chi/WoodruffSKRSW24,DBLP:conf/chi/LeeSTDRBW25,DBLP:conf/chi/TankelevitchKSS24}. To address these risks, an extensive body of prior work in AI-assisted decision making has explored design interventions such as explanations, confidence cues, and cognitive forcing functions that encourage users to engage more critically with AI-generated recommendations~\citep{DBLP:journals/pacmhci/BucincaMG21,DBLP:conf/chi/BucincaSPDG25,kazemitabaar2025exploring}.

%%%%%% GenAI in knowledge work today introduces lengthy outputs
However, reviewing AI output has become increasingly demanding as GenAI systems produce longer, more open-ended, and more procedural responses. In many current workflows, users are no longer evaluating a single prediction or recommendation, but instead must interpret multi-step plans, lengthy drafts, or structured reasoning produced by the system. This shifts the user's task from verifying an outcome to assessing whether the AI has adopted an appropriate approach in the first place, a form of judgment that is cognitively demanding and susceptible to overreliance.

%%%%%%%% GenAI outputs structured using plans
\looseness-1A common design pattern that has emerged in GenAI interfaces is the explicit generation of an execution plan: a structured description of the steps the system proposes to follow in order to fulfill a user's request. In principle, such plans are intended to support transparency and enable users to verify that the system has correctly interpreted their intent. In practice, however, AI-generated plans are themselves complex artifacts that users may accept uncritically, particularly when they appear coherent and plausible, an effect well documented in studies of automation bias and AI-assisted decision making~\citep{DBLP:journals/pacmhci/BucincaMG21,bansal2021does}. Despite their growing prevalence in generative AI interfaces, AI-generated plans have received relatively little attention as primary objects of systematic user evaluation in AI-assisted knowledge work, with prior research primarily focusing on the review of predictions, recommendations, or explanations rather than procedural plans.

%%%%%%%%%%%%%%%% CFFs and the gap between AI-assisted knowledge work and 
Cognitive forcing functions (CFFs) are design interventions that intentionally constrain or interrupt interaction to require users to engage in specific reasoning or verification processes before proceeding further. They are well suited to support critical engagement during AI output review, and have been extensively studied in AI-assisted decision-making contexts to reduce automation bias and promote reflective reasoning~\citep{DBLP:journals/pacmhci/BucincaMG21,woods1994behind}. Yet it remains unclear how different types of CFFs function when applied specifically to AI-generated plans, and which critical thinking skills they should target in this context. In particular, prior work has not examined plan-centered CFFs for mitigating overreliance on AI in lengthy, open-ended knowledge work tasks.

%%%%%%%%%%%%%%%%%%%%%%%%%%%%%%%%%%%%%%%%%%%
% !TEX root =  main.tex
%%%%%%%%%%%%%%%%%%%%%%%%%%%%%%%%%%%%%%%%%%%%%%%%%%%%%%%%%%
%%%%%%%%%%%%%%%%%%%%%%%%%%%%%%%%%%%%%%%%%%%%%%%%%%%%%%%%%%

\begin{figure*}
    \includegraphics[width=\textwidth]{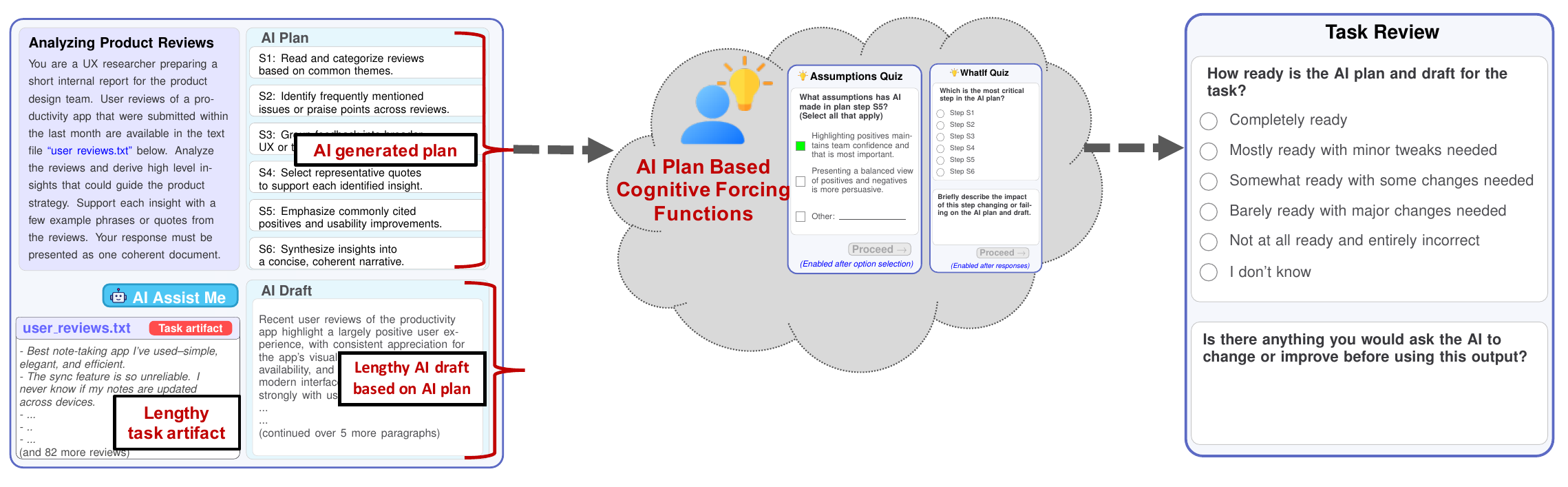}
    \caption{AI-assisted knowledge work with AI plan–based cognitive forcing functions. We illustrate a representative knowledge work task involving the analysis of product reviews. The interface presents a task description along with a lengthy task artifact comprising multiple product reviews. Upon clicking the ``AI Assist Me'' button, the system generates an AI execution plan and a corresponding draft. Before users can review the draft and decide whether to accept or reject the AI output, they are presented with cognitive forcing functions (CFFs) centered on the AI-generated plan and instantiated as short evaluative questions. After engaging with these questions, users proceed to the task review stage, where they assess the readiness of the AI output and provide follow-up instructions.}
    \Description{}
    \label{fig1:cff_workflow}
\end{figure*}
%%%%%%%%%%%%%%%%%%%%%%%%%%%%%%%%%%%%%%%%%%%%%%%%%%%

%%%%%% our contributions and research questions
In this study, we investigate the effectiveness of CFFs when applied to AI execution plans for AI-assisted writing tasks. More specifically, we ground our exploration of CFF design within frameworks of critical thinking and metacognitive monitoring. Figure~\ref{fig1:cff_workflow} is a schematic overview of the CFF interface used in our study. We compare three CFFs that target different critical thinking skills (as categorised by Halpern \cite{halpern1998teaching,halpern2013thought}) during the review of AI output: (i) \assumptioncff{}, where users are provoked to identify implicit assumptions behind the AI's plan, targeting the argument analysis skill; (ii) \whatifcff{}~, where users reason about how the outcome would change if a key step in the AI-plan were altered, targeting the hypothesis testing skill; and (iii) \bothcff{} together. A group with no CFF (\nocff{}) serves as control. 

We evaluate these interventions in realistic writing  scenarios assisted by AI, using both a large-scale experimental study (with $214$ participants) and think-aloud interviews (with $12$ participants). Specifically, our work addresses the following research questions (RQs) regarding how different cognitive forcing functions support users in evaluating AI-generated execution plans and corresponding outputs in AI-assisted knowledge work:
\begin{enumerate}[label={[RQ\arabic*]}]
   \item How do different CFFs influence knowledge workers' likelihood to revise their initial assessment of the readiness of AI-generated plans and outputs in knowledge work tasks?
    %%%%%%
    \item How do CFFs affect knowledge workers’ overreliance on AI-generated plans and responses?
    %%%%%%%%%
    \item How does the cognitive load of different CFFs compare to each other?
    %%%%%%%%%
    \item How do knowledge workers perceive the helpfulness and trustworthiness of different CFFs in mitigating AI overreliance and promoting critical thinking?
\end{enumerate}

Together, these research questions allow us to empirically compare plan-based cognitive forcing functions and derive design implications for supporting critical evaluation and trust calibration in AI-assisted knowledge work.

% !TEX root =  main.tex
%%%%%%%%%%%%%%%%%%%%%%%%%%%%%%%%%%%%%%%%%%%%%%%%%%%%%%%%%%
%%%%%%%%%%%%%%%%%%%%%%%%%%%%%%%%%%%%%%%%%%%%%%%%%%%%%%%%%%

\section{Related Work}\label{sec:related_work}
In this section, we situate our work within prior research on the cognitive impacts of GenAI in knowledge work. We also examine prior approaches to mitigating overreliance in AI-assisted tasks and design strategies for CFFs that support critical engagement during AI output review.

\subsection{Impact of GenAI on Perceived Productivity, Cognitive Effort, and Overreliance in Knowledge Work}\label{sec:related_work.impact}

%%%%% overreliance on AI responses
The introduction of AI assistance significantly affects knowledge workers' (perceived) productivity, cognitive effort, and skill development. While GenAI systems can increase output and reduce routine effort, they also risk encouraging cognitive offloading and overreliance, particularly when users defer judgment to fluent or authoritative AI outputs~\citep{DBLP:conf/chi/LeeSTDRBW25}. Several studies show that humans often over-trust AI suggestions, leading them to accept incorrect outputs with minimal scrutiny~\citep{passi2022overreliance}. Buçinca et al. observed that, in decision making tasks people frequently ``follow the AI's suggestions even when those suggestions are wrong and the person would have made a better choice on their own''~\citep{DBLP:journals/pacmhci/BucincaMG21}.

%%%%%%%%%%%%%%% AI explanations to mitigate overreliance
Interestingly, providing AI explanations alone does not mitigate this problem. Prior work shows that explainable AI interfaces can fail to reduce overreliance and may even exacerbate it by increasing users' confidence in the system~\citep{vaccaro2024combinations,DBLP:journals/pacmhci/BucincaMG21}. In such cases, the presence of an explanation functions as a heuristic signal of competence rather than a prompt for deeper evaluation~\citep{DBLP:journals/pacmhci/BucincaMG21}. This aligns with broader findings in cognitive psychology and human–AI interaction showing that people often act as ``cognitive misers,'' favoring low-effort, intuitive judgments over analytical scrutiny when interacting with intelligent systems\citep{DBLP:journals/pacmhci/BucincaMG21,fan2025beware,DBLP:conf/chi/TankelevitchKSS24,kahneman2011thinking,fiske1984social}. As a result, metacognitive engagement including, monitoring, questioning, and verifying one's own reasoning, often decreases in AI-assisted settings, particularly when the system appears confident, fluent, or competent.

%%%% metacognitive demands of AI responses 
Recent work suggests that GenAI fundamentally changes the nature of cognitive effort in knowledge work, shifting user roles from direct task execution toward monitoring, verification, and oversight of AI-generated outputs~\citep{DBLP:journals/corr/abs-2508-21036}. Tankelevitch et al. characterize this shift as an increase in metacognitive demands: users must monitor, verify, and correct AI outputs rather than directly performing the task themselves~\citep{DBLP:conf/chi/TankelevitchKSS24}. However, empirical studies show that these demands are unevenly met. Lee et al. found that higher confidence in GenAI systems correlates with reduced critical thinking effort, whereas confidence in one's own abilities predicts sustained scrutiny~\citep{DBLP:conf/chi/LeeSTDRBW25}. Similarly, Swaroop et al. identify a class of ``over-reliers'' who expend less effort, feel less agency, and experience reduced perceived choice when using AI decision support~\citep{DBLP:conf/iui/SwaroopBGD25}. Together, these findings highlight an implicit dichotomy: \emph{although AI-assisted knowledge work requires sustained metacognitive oversight, users often disengage precisely when such oversight is most needed.}

%%%%%%%%%%%%%%%%%%%%%%%%%%%%%%%%%%%%%%%%%%%%%%%%%%%%%%
%%%%%%%%%%%%%%%%%%%%%%%%%%%%%%%%%%%%%%%%%%%%%%%%%%%%%%
\subsection{Cognitive Forcing Functions in AI-Assisted Tasks}\label{sec:related_work.cffs}

%%%%% what are CFFs
Cognitive Forcing Functions (CFFs) are interventions designed to deliberately slow down or adjust a user's reasoning process – ``to disrupt heuristic reasoning and thus cause the person to engage in analytical thinking''~\citep{DBLP:journals/pacmhci/BucincaMG21}. Originating in safety-critical domains like medicine and aviation (e.g. checklists, ``diagnostic time-outs,'' or requiring an alternative hypothesis before accepting a diagnosis), CFFs have been adopted in AI-assisted decision-making contexts as a means of mitigating automation bias and promoting reflective reasoning.

%%%%%% CFFs and mitigation of overreliance phenomena
\looseness-1In human–AI interaction research, CFFs have been shown to reduce overreliance by compelling users to actively consider alternatives, question AI recommendations, or express reasons for acceptance or rejection. For example, Buçinca et al. demonstrated that requiring users to explicitly consider why an AI suggestion might be wrong significantly reduced overreliance compared to standard explainable AI interfaces. However, their findings also revealed a critical design tradeoff: interventions that most effectively reduced overreliance were often perceived as frustrating, complex, or cumbersome~\citep{DBLP:journals/pacmhci/BucincaMG21}. In other words, stricter forcing functions improved objective decision quality at the cost of subjective usability.  

%%%%%% design tradeoff in CFFs between cogntive engagement and cogntive load
This tradeoff has emerged as a central challenge in CFF design. While CFFs must introduce sufficient friction to prompt critical engagement, excessive disruption can increase cognitive load, reduce task performance, or reduce user trust~\citep{DBLP:journals/corr/abs-2501-16627,kazemitabaar2025exploring,DBLP:journals/pacmhci/BucincaMG21}. As a result, \emph{effective CFFs must be carefully calibrated to induce reflection without overwhelming users or disrupting the continuity of workflow}.

%%%%%%%%%%%%%%%%%%%%%%%%%%%%%%%%%%%%%
\subsection{Design Space of Cognitive Forcing Functions}\label{sec:related_work.design}

%%%%% explanations as CFFs
Prior work has explored a range of approaches for implementing CFFs in AI-assisted tasks. One prominent category focuses on explanation-driven interventions, where the design of explanations is used to prompt deeper reasoning~\citep{DBLP:journals/pacmhci/VasconcelosJGGBK23,DBLP:conf/chi/BucincaSPDG25,DBLP:journals/pacmhci/JongPTB25}. For example, partial or minimal explanations deliberately withhold aspects of the AI's rationale, forcing users to actively interpret or reconstruct missing information~\citep{DBLP:journals/pacmhci/JongPTB25}. Similarly, contrastive explanations highlight why an AI chose one option over a plausible alternative, aligning with users' natural tendency to seek comparative reasoning~\citep{DBLP:conf/chi/BucincaSPDG25}. These approaches have been shown to reduce overreliance and, in some cases, support longer-term learning without sacrificing immediate task accuracy.

%%%%% limitations of explanation based CFFs: grounding in critical thinking frameworks
However, explanation-based interventions also introduce limitations, particularly in knowledge work settings where AI outputs are lengthy, procedural, or open-ended. In such contexts, layering additional explanations onto already complex outputs risks increasing cognitive load and disrupting users' review processes. More fundamentally, many explanation-based CFFs are designed in an ad hoc manner, targeting intuitive notions of ``better explanations'' rather than being explicitly grounded in theories of critical thinking or metacognitive monitoring.

%%%%% CFFs grounded in CT/metacognitive frameworks
An alternative line of work draws more directly from cognitive science and education, using reflection prompts and process-oriented scaffolds to support reasoning. For example, Fischer et al. propose a taxonomy of Socratic questions designed to elicit critical reflection in machine-assisted decision making\citep{fischer2025taxonomy}. In educational contexts, Kazemitabaar et al. show that guided, step-by-step scaffolding, requiring learners to articulate their reasoning before receiving AI-generated code, can effectively maintain cognitive engagement and prevent passive copying~\citep{kazemitabaar2025exploring}. Notably, these interventions are grounded in well-established theories of learning and metacognition, rather than surface-level interface heuristics.

%%%%%%%% AI-plan based CFFs
Building on this perspective, AI-generated execution plans represent a promising but underexplored avenue for designing CFFs in knowledge work. Plans externalize the AI's proposed approach and procedural structure, making them a natural focus for process-oriented interventions. By centering CFFs on AI-generated plans, rather than on final outputs or post hoc explanations, designers can prompt users to engage in critical thinking skills such as argument analysis and hypothesis testing at a point where high-level reasoning about approach and assumptions is most relevant with reduced cognitive load. Despite the growing prevalence of AI-generated plans in generative AI interfaces, \emph{prior work has not systematically examined plan-based CFF designs for supporting critical evaluation and trust calibration in AI-assisted knowledge work}.

% !TEX root =  main.tex
%%%%%%%%%%%%%%%%%%%%%%%%%%%%%%%%%%%%%%%%%%%%%%%%%%%%%%%%%%
%%%%%%%%%%%%%%%%%%%%%%%%%%%%%%%%%%%%%%%%%%%%%%%%%%%%%%%%%%

\section{Developing Candidate CFFs for AI Plans}\label{sec:method}
In this section, we describe how we derived candidate cognitive forcing functions centered on AI-generated execution plans. Our goal was not to exhaustively enumerate possible CFFs, but to systematically select contrasting interventions grounded in established critical thinking skills and suited to plan evaluation in knowledge work.

%%%%%%%%%%%%%%%%%%%%%%%%%%%%%%%%
\begin{table}[t!]
\centering
\caption{\looseness-1Comparison of major critical thinking (CT) and metacognitive frameworks reviewed for CFF design. The second column indicates whether each framework is outcome- or process-oriented in its approach to cultivating critical thinking. The third column presents the key ideas behind the framework. The final column outlines their relevance to the evaluation phase of AI-assisted workflows.}
\label{tab:ct_frameworks}
\begin{tabularx}{\textwidth}{
>{\raggedright\arraybackslash}p{0.15\textwidth}  % Framework
>{\raggedright\arraybackslash}p{0.10\textwidth}  % Orientation
>{\raggedright\arraybackslash}p{0.27\textwidth}  % Focus/Key Components
>{\raggedright\arraybackslash}p{0.40\textwidth}  % Relevance to AI Monitoring
}
\toprule
\textbf{Framework} & \textbf{Orientation} & \textbf{Focus/Key Components} & \textbf{Relevance}\\
\midrule
%%%%%%%
Bloom’s Taxonomy~\citep{bloom1956taxonomy,anderson2001taxonomy} &
Outcome-oriented &
Hierarchical model of cognitive skills (remembering–creating). &
Defines broad learning outcomes but lacks procedural mechanisms for reflection.\\
\addlinespace

%%%%%%%%%%%%%
Facione’s Delphi Report~\citep{facione1990critical} &
Outcome-oriented &
Six CT skills (interpretation, analysis, evaluation, inference, explanation, self-regulation) and affective dispositions. &
Describes \emph{what} skilled thinkers do but not \emph{how} to elicit those processes during task monitoring.\\
\addlinespace

%%%%%%%%%%%%%
Paul \& Elder~\citep{paul2000miniature} &
Outcome-oriented &
Elements of thought and intellectual standards (clarity, accuracy, relevance, logic). &
Useful for evaluating quality of reasoning; less explicit on self-monitoring mechanisms.\\
\addlinespace

%%%%%%%%%%%%%
Ennis~\citep{ennis1996critical} &
Outcome-oriented &
Comprehensive taxonomy of CT abilities and dispositions. &
Covers full range of CT behaviors but highly granular for brief interventions.\\
\addlinespace

%%%%%%%%%%%%%
Winne \& Hadwin~\citep{winne2011srl} &
Process-oriented &
SRL model (COPES: Conditions, Operations, Products, Evaluations, Standards). &
Models regulation across full learning cycles, not single-task reasoning.\\
%%%%%%%%%%%%%
\addlinespace

%%%%%%%%%%%%%
\rowcolor{green!20}
Halpern~\citep{halpern1998teaching,halpern2013thought} &
Process-oriented &
Five CT skills (verbal reasoning, argument analysis, hypothesis testing, likelihood and uncertainty, decision making); plan–monitor–evaluate cycle. &
Directly aligns with reflective monitoring demands in AI-assisted reasoning.\\

\bottomrule
\end{tabularx}
\end{table}

%%%%%%%%%%%%%%%%%%%%%%%%%%%%%%%%%

\subsection{Interpreting Critical Thinking Frameworks as CFF Design Space}\label{sec:method.ct}

%%%%% Lack of formal frameowrks to ground the design of CFFs in knowledge work
\looseness-1Prior work has proposed many CFFs to mitigate overreliance on AI outputs, but these interventions are often described at the level of interface mechanics such as, asking users to justify a decision, presenting explanations or confidence cues, or delaying the display of an AI recommendation, rather than in terms of the cognitive processes they are intended to elicit. As a result, it can be difficult to compare CFFs across studies, to understand which aspects of ``critical engagement'' they target, and to reason about why one forcing function may work better than another in a given setting. This challenge becomes even more pronounced in AI-assisted knowledge work, where evaluation is not limited to a single prediction or recommendation, but may involve assessing multi-step execution plans and long-form drafts. While some recent work has begun to ground cognitive engagement techniques in established educational frameworks—for example, Kazemitabaar et al's use of Bloom's Taxonomy to structure engagement with AI-generated code in novice programming contexts~\citep{kazemitabaar2025exploring}—such approaches focus on learning-oriented outcomes and skill acquisition rather than on supporting calibrated evaluation and trust in professional knowledge work. Consequently, how critical thinking and metacognitive frameworks can inform the design of plan-centered CFFs for AI-assisted knowledge work remains an open question.

%%%%% Our apporach to design
To move beyond ad hoc design, we sought a principled way to characterize dimensions of thinking that could be induced during the review of AI-generated plans. Specifically, we aimed to (1) identify cognitive and metacognitive processes that are relevant to evaluating an AI-generated plan, (2) map these processes onto concrete, lightweight interaction prompts that can function as CFFs, and (3) select a small set of contrasting candidate CFFs suitable for controlled experimental comparison.

%%%%%%% exploration of CT and metacognitive frameworks
We therefore surveyed a set of influential frameworks from the educational, critical thinking, and metacognition literature, including Bloom's Taxonomy~\citep{bloom1956taxonomy}, Facione's Delphi Report~\citep{facione1990critical}, Paul and Elder's critical thinking model~\citep{paul1997brief,paul2000miniature}, Ennis's taxonomy of abilities and dispositions~\citep{ennis1996critical}, Halpern's critical thinking framework~\citep{halpern1998teaching,halpern2013thought}, and Winne and Hadwin's self-regulated learning (SRL) framework~\citep{winne2011srl}. These frameworks vary in the extent to which they emphasize (a) cognitive outcomes (e.g., achieving higher-order thinking) versus cognitive processes (e.g., explicit steps of monitoring and evaluation), and (b) whether they provide concrete mechanisms that can be translated into lightweight prompts based on AI plans during the review of AI output.

%%%%%%%% description of the frameworks
Table~\ref{tab:ct_frameworks} summarizes our comparison. Specifically, we found that outcome-oriented frameworks (e.g., Bloom's Taxonomy) help articulate broad goals for thinking but provide limited guidance for designing specific interventions during plan review. In contrast, process-oriented frameworks more directly support the design of CFFs because they characterize the component skills and monitoring processes involved in evaluation. Based on this analysis, we used \emph{Halpern's critical thinking framework} to operationalize which critical thinking skills a plan-based CFF should elicit, and we used metacognitive monitoring concepts (as emphasized in SRL frameworks) to justify why prompting these skills during review can support calibrated trust. This framing allowed us to treat critical thinking skills as a design space for plan-centered CFFs: rather than creating prompts in an ad-hoc manner, we selected candidate interventions that (1) target distinct critical thinking skills, (2) align naturally with the structure of execution plans (e.g., explicit steps, implicit assumptions, dependency between steps), and (3) can be implemented in a lightweight manner that fits within the flow of AI-assisted knowledge work. In the following subsections, we describe the resulting candidate CFFs and how they map onto specific critical thinking skills.

% !TEX root =  main.tex
%%%%%%%%%%%%%%%%%%%%%%%%%%%%%%%%%%%%%%%%%%%%%%%%%%%%%%%%%%
%%%%%%%%%%%%%%%%%%%%%%%%%%%%%%%%%%%%%%%%%%%%%%%%%%%%%%%%%%

\subsection{Experimental CFFs Derived from Halpern's Framework}\label{sec:method.cff}
Building on the analysis in the previous subsection (Section~\ref{sec:method.ct}), we derived candidate CFFs by operationalizing specific critical thinking skills from Halpern's framework in the context of AI-generated execution plans. Rather than attempting to cover the full breadth of critical thinking skills described by Halpern, our goal was to select a small number of contrasting skills that are (1) central to evaluating procedural approaches, (2) well supported by the structure of AI-generated execution plans, and (3) suitable for lightweight prompting during AI output review.

Halpern's framework characterizes critical thinking as the deliberate application of skills such as argument analysis, hypothesis testing, and decision evaluation, guided by metacognitive monitoring. Among these, we identified argument analysis and hypothesis testing as particularly relevant to plan evaluation. Execution plans make explicit the steps an AI proposes to take and implicitly encode assumptions about goals, constraints, and causal relationships between steps. This structure enables users to reason about whether the plan is well founded and how alternative approaches might affect outcomes, without requiring detailed inspection of the final AI-generated draft. Based on this mapping, we designed two primary plan-based CFFs, each targeting a distinct critical thinking skill from Halpern's framework, as well as a combined variant to explore potential interactions between skills. Table~\ref{tab:ct_skills_mapping} summarizes the three CFF designs and a baseline condition with no plan-based CFF, along with their correspondence to Halpern's five critical thinking skills. We note that three critical thinking skills in Halpern's framework, namely, verbal reasoning, reasoning under uncertainty, and decision making and problem solving, are engaged across all conditions. This is a consequence of the (AI-assisted) writing task structure and execution plans. More specifically, execution plans explicitly verbalize the AI's reasoning process, naturally invoking verbal reasoning. Uncertainty is inherent due to the stochastic nature of AI-generated outputs. Finally, decision making and problem solving are engaged whenever users are asked to assess, critique, or form judgments about the AI's response. As a result, these skills are not uniquely attributable to any specific CFF and are therefore marked as present across all four conditions in Table~\ref{tab:ct_skills_mapping}. We detail each CFF condition below.

%%%%%%%%%%%%%%%%%%%%%%%%%%%%%%%%
% !TEX root = main.tex
%%%%%%%%%%%%%%%%%%%%%%%%%%%%%%%%
%%%%%%%%%%%%%%%%%%%%%%%%%%%%%%%%

\begin{table}[t!]
\centering
\renewcommand{\arraystretch}{1.2}
\setlength{\tabcolsep}{4pt}

\begin{tabularx}{\textwidth}{
>{\raggedright\arraybackslash}p{0.12\textwidth}   
>{\centering\arraybackslash}p{0.14\textwidth}   % Column 2: Verbal Reasoning
>{\centering\arraybackslash}p{0.14\textwidth}   % Column 3: Argument Analysis
>{\centering\arraybackslash}p{0.14\textwidth}   % Column 4: Hypothesis Testing
>{\centering\arraybackslash}p{0.14\textwidth}   % Column 5: Likelihood & Uncertainty
>{\centering\arraybackslash}p{0.18\textwidth}   % Column 6: Decision Making
}
\hline
%%%%%%%%%%%%%%%%%%
\textbf{CFF} & 
\textbf{Verbal Reasoning} & 
\textbf{Argument Analysis} & 
\textbf{Hypothesis Testing} & 
\textbf{Likelihood \& Uncertainty} & 
\textbf{Decision Making \& Problem Solving} \\
\hline
%%%%%%%%%%%%%%%%%%
\nocff & \yestick & \cellcolor{red!20}{\notick} & \cellcolor{red!20}{\notick} & \yestick & \yestick \\ 
\hline
%%%%%%%%%%%%%%%%%%
\assumptioncff & \yestick & \cellcolor{green!20}{\yestick} & \cellcolor{red!20}{\notick} & \yestick & \yestick \\ 
\hline
%%%%%%%%%%%%%%%%%%
\whatifcff & \yestick & \cellcolor{red!20}{\notick} & \cellcolor{green!20}{\yestick} & \yestick & \yestick \\ 
\hline
%%%%%%%%%%%%%%%%%%
\bothcff & \yestick & \cellcolor{green!20}{\yestick} & \cellcolor{green!20}{\yestick} & \yestick & \yestick \\ 
\hline
\end{tabularx}

\caption{Mapping of CFFs to critical thinking (CT) skills based on Halpern’s framework~\citep{halpern1998teaching,halpern2013thought}. Each CFF variant (\nocff, \assumptioncff, \whatifcff, and \bothcff) is evaluated against five CT skill domains. A tick (\yestick) indicates the skill domain that the CFF explicitly supports, whereas a cross (\notick) indicates lack of explicit support. Three of the CT skills are supported across all conditions due to the structure of the AI-assisted tasks and execution plans. Further details of the CFFs are presented in Section~\ref{sec:method.cff}.}
\label{tab:ct_skills_mapping}
\end{table}

%%%%%%%%%%%%%%%%%%%%%%%%%%%%%%%%

%%%%%%%%%%%% description of Assumptions
\paragraph{\assumptioncff{}.} The assumption-based CFF targets argument analysis, which in Halpern's framework involves the ability to identify and evaluate the premises underlying a conclusion or proposed course of action. In the context of AI-generated plans, these premises often take the form of implicit assumptions about task requirements, source reliability, or the appropriateness of specific strategies. 

\begin{figure*}
\centering
    %% cff for task: task
    \begin{subfigure}[c]{0.49\textwidth}
       \centering
       \includegraphics[width=\textwidth]{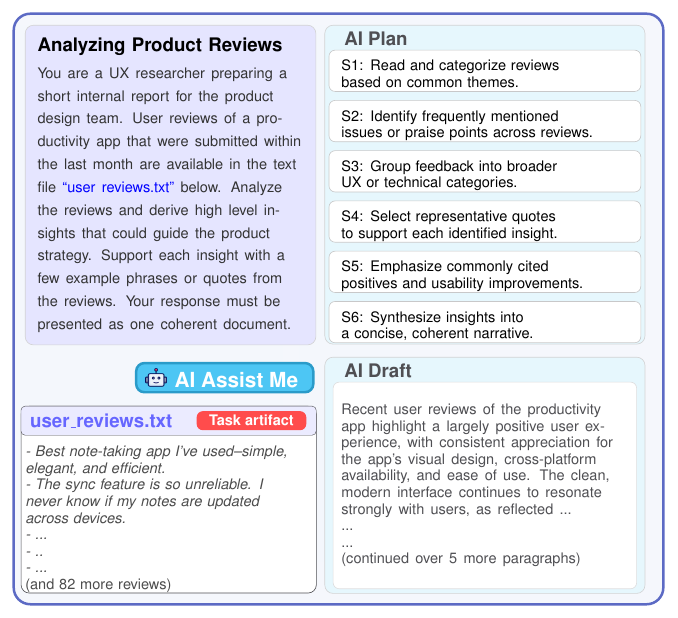}
        \caption{Knowledge work task (Task 2)}
        \label{fig3:cff.project}
    \end{subfigure}
    \hspace{1em}
    %% cff: none
    \begin{subfigure}[c]{0.45\textwidth}
        \centering
        \includegraphics[width=\textwidth]{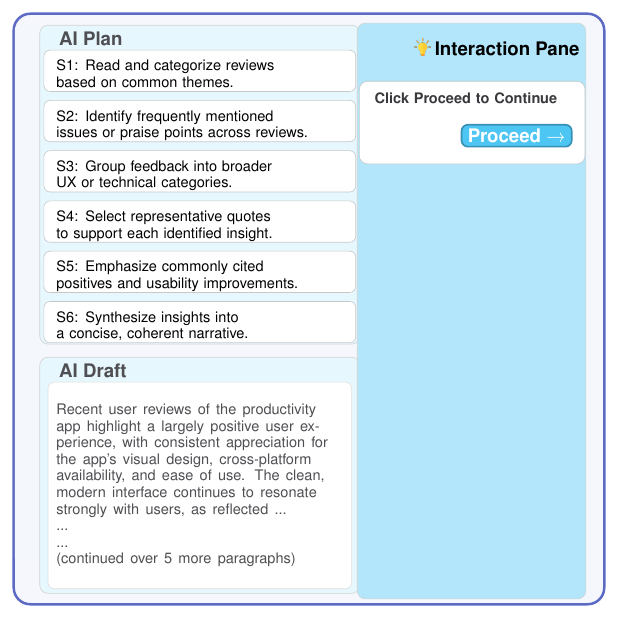}
        \caption{CFF method: \nocff{}}
        \label{fig3:cff.none}
    \end{subfigure}
    \\
   \vspace{1mm}
    %% cff: assumptions
    \begin{subfigure}[c]{0.45\textwidth}
          \centering
        \includegraphics[width=\textwidth]{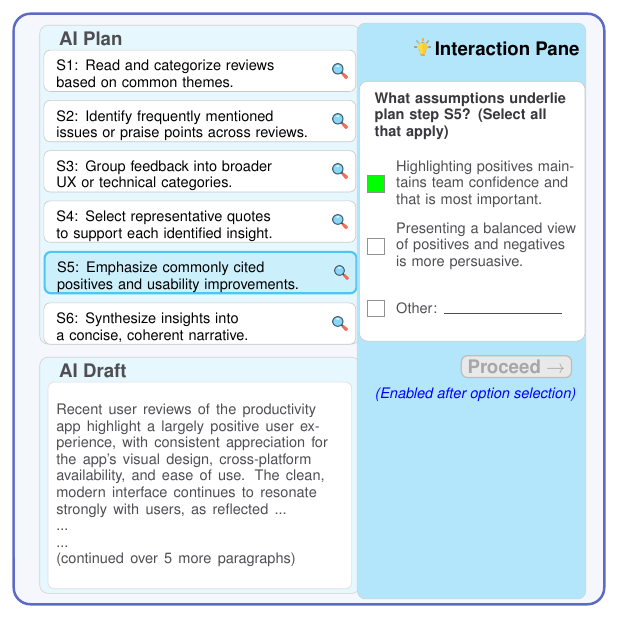}
        \caption{CFF method: \assumptioncff{}}
        \label{fig3:cff.assumption}
    \end{subfigure}
    \hspace{1em}
    %%% cff: whatif
     \begin{subfigure}[c]{0.45\textwidth}
          \centering
       \includegraphics[width=\textwidth]{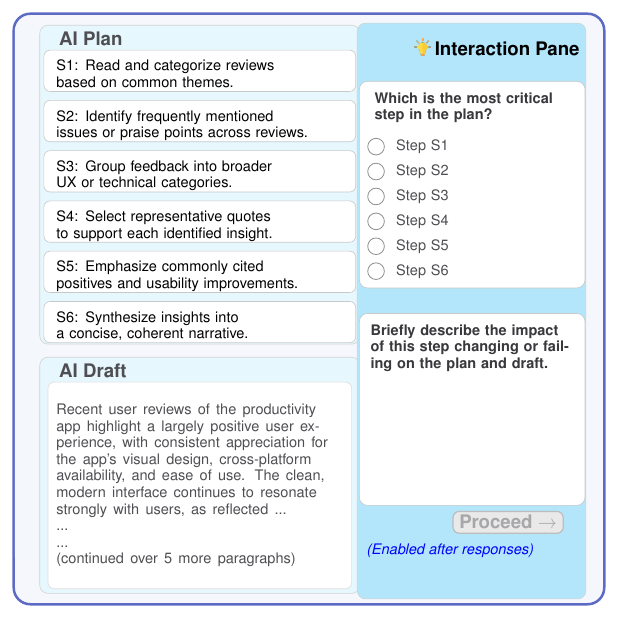}
        \caption{CFF method: \whatifcff{}}
        \label{fig3:cff.whatif}
    \end{subfigure}
   %%%%%%%%%%%%
   %\vspace{-2mm}
    \caption{\looseness-1~Illustration of the cognitive forcing function (CFF) conditions evaluated in our study, shown for knowledge work Task 2 in (a). (b) shows the \nocff{} condition, which comprises a simple ``Proceed'' button and takes users directly to the final assessment stage of the task. (c) shows the \assumptioncff{} condition, in which users are prompted to identify assumptions underlying a selected step of the AI-generated plan (illustrated here for Step 5). Users may explore multiple plan steps, but must respond to at least one prompt before proceeding to the final assessment stage. (d) shows the \whatifcff{} condition, which consists of two prompts: the first asks users to identify what they perceive as the most critical step in the plan, and the second asks them to reason about how the outcome might change if that step were to fail or be altered. In both the \assumptioncff{} and \whatifcff{} conditions, there are no correct or incorrect answers; the prompts are designed to encourage critical reflection rather than correctness. In the \bothcff{} condition, users first engage with the \assumptioncff{} prompt and then with the \whatifcff{} prompt before proceeding to the final assessment stage. Additional details are provided in Section~\ref{sec:method.cff}.
     }
    \label{fig3:cff_methods}
\end{figure*}

This CFF prompts users to identify assumptions underlying one or more steps in the AI-generated plan. By surfacing these assumptions, the intervention encourages users to reflect on whether the plan's reasoning is appropriate for the specific task context. As execution plans decompose the AI's approach into discrete steps, this prompt can be applied in a focused and lightweight manner, supporting critical evaluation without requiring users to scrutinize the full AI-generated output. More specifically, upon selecting a specific plan step, users are asked to infer what assumption the AI might have made about the project that led to that decision. Users can choose from multiple-choice options or enter their own assumptions from the AI's perspective. Although they may examine multiple steps, completion of at least one is required before proceeding to the feedback stage. Figure~\ref{fig3:cff.assumption} presents an example of this CFF for the knowledge work task shown in Figure~\ref{fig3:cff.project}.

%%%%%%%%%%%% description of WhatIf
\paragraph{\whatifcff{}.} The what-if CFF targets hypothesis testing, which in Halpern's framework involves considering alternative possibilities and reasoning about their potential consequences. This skill is central to evaluating whether a proposed approach is robust or whether alternative strategies might better serve the task goals.

This CFF prompts users to consider how the outcome might change if a key step in the AI-generated plan were altered, removed, or replaced. This framing encourages users to treat the plan as a tentative hypothesis rather than a fixed procedure, and to cognitively simulate alternative courses of action. As with the assumption-based CFF, the procedural structure of execution plans supports this form of reasoning at a high level, allowing users to engage in hypothesis testing without incurring the cognitive cost of rewriting or re-evaluating the entire draft. More specifically, users first have to identify the most critical step and are then asked to reason about what would happen if that step were altered or failed. Figure~\ref{fig3:cff.whatif} presents an example of this CFF for the knowledge work task shown in Figure~\ref{fig3:cff.project}.

%%%%%%%%%%%%% description of Both
\paragraph{\bothcff{}.} To explore whether engaging multiple critical thinking skills in sequence would yield additional benefits, we also designed a combined CFF that presents the assumption-based prompt (\assumptioncff{}) followed by the what-if prompt (\whatifcff{}). Presenting the prompts in this sequence mirrors common analytical workflows in which assumptions are first presented and assessed before considering alternative courses of action. This condition allows us to examine whether prompting both argument analysis and hypothesis testing leads to deeper critical engagement, or whether stacking CFFs introduces diminishing returns due to increased cognitive load. Figure~\ref{fig3:cff.assumption} and \ref{fig3:cff.whatif} when presented one after the other, presents an example of this CFF for the knowledge work task shown in Figure~\ref{fig3:cff.project}.

%%%%%%%%%%%% description of None
\paragraph{\nocff{}.} In the control condition (\nocff{}), after reviewing the AI-generated plan and draft, users simply click a ``Proceed'' button before submitting their evaluation. Figure~\ref{fig3:cff.none} presents an example of this CFF for the knowledge work task shown in Figure~\ref{fig3:cff.project}.

%%%%%%%%%%%%%%%%%%%%%%%%%%%%%%%%%%%%%%%%%%%%%%%%%%%%%%%%
% !TEX root = main.tex
%%%%%%%%%%%%%%%%%%%%%%%%%%%%%%%%
%%%%%%%%%%%%%%%%%%%%%%%%%%%%%%%%

%%%%%%%%%%%%%%%%%%%%%%%%%%%%%%%%%%%%%%%%%%%%%%%%%%%%%%%%

% !TEX root =  main.tex
%%%%%%%%%%%%%%%%%%%%%%%%%%%%%%%%%%%%%%%%%%%%%%%%%%%%%%%%%%
%%%%%%%%%%%%%%%%%%%%%%%%%%%%%%%%%%%%%%%%%%%%%%%%%%%%%%%%%%

\section{Experiment and Interview Design}
\label{sec:setup}

%%%%%%%%%%%%%%%%%%%%%
% !TEX root = main.tex
%%%%%%%%%%%%%%%%%%%%%%%%%%%%%%%%
%%%%%%%%%%%%%%%%%%%%%%%%%%%%%%%%

\begin{figure*}
\centering
%%%%%%%%%%%%%%%
    \begin{subfigure}[c]{0.95\textwidth}
        \includegraphics[width=\textwidth]{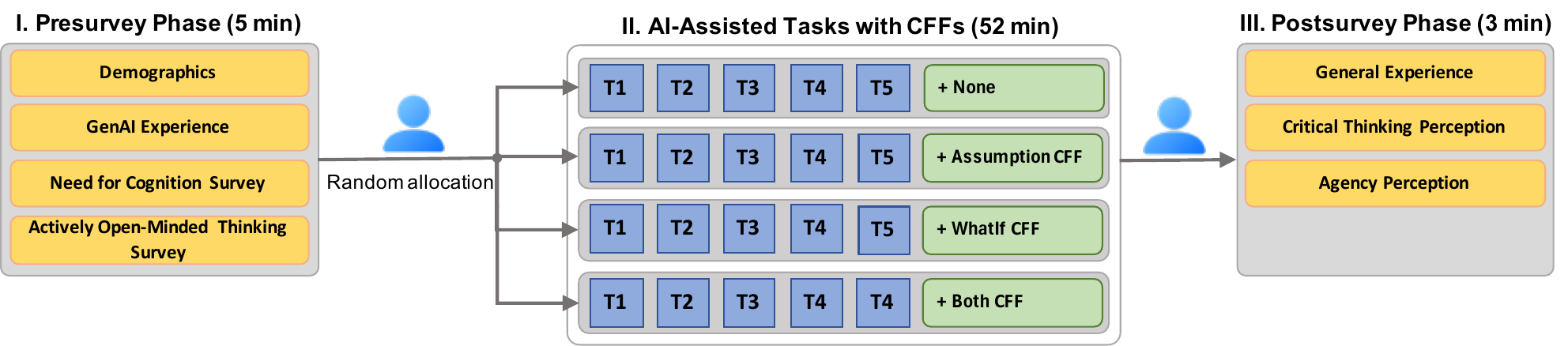}
    \caption{Online unmoderated comparative study pipeline. Participants first completed the presurvey and were then randomly assigned to one of four cognitive forcing function (CFF) conditions. They subsequently engaged in five AI-assisted knowledge work tasks (T1–T5) in sequence under their assigned CFF condition, and finally completed the postsurvey.}
    \label{fig:study_design_quant}
    \end{subfigure}
    \\
%%%%%%%%%%%%%%%
    \begin{subfigure}[c]{0.95\textwidth}
        \includegraphics[width=\textwidth]{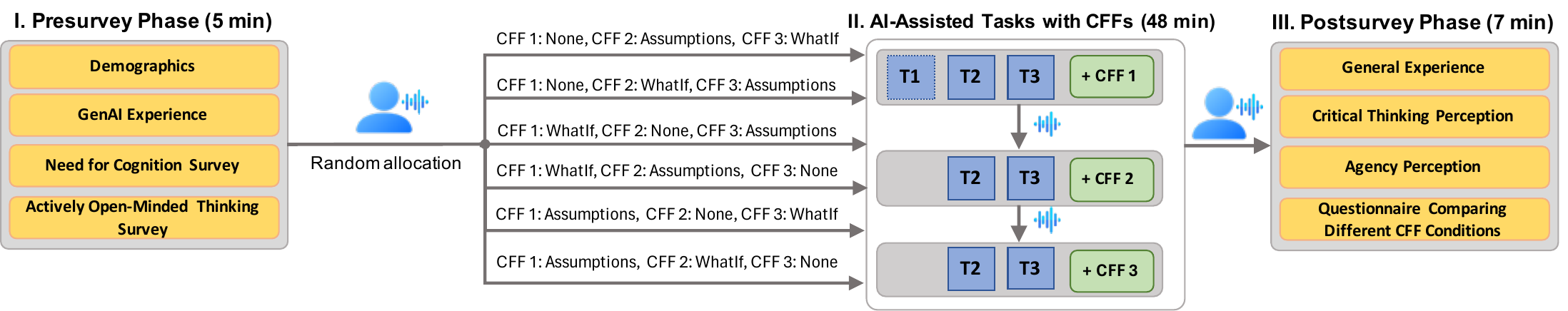}
    \caption{Interview study pipeline. Participants first completed the presurvey and then engaged in three AI-assisted knowledge work tasks (T1–T3). Participants were assigned to one of six cognitive forcing function (CFF) sequences and completed tasks T1–T3 under the first CFF in the sequence, with T1 serving as a warm-up task. They then revisited tasks T2 and T3 under the second and third CFF conditions in the sequence. Throughout task engagement, participants verbalized their thought processes using a think-aloud protocol. After completing tasks under all three CFF conditions (\nocff{}, \assumptioncff{}, \whatifcff{}), participants completed a postsurvey. In addition to postsurvey questions similar to those in the online unmoderated study, participants also answered five interview-style questions comparing their experiences across the three CFF conditions.}
    \label{fig:study_design_qual}
    \end{subfigure}
    \\
%%%%%%%%%%%%%%%
    \begin{subfigure}[c]{0.95\textwidth}
        \includegraphics[width=\textwidth]{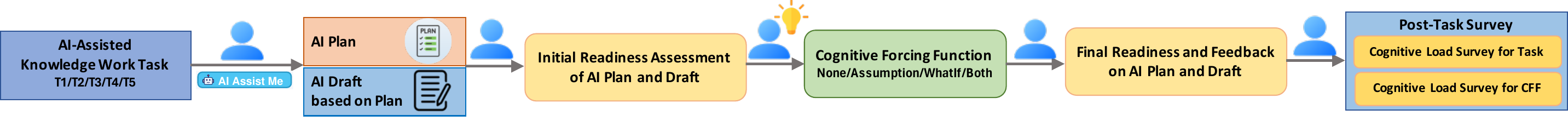}
    \caption{Interaction flow within a single AI-assisted knowledge work task. All knowledge work tasks (T1–T5) followed the same interaction flow. Participants were first presented with the task description and its associated task artifact, and were then prompted to click the ``AI Assist Me'' button. This revealed the AI-generated plan, followed by the corresponding draft. Participants next provided an initial assessment of the readiness of the AI output for the task. They then engaged with the cognitive forcing function (CFF) corresponding to their assigned condition. After completing the CFF, participants reported their final readiness assessment and provided feedback on the AI output. Finally, participants completed a post-task survey including questions on perceived cognitive load of the task and specific CFF condition.}
    \label{fig:study_design_task}
    \end{subfigure}
%%%%%%%%%%%%%%%    
\caption{\looseness-1Overview of study procedures for the online unmoderated study and interview conditions. Across both study variants, participants first completed a presurvey, then engaged in a sequence of AI-assisted knowledge work tasks (T1–T5) with a specific CFF condition, and finally completed a postsurvey. The presurvey included questions on participant demographics, experience with generative AI tools, need for cognition, and actively open-minded thinking. The postsurvey included questions on the participants' overall study experience, their perceived critical thinking while engaging with different cognitive forcing function (CFF) conditions, and their sense of agency when completing the tasks. Each study session lasted approximately one hour, with the breakdown of time across phases shown in the corresponding figures. Panel (a) shows the detailed pipeline for the online unmoderated study, panel (b) shows the pipeline for the interview study, and panel (c) shows the interaction flow within a single AI-assisted knowledge work task.}
\Description{}
\label{fig:study_design}
\end{figure*}
%%%%%%%%%%%%%%%%%%%

\label{sec:setup.studydesign}
\looseness-1To evaluate the CFFs, we conducted two studies. The first was a large-scale ($n=214$) unmoderated comparative online experiment, and the second was a moderated interview ($n=12$) with a separate participant sample. Both studies followed a three-phase structure: a presurvey phase, an AI-assisted task completion phase, and a postsurvey phase. Figure~\ref{fig:study_design_quant} presents the phases of the online unmoderated study, and Figure~\ref{fig:study_design_qual} presents the phases of the interview study.

\paragraph{Presurvey.} In the presurvey phase, we collected information on participants' demographics, familiarity with AI tools, and individual cognitive dispositions. The latter was measured using two validated psychometric instruments: a $6$-item short form of the \emph{Need for Cognition scale}~\citep{lins2020very} with each item rated on a scale of 1--5, and the $13$-item \emph{Actively Open-Minded Thinking} scale~\citep{stanovich2023actively} with each item rated on a scale of 1--6. These measures helped contextualize participants' baseline tendencies toward critical thinking and reflective judgment. The full set of presurvey questions are presented in Appendix D.

\paragraph{AI-assisted task completion.} Following the presurvey, participants in the comparative online experiment were asked to complete five AI-assisted knowledge work tasks. All tasks were based on the writing domain, chosen for their accessibility and representativeness of common AI-supported tasks~\citep{brachman2025current}. Each task involved an AI-generated plan and corresponding draft. Tasks covered a range of task types including synthesis, analysis, summarization, and restructuring, and were designed to capture realistic variations in the cognitive demands of AI-assisted writing tasks. To evaluate participants' calibration and reliance on AI responses, selected plans and drafts were seeded with synthetic ``bugs'' (logical or factual inconsistencies, detailed in Section~\ref{sec:setup.tasks}). After completing each task, participants were prompted to provide written feedback on the AI-generated plan and draft, including any issues they identified. To encourage thoughtful engagement, participants were informed that \emph{high-quality feedback would be eligible for a performance-based bonus}. Participants were randomly assigned to one of four CFF conditions (\nocff{}, \assumptioncff{}, \whatifcff{}, or \bothcff{}), which were automatically triggered during the task. After each task, participants completed a brief cognitive load questionnaire based on the validated \emph{NASA Task Load Index (NASA-TLX)}~\citep{hart1988development}, assessing both the task and the associated CFF. The five tasks used in the studies are described in Section~\ref{sec:setup.tasks}. Further details of each task, exact prompts used for \assumptioncff{} and \whatifcff{} conditions, and the post-task survey questions are presented in Appendix E.

In the interviews, a separate set of participants completed three of the five tasks (Task 1, Task 2, and Task 3). The first served to familiarise participants with the interface. The remaining two were used for the study where Task 2 included a synthetic error, and Task 3 did not -- in order to cover both cases. Each participant experienced three CFF conditions (\nocff{}, \assumptioncff{}, and \whatifcff{}). For the first CFF they experienced, they completed Tasks 1,2, and 3. For the second and third CFF, they completed Tasks 2 and 3 again. Across all the participants in the interview study, we counterbalanced the order in which the three CFFs were presented to mitigate sequence effects. During a task, participants were asked to think aloud, verbalizing their reasoning and decision-making process as they completed the tasks. Figure~\ref{fig:study_design_task} illustrates the stages of an AI-assisted task incorporating the CFF.

\paragraph{Postsurvey.} After completing all assigned tasks, participants completed a post-survey evaluating their overall experience with the AI-assisted workflow and the CFFs. The questionnaire included items assessing perceived support for critical thinking, agency, and confidence in evaluating AI-generated outputs. As existing validated instruments do not directly capture users' subjective experiences with plan-centered CFFs in AI-assisted knowledge work, these items were developed specifically for this study. For participants in the interview study, the post-survey included a comparative analysis of the three CFF types. The full set of these comparative questions are presented in Appendix G. Participants were asked which CFF they found most helpful, which they perceived as hindering, which best supported critical reflection on AI responses, and which they would prefer to see integrated into everyday AI-assisted tools. The full set of postsurvey questions are presented in Appendix F.

%%%%%%%%%%%%%%%%%%%%%%%%%%
% !TEX root = main.tex
%%%%%%%%%%%%%%%%%%%%%%%%%%%%%%%%
%%%%%%%%%%%%%%%%%%%%%%%%%%%%%%%%

\begin{table}[t!]
\centering
    \begin{tabularx}{\textwidth}{
>{\centering\arraybackslash}p{0.10\textwidth}   
>{\raggedright\arraybackslash}p{0.24\textwidth} 
>{\raggedright\arraybackslash}p{0.10\textwidth}   
>{\raggedright\arraybackslash}p{0.10\textwidth}   
>{\raggedright\arraybackslash}p{0.18\textwidth} 
>{\raggedright\arraybackslash}p{0.12\textwidth}    
}
    \hline
    %%%%%%%
    \textbf{Task} & \textbf{Description} & \textbf{Primary Theme} & \textbf{Subtheme} & \textbf{Information Processing Demand} &\textbf{Synthetic Error Type} \\ 
    \hline
    %%%%%%%
    Task $1$ (T1) & \emph{Notice synthesis}: Warm-up task involving the creation of a library notice for patrons based on a short input document & Information & Summarize & Artifact: $22$ words,\newline AI draft: $45$ words & None \\ 
    \hline
    %%%%%%%
    Task $2$ (T2) & \emph{Analyzing product reviews}~\citep{DBLP:conf/emnlp/BrazinskasLT21}: Analyzing user feedback on a productivity app to identify key themes, usability issues, and design implications & Information & Analyze & Artifact: $971$ words, AI draft: $521$ words & Incorrect plan step \\ 
    \hline
    %%%%%%%
    Task $3$ (T3) & \emph{Persuasive essay synthesis}~\citep{anthropic2024persuasiveness,singh2024measuring}: Composition of an argumentative essay on a contemporary policy issue using an evidence-based source document & Create & Artifact & Artifact: $944$ words, AI draft: $651$ words & None \\ 
    \hline
    %%%%%%%
    Task $4$ (T4) & \emph{Restructuring meeting minutes}~\citep{huetal2023meetingbank}: Reorganization and refinement of city-council meeting minutes to improve clarity, coherence, and readability. & Advice & Improve & Artifact: $592$ words, AI draft: $510$ words & Missing plan step \\ 
    \hline
    %%%%%%%
    Task $5$ (T5) & \emph{News article summarization}~\citep{seeetal2017get}: Generation of a concise, curator-focused news brief from a long-form article & Information & Summarize & Artifact: $465$ words, AI draft: $118$ & Overplanning with irrelevant plan steps \\ 
    \hline
    %%%%%%%
    \end{tabularx}
\caption{Overview of the five knowledge work tasks in the writing domain used in our study. Columns three and four classify each task according to the thematic taxonomy proposed by Brachman et al.~\citep{brachman2025current}, focusing on themes whose anticipated future use of large language models (LLMs) exceeds their current adoption. Column five presents the word count of the task artifact and AI draft serving as a proxy for the information processing demands imposed on users and motivating the use of LLM assistance. Note that Task 1 has the lowest word count for both the artifact and the draft, as it was designed primarily as a warm-up activity to familiarize participants with the study platform and task structure. For Tasks~2–5, the AI-generated execution plan consisted of six steps, whereas the plan for Task~1 consisted of four steps. Finally, in column six we present the synthetic bugs induced in the AI generated plan (based on which the draft was created) for each task. We restricted the bugs within the AI-generated plans only and avoided inducing inconsistencies between the plan and the corresponding draft. Figure~\ref{fig3:cff.project} illustrates Task 2. Detailed task descriptions, artifacts, and AI plan and draft are presented in Appendix E.4.} 
\label{tab:writing_projects}
\end{table}
%%%%%%%%%%%%%%%%%%%%%%%%%%

% !TEX root =  main.tex
%%%%%%%%%%%%%%%%%%%%%%%%%%%%%%%%%%%%%%%%%%%%%%%%%%%%%%%%%%
%%%%%%%%%%%%%%%%%%%%%%%%%%%%%%%%%%%%%%%%%%%%%%%%%%%%%%%%%%

\subsection{Tasks and Synthetic Errors}\label{sec:setup.tasks}
\looseness-1Table~\ref{tab:writing_projects} presents the five knowledge work tasks used in the user study, along with the synthetic bugs introduced in the AI plans.

\paragraph{Task description.} All tasks were based on the writing domain, representing tasks for which knowledge workers frequently seek AI assistance. Specifically, we selected task themes where the perceived future use of AI exceeds its current use, as reported by Brachman et~al.~\citep{brachman2025current}. The tasks covered the following task types: synthesis, analysis, summarization, and restructuring. To ensure external validity, the tasks were constructed using materials adapted from real-world datasets, such as the CNN/DailyMail news corpus~\citep{seeetal2017get}, the MeetingBank dataset~\citep{DBLP:conf/emnlp/BrazinskasLT21}, and SelSum product review dataset~\citep{DBLP:conf/emnlp/BrazinskasLT21}. The first task was intentionally kept simple to help participants familiarize themselves with the interface before proceeding to more demanding tasks.

Each task included a task artifact that supplemented the task description. These artifacts were intentionally designed to be information-rich and occasionally lengthy, for example, unstructured meeting transcripts, lists of factual statements, or long-form articles. Similarly, the required deliverables were substantial, such as an essay or extended report, reflecting authentic knowledge work outputs. 

During each task, participants were required to use the AI tool by clicking the ``AI Assist Me'' button which triggered the generation of the AI-execution plan followed by a draft based on that plan. Participants did not complete the tasks manually without AI assistance; this was a deliberate design choice to ensure that all participants evaluated the same AI-generated plans and drafts, enabling us to systematically examine how different cognitive forcing functions influenced reliance on and evaluation of AI outputs. An example is shown in Figure~\ref{fig3:cff.project}, which illustrates Task~2 along with its AI-generated plan and draft. After the AI response was presented, participants were asked two preliminary questions, one assessing their understanding of the AI-generated plan and another assessing how ready they thought the plan and draft were. Following these, the assigned CFF was triggered, after which participants were again asked about readiness, provided written feedback to the AI tool, and rated their confidence in the AI-generated plan and draft before completing the post task survey questions on the cognitive load of the task and the CFF. Figure~\ref{fig:study_design_task} presents the stages within a knowledge-work task.
% \asnote{This suggests that they had the option not to click it, and may have done the task entirely manually, but I don't think any participants did this, right? If not, we should rewrite this to make clearer that this step was always taken.} 

\paragraph{Error categories.} To evaluate users' ability to detect errors and check for underreliance and overreliance, we introduced synthetic bugs into the AI-generated plans. We maintained consistency between each plan and its corresponding draft, that is, the draft is consistent with the plan; an error in the plan is carried over into the draft, and if the plan is error-free, so is the draft. We note that in production systems, errors might be introduced (or removed) in the draft during the plan execution step; we deliberately avoided evaluating this scenario to limit the scope and complexity of the errors experienced by participants in the experiment.

We introduced errors of the following categories:
\begin{table}[H]  
\centering   
\begin{tabular}{ll}  
\toprule  
\textbf{Error Category} & \textbf{Description} \\  
\midrule  
None & No error in the plan \\  
Missing step & Omission of an essential plan step \\  
Over-planning & Inclusion of unnecessary or irrelevant plan steps \\  
Incorrect step & A plan step that was conceptually inappropriate or logically flawed \\  
Misordered steps & Plan steps presented in an illogical or inefficient sequence \\  
\bottomrule  
\end{tabular}  
% \caption{Error Categories}  
\end{table}  

These error categories were selected to reflect common and well-documented failure models of GenAI systems when producing structured outputs. Prior work has shown that LLMs frequently omit critical steps in multi-step reasoning~\citep{DBLP:conf/nips/DziriLSLJLWWB0H23}, introduce hallucinated content~\citep{DBLP:journals/csur/JiLFYSXIBMF23}, generate conceptually incorrect reasoning steps~\citep{DBLP:conf/acl/TyenMCCM24,DBLP:conf/nips/ValmeekamMSK23,DBLP:conf/fat/BenderGMS21}, or produce planning steps that might be misordered or inefficient~\citep{DBLP:conf/nips/ValmeekamMSK23,DBLP:conf/fat/BenderGMS21}. Although no single prior work proposes a comprehensive taxonomy of plan-level errors, these categories align with recurring patterns reported in studies of hallucination, or reasoning errors in LLM-generated content.

Errors were designed to be contextually plausible within each task to preserve ecological validity. More specifically, rather than inserting overt or trivial mistakes, we constructed errors that could reasonably arise from misinterpretations of task requirements, overgeneralization, or incomplete reasoning by the AI, thereby requiring participants to engage in meaningful evaluation of the plan. Two of the five tasks (Tasks~1 and~3) did not contain any synthetic errors, while the remaining three each contained exactly one error, as indicated in the ``Synthetic Error Type'' column of Table~\ref{tab:writing_projects}. The AI-generated plan and draft were fixed prior to the study to ensure consistency across participants. For the interview study, Tasks~1,~2, and~3 were used—Task~1 serving as a warm-up, with one of the remaining tasks containing a synthetic bug and the other not. This design allowed for comparative evaluation of user responses across both error-present and error-free conditions. We note as a limitation that different tasks contained different types of errors. Our study was not designed to compare overreliance across specific error categories, and thus we do not make claims about differential detectability or impact of particular error types. While we attempted to balance the perceived difficulty of errors across tasks, it is possible that error type influenced participant behavior in ways we could not fully control.

\paragraph{Implementation details of all tasks and CFF prompts.} Full details of all five knowledge work tasks, including task descriptions and associated artifacts, are provided in Appendix E.4. The prompts used for the \assumptioncff{} and \whatifcff{} conditions are also presented for each task. For each task, the AI-generated execution plan and corresponding draft were produced using GPT-4o and subsequently manually inspected. For tasks requiring synthetic errors, the prompt used to generate the plan included an explicit instruction specifying the type of error to be induced. All participants in both studies were shown the same plan and draft for each task; specifically, AI outputs were not generated dynamically during participant interaction. For the \assumptioncff{}, questions corresponding to each plan step were generated using GPT-4o, based on a prompt instructing the model to articulate the assumptions underlying that step along with one distractor option. In contrast, the \whatifcff{} prompts were template-based and not generated by GPT-4o.

%%%%%%%%%%%%%%%%%%%%%%%%%%%%%%%%%%%%%%%%%%%%%%%%%%%%%%%%
% !TEX root =  main.tex
%%%%%%%%%%%%%%%%%%%%%%%%%%%%%%%%%%%%%%%%%%%%%%%%%%%%%%%%%%
%%%%%%%%%%%%%%%%%%%%%%%%%%%%%%%%%%%%%%%%%%%%%%%%%%%%%%%%%%

\begin{figure*}
    %%%% nfc
    \begin{subfigure}[c]{0.44\textwidth}
       \centering
       \includegraphics[width=\textwidth]{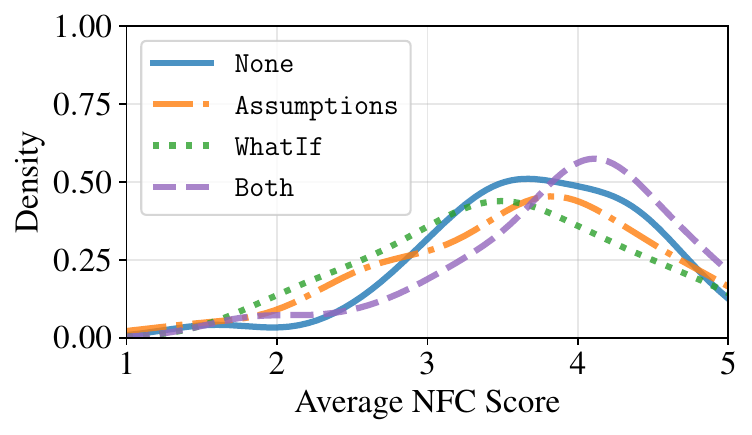}
        \caption{Need for Cognition Scores (NFC)}
        \label{fig:presurvey.nfc}
    \end{subfigure}
    \hspace{3em}
    %%%% aot
    \begin{subfigure}[c]{0.44\textwidth}
       \centering
       \includegraphics[width=\textwidth]{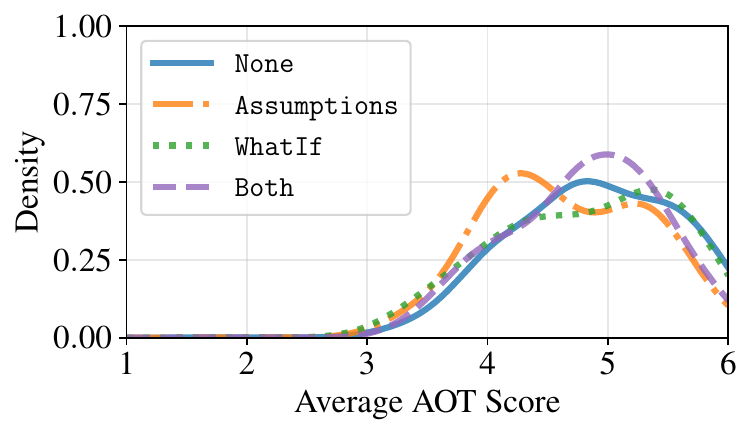}
        \caption{Actively Open Minded Thinking Scores (AOT)}
        \label{fig:presurvey.aot}
    \end{subfigure}
    \\
     %%%% genAi usage
    \begin{subfigure}[c]{0.44\textwidth}
       \centering
       \includegraphics[width=\textwidth]{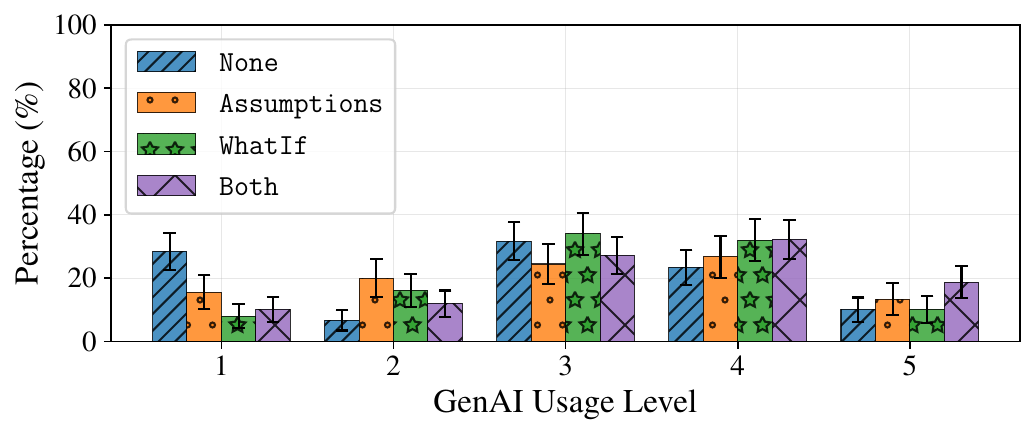}
        \caption{Usage of GenAI tools}
        \label{fig:presurvey.genaiusage}
    \end{subfigure}
    \hspace{3em}
     %%%% genAi tru
    %%%% genAi trust
    \begin{subfigure}[c]{0.44\textwidth}
       \centering
       \includegraphics[width=\textwidth]{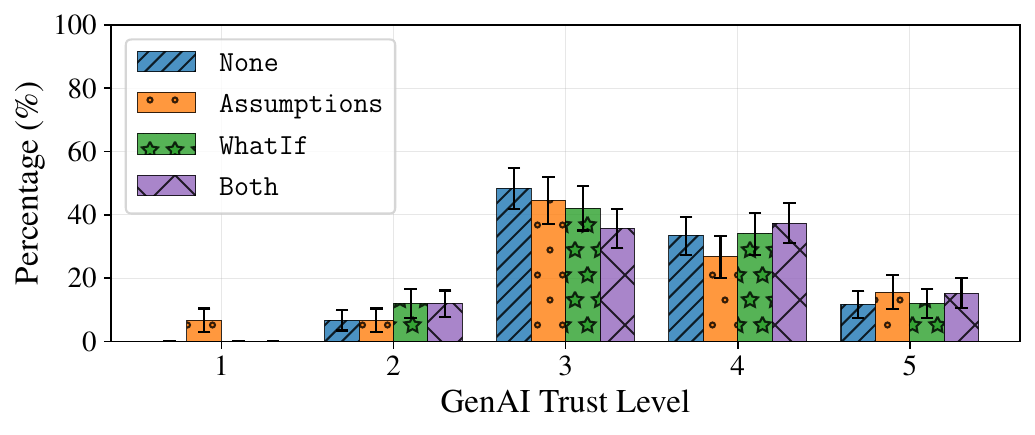}
        \caption{Trust in GenAI tools}
        \label{fig:presurvey.genaitrust}
    \end{subfigure}
    %%%%%%%%%%%%%%
    \caption{Distribution of participants' need for cognition scores, thinking disposition measured via the actively open-minded thinking scale, and their self-reported usage of and trust in GenAI tools collected via the presurvey.}
    \Description{}
    \label{fig:presurvey}
\end{figure*}
%%%%%%%%%%%%%%%%%%%%%%%%%%%%%%%%%%%%%%%%%%%%%%%%%%%%%%%%

% !TEX root =  main.tex
%%%%%%%%%%%%%%%%%%%%%%%%%%%%%%%%%%%%%%%%%%%%%%%%%%%%%%%%%%
%%%%%%%%%%%%%%%%%%%%%%%%%%%%%%%%%%%%%%%%%%%%%%%%%%%%%%%%%%

\subsection{Study Logistics and Participant Demographics}\label{sec:setup.demographics}
We obtained an IRB approval before conducting the user study from the Ethics Committee of \emph{Anonymized Institution}. Data collection was conducted between July and August~2025 through two complementary user studies. The first was a large-scale comparative online experiment designed to evaluate the effectiveness of the CFFs, while the second comprised a smaller set of moderated interviews aimed at gathering qualitative insights into participants' experiences and reasoning processes when using the CFFs. Participants for both studies were recruited through the \emph{Prolific} platform.

\paragraph{Logistics for  study.} Participants in the comparative online study were compensated at a rate of GBP \pounds12.50 per hour. Participation was fully anonymous and each study session lasted approximately one hour. Each participant was randomly assigned to one of the four CFF conditions: \nocff{}, \assumptioncff{}, \whatifcff{}, or \bothcff{}. The study was administered through a custom-designed web application that integrated all three phases of the experiment—the presurvey, the AI-assisted knowledge work projects with their corresponding CFF interventions, and the post-task survey. This setup enabled large-scale data collection under consistent experimental conditions while minimizing researcher interference.

\paragraph{Logistics for interviews.} Participants in the moderated study were compensated at a rate of GBP \pounds14 per hour. Each study session lasted one hour. They interacted with the same web application but completed their tasks while being interviewed by a researcher via Microsoft Teams. All identifiable data were removed prior to analysis. The interviews provided qualitative insights into how participants engaged with and reflected upon the different CFFs during the task. Each participant was exposed to all three CFFs—\nocff{}, \assumptioncff{}, and \whatifcff{}—presented in randomized order. They completed three knowledge work projects (Tasks~1–3), with Task~1 serving as a warm-up and Tasks~2 and~3 completed under different CFF conditions.

%%%%%%%%%%%%%%%%%%%%%%%%%
% !TEX root =  main.tex
%%%%%%%%%%%%%%%%%%%%%%%%%%%%%%%%%%%%%%%%%%%%%%%%%%%%%%%%%%
%%%%%%%%%%%%%%%%%%%%%%%%%%%%%%%%%%%%%%%%%%%%%%%%%%%%%%%%%%

\begin{figure*}
\centering
%%%%%%%
    \begin{subfigure}[c]{0.48\textwidth}
        \includegraphics[width=\textwidth]{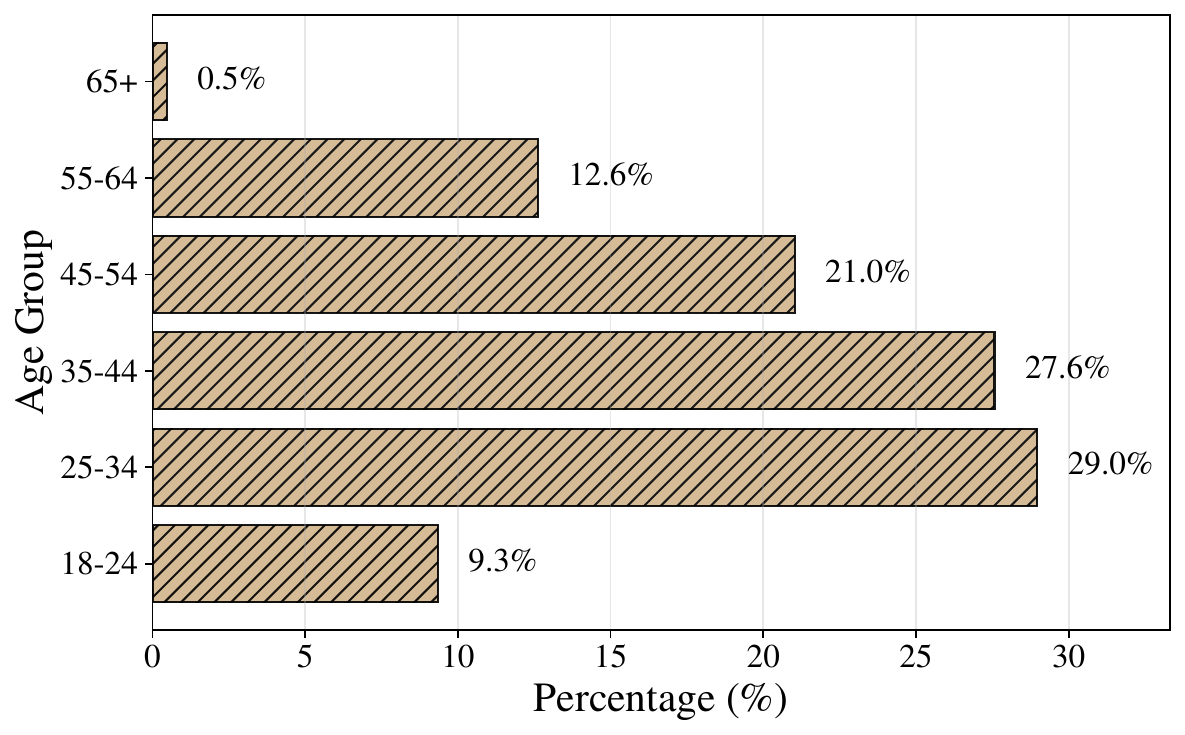}
    \caption{Age distribution}
    \label{fig:presurvey_age_industry.age}
    \end{subfigure}
%%%%%%%
\hspace{1em}
    \begin{subfigure}[c]{0.48\textwidth}
        \includegraphics[width=\textwidth]{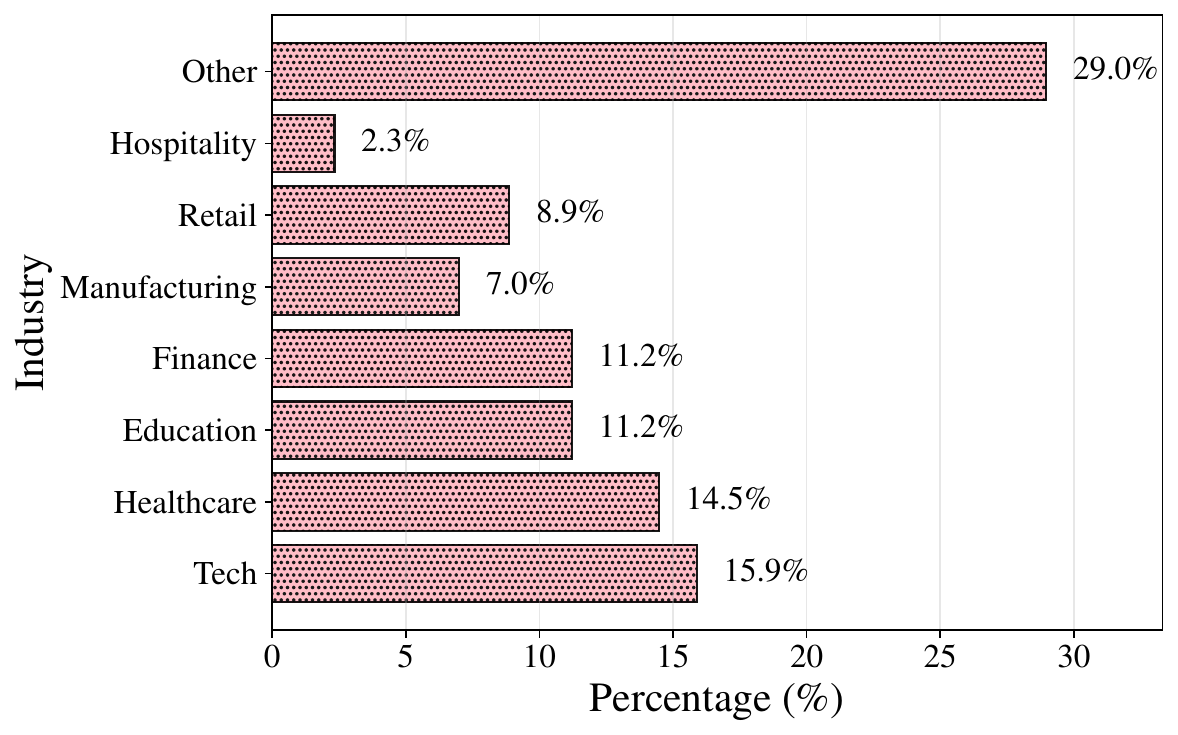}
    \caption{Work industry distribution} 
    \label{fig:presurvey_age_industry.industry}
    \end{subfigure}
\caption{Demographic distribution of participants by age and work industry in the online comparative experiment.}
\Description{}
\label{fig:presurvey_age_industry}
\end{figure*}
%%%%%%%%%%%%%%%%%%%%%%%%

\paragraph{Demographics for comparative online experiment.} A total of $214$ participants took part in the comparative online experiment. Participants were distributed across the four CFF conditions as follows: $60$ in \nocff{}, $45$ in \assumptioncff{}, $50$ in \whatifcff{}, and $59$ in \bothcff{}. Participants were assigned to conditions in a round-robin fashion. Minor differences in final group sizes arose due to participant attrition prior to completion. The age distribution of participants is presented in Figure~\ref{fig:presurvey_age_industry.age}. Gender distribution was approximately balanced, with $47.66\%$ identifying as male, $51.87\%$ as female, and $0.47\%$ preferring not to disclose their gender. In terms of country of residence, $71.03\%$ of participants were based in the UK and $28.97\%$ in the USA. The distribution of participants by their work industry is presented in Figure~\ref{fig:presurvey_age_industry.industry}. The majority of participants selected ``Other'' as their work industry, which included a diverse range of categories such as entertainment, creative arts, and unemployment. 

We collected information on participants' use of generative AI (GenAI) tools and their initial trust in these tools, measured on a five-point Likert scale ($1$ = low, $5$ = high). Figures~\ref{fig:presurvey.genaiusage} and~\ref{fig:presurvey.genaitrust} show the distribution of responses across CFF groups. A \emph{Kruskal–Wallis} test~\citep{kruskal1952use} indicated no significant differences between the four CFF groups for either GenAI tool usage ($ H = 6.01, p = 0.11$) or trust ($H =  0.80, p = 0.85$).

Finally, we collected participants' cognitive disposition measures: the Need for Cognition (NFC) score, using the six-item NFC scale~\citep{lins2020very} with each item rated on a scale of 1--5, and the Actively Open-Minded Thinking (AOT) score, using the $13$-item AOT scale~\citep{stanovich2023actively} with each item rated on a scale of 1--6. Figures~\ref{fig:presurvey.nfc} and~\ref{fig:presurvey.aot} present the distribution of these scores across CFF groups. There was no significant difference across CFF groups for AOT ($H = 3.54, p = 0.32$), or NFC ($H = 7.82, p = 0.05$).

\paragraph{Demographics for interview.} A total of $12$ participants took part in the qualitative interview study. Each participant completed a warm-up project (Task~1) followed by two additional tasks (Tasks~2 and~3). Tasks~2 and~3 were completed under three CFF conditions—\nocff{}, \assumptioncff{}, and \whatifcff{}—with all six possible CFF order combinations represented, each tested by at least two participants to mitigate order effects. Gender distribution of the participants was as follows: $5$ participants ($41.7\%$) identified as male and $7$ participants ($58.3\%$) identified as female. In terms of country of residence, majority ($91.67\%$) were based in the UK and only one participant ($8.34\%$) was based in the USA. Participants represented a diverse range of professional backgrounds, including technology, arts, teaching, healthcare, hospitality, and finance. Participants also reported their use of generative AI tools and their level of trust in these tools on a five-point Likert scale ($1$ = low, $5$ = high). For generative AI usage, $58.33\%$ of participants rated $3$ or higher, while $91.67\%$ rated $3$ or higher for trust. The average Need for Cognition (NFC) score among participants was $3.51$ out of $5$, and the average Actively Open-Minded Thinking (AOT) score was $4.67$ out of $6$.

% !TEX root =  main.tex
%%%%%%%%%%%%%%%%%%%%%%%%%%%%%%%%%%%%%%%%%%%%%%%%%%%%%%%%%%
%%%%%%%%%%%%%%%%%%%%%%%%%%%%%%%%%%%%%%%%%%%%%%%%%%%%%%%%%%

\subsection{Analysis}\label{sec:setup.analysis}

%%%%%%%%%%%%%%%%%%%%%%%%
\paragraph{RQ1: Effect of CFFs on revision of initial readiness of AI response.} 
To evaluate how different CFFs influenced the likelihood that participants would review their initial readiness assessment of the AI response, we analyzed the measure ``Change in Readiness Rating''. Readiness captures participants' overall judgment of whether an AI response is suitable for use, and revising this judgment after engaging with a CFF reflects recalibration of trust prompted by structured critical reflection rather than reliance on first impressions alone. We therefore treat changes in readiness following the CFF as a behavioral indicator of how effectively the intervention engages participants in reviewing and re-evaluating the AI-generated plan and draft.

Participants rated the readiness of the AI-generated plan and draft on a five-point Likert scale ($1$: not ready at all, $5$: completely ready, with an additional ``I don't know'' option) before and after completing the CFF. A binary change rate variable was calculated for each participant ($1$ = changed rating, $0$ = unchanged). We report the percentage of participants who changed their assessment per CFF condition and per task. Statistical differences between groups were analyzed pairwise using the nonparametric Chi-square test~\citep{cochran1952chi2}. Corresponding effect sizes were also calculated using Cramer's V~\citep{cramer1999mathematical}. For multiple pairwise comparisons, we controlled the false discovery rate using the Benjamini–Hochberg procedure~\citep{benjamini1995controlling} at $q = 0.05$; henceforth, we refer to the resulting corrected p-values as $p_{\text{BH}}$.

Additionally, we used logistic mixed‑effects models to account for individual participant differences and repeated measures, with results reported as odd-ratios (ORs) and p-values.

%%%%%%%%%%%%%%%%%%%%%%%%
\paragraph{RQ2: Effect of CFFs on overreliance on AI responses.} 
To evaluate whether CFFs reduced users' overreliance on AI-generated outputs, we derived an accuracy measure: the proportion of instances where participants correctly trusted or dismissed the AI response. Specifically, participants' written feedback to the AI responses (collected after each CFF intervention) was manually coded against the pre-defined errors listed in Table~\ref{tab:writing_projects}. If a bug was present but unrecognized, or if no bug was present but one was assumed, the instance was scored as $0$ (not accurate) and $1$ (accurate) otherwise. We report the average accuracy per CFF condition and per task.

To more directly capture miscalibrated trust, we further decomposed inaccurate responses into underreliance and overreliance. For error-free AI outputs (Tasks 1 and 3), feedback was labeled as underreliant if participants incorrectly rejected the AI-generated plan or draft, and non-underreliant otherwise. For AI outputs containing synthetic errors (Tasks 2, 4, and 5), feedback was labeled as overreliant if participants incorrectly accepted the AI-generated plan or draft despite the presence of errors, and non-overreliant otherwise. Higher underreliance and overreliance rates indicate poorer calibration of trust.

Statistical differences in accuracy, underreliance, and overreliance rates between CFF conditions were analyzed using pairwise nonparametric Chi-square tests. Effect sizes were computed using Cramér's V. For multiple pairwise comparisons, we controlled the false discovery rate using the Benjamini–Hochberg procedure at $q = 0.05$, as per RQ1.

Additionally, we used logistic mixed‑effects models to account for individual participant differences and repeated measures, with results reported as odd-ratios (ORs) and p-values.

%%%%%%%%%%%%%%%%%%%%%%%%
\paragraph{RQ3: Cognitive load associated with CFFs.} To evaluate the cognitive effort required by different CFFs, we analyzed responses to the NASA Task Load Index (NASA-TLX) questionnaire~\citep{hart1988development}. Participants rated the workload for each task and CFF on a five-point Likert scale ($1$ = low, $5$ = high). The ``Overall Cognitive Load'' score was calculated as the mean of the TLX subscales, excluding physical demand (not applicable to this task context). We report the average cognitive load per CFF condition and per task. To further analyze the cognitive demand imposed by the CFFs themselves, we also analyzed the ``Mental Demand'' subscale for each CFF. Statistical comparisons were performed using the nonparametric \emph{Kruskal–Wallis test}, followed by pairwise \emph{Mann–Whitney U tests} with corresponding effect sizes ($r$). For multiple pairwise comparisons, we controlled the false discovery rate using the Benjamini–Hochberg procedure at $q = 0.05$, as per RQ1.

%%%%%%%%%%%%%%%%%%%%%%%%
\looseness-1\paragraph{RQ4: Helpfulness of CFFs for promoting critical thinking.} To investigate how participants perceived the usefulness of CFFs in supporting critical thinking and reducing overreliance, we analyzed two complementary measures. First, participants rated the ``Perceived Helpfulness'' of each CFF on a five-point Likert scale, averaged across the five tasks. Second, we analyzed open-ended responses to the post-survey question asking how the CFFs influenced their critical thinking. These textual responses were coded using Halpern's critical thinking rubric (see Table~\ref{tab:ct_skills_mapping}) and analyzed per CFF condition. Quantitative comparisons of helpfulness ratings were conducted using the nonparametric \emph{Kruskal–Wallis test}, followed by pairwise \emph{Mann–Whitney U tests} with corresponding effect sizes ($r$). For multiple pairwise comparisons, we controlled the false discovery rate using the Benjamini–Hochberg procedure at $q = 0.05$, as per RQ1.

%%%%%%%%%%%%%%%%%%%%%%%%
\paragraph{Analysis of metacognitive reflection and critical thinking.} Finally, we analyzed qualitative data from the interviews to understand how different CFFs influenced reflection, evaluation, and decision-making during AI-assisted tasks. The interviews also offered within-subjects comparative perspectives, as unlike the participants in the comparative online experiment, who each only interacted with a single CFF, these participants experienced all three CFF conditions—\nocff{}, \assumptioncff{}, and \whatifcff{} and discussed their preferences, perceived benefits, and limitations. This contrast provided a richer understanding of how participants evaluated the CFFs relative to each other.

% !TEX root =  main.tex
%%%%%%%%%%%%%%%%%%%%%%%%%%%%%%%%%%%%%%%%%%%%%%%%%%%%%%%%%%
%%%%%%%%%%%%%%%%%%%%%%%%%%%%%%%%%%%%%%%%%%%%%%%%%%%%%%%%%%

\section{Results}\label{sec:results}
\looseness-1In this section, we present the results of our study centered around the research questions (RQs) presented in Section~\ref{sec:intro}. We first report how different CFFs influenced participants' likelihood of revising their readiness assessments of AI-generated plans and drafts (Section~\ref{sec:results.rq1}). We then report results on accuracy of feedback and overreliance on AI outputs (Section~\ref{sec:results.rq2}), followed by analyses of cognitive load and subjective perceptions of the CFFs (Sections~\ref{sec:results.rq3} and \ref{sec:results.rq4} respectively).

\subsection{RQ1: Change of AI Response Readiness}\label{sec:results.rq1}
\begin{figure*}
    %%%% nfc
    \begin{subfigure}[c]{0.47\textwidth}
       \centering
       \includegraphics[width=\textwidth]{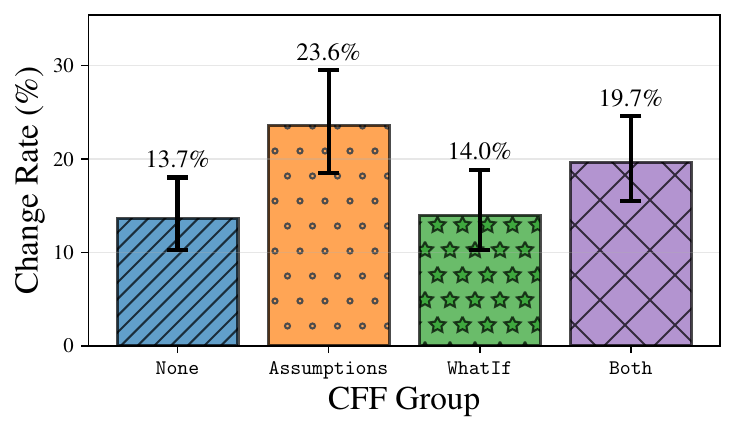}
        \caption{Change Rate per CFF Group}
        \label{fig:rq1.changerate.alg}
    \end{subfigure}
    \hspace{2em}
    %%%% aot
    \begin{subfigure}[c]{0.47\textwidth}
       \centering
       \includegraphics[width=\textwidth]{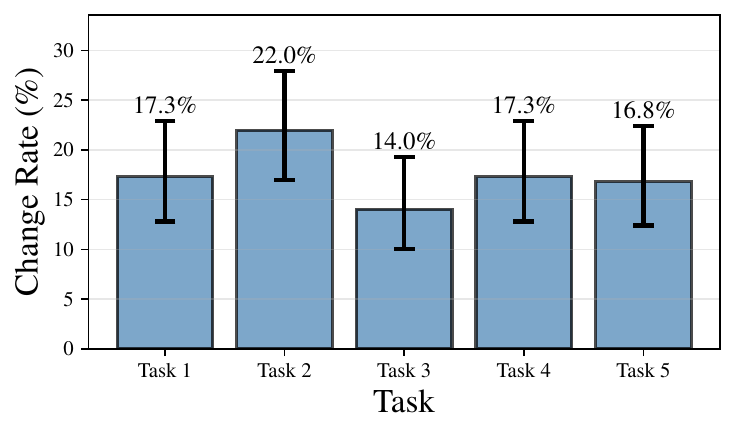}
        \caption{Change Rate per Project}
        \label{fig:rq1.changerate.task}
    \end{subfigure}
    %%%%%%%%%%%%%%
    %%%%%%%%%%%%%%%%
    \caption{\looseness-1Change rate in participants’ ratings of AI readiness for the knowledge work project. Readiness was assessed on a $5$-point Likert scale ($5$ = completely ready, $1$ = not ready at all) with an additional ``I don’t know'' option. Ratings were collected before and after applying the cognitive forcing function and change rate indicates the percentage of participants who changed their response. (a) shows results by CFF group  and (b) shows results by knowledge work project. Error bars represent standard error.}
    \Description{}
    \label{fig:rq1_opinion_change}
\end{figure*}

To address the first research question on how the different CFFs influenced users' likelihood to revise their initial assessment of the AI-generated plan and draft, we analyzed the raw change rate, as defined in Section~\ref{sec:setup.analysis}. The change rate captures whether participants revised their readiness rating after interacting with the assigned CFF, regardless of direction. Figure~\ref{fig:rq1.changerate.alg} shows the average change rate per CFF group (averaged across all tasks), and Figure~\ref{fig:rq1.changerate.task} shows the change rate per task (averaged across all CFF groups).

Across all tasks, \assumptioncff{} led to the highest mean rate of readiness revision, followed by the \bothcff{}, \whatifcff{}, and \nocff{} conditions. Compared to the \nocff{} condition, the \assumptioncff{} condition showed a significantly higher likelihood of revising readiness ratings ($\chi^2 = 7.89$, $p_{\text{BH}} = 0.03$, $V = 0.12$). \assumptioncff{} also differs significantly from \whatifcff{} ($\chi^2 = 6.54$, $p_{\text{BH}} = 0.03$, $V = 0.12$). No other pairwise differences between CFF conditions reached statistical significance after controlling for multiple comparisons. When evaluated per task, change rates were generally higher for tasks containing synthetic errors (Tasks~2,~4, and~5) than for error-free tasks (Tasks~1 and~3). Task~2 led to the highest overall change rate, while Task~3 had the lowest change rate. However, differences in change rates between tasks were not statistically significant.

%%%%%%%%%%%%%%%%%%%%%%%%%%%%%%%%%%%%%%%%%%%%%%%%%%%%%%%%
% !TEX root =  main.tex
%%%%%%%%%%%%%%%%%%%%%%%%%%%%%%%%%%%%%%%%%%%%%%%%%%%%%%%%%%
%%%%%%%%%%%%%%%%%%%%%%%%%%%%%%%%%%%%%%%%%%%%%%%%%%%%%%%%%%

% \input{figs/fig_rq1_switchrates}
%%%%%%%%%%%%%%%%%%%%%%%%%%%%%%%%%%%%%%%%%%%%%%%%%%%%%%%%

\paragraph{Robustness analysis with mixed-effects modeling.}
To assess whether the observed readiness-revision patterns persisted after accounting for individual participant differences and repeated measures, we conducted a generalized linear mixed-effects analysis (GLMM) with participant as a random intercept and presurvey covariates as fixed effects (Appendix B). Reported p-values are based on model-level tests of fixed effects, reflect planned contrasts between the CFF groups, and are not corrected for multiple comparisons. Consistent with the descriptive analyses above, the \assumptioncff{} condition exhibited the highest model-adjusted probability of readiness revision, while the \nocff{} condition exhibited the lowest. However, after controlling for individual differences and task effects, pairwise differences between CFF conditions were no longer statistically significant.

Instead, participants' cognitive disposition and prior experience with GenAI were the strongest predictors of opinion change: higher \emph{Actively Open-Minded Thinking} (AOT) scores were associated with significantly lower odds of revising readiness judgments ($OR = 0.60, p < 0.001$), as was greater familiarity with GenAI tools ($OR = 0.70, p = 0.048$). These results suggest that while plan-centered CFFs influence readiness revision behavior at the group level, individuals' baseline cognitive styles and experience significantly effect their likelihood of revising judgments about AI outputs. Detailed GLMM analysis of readiness revision is presented in Appendix B.

% !TEX root =  main.tex
%%%%%%%%%%%%%%%%%%%%%%%%%%%%%%%%%%%%%%%%%%%%%%%%%%%%%%%%%%
%%%%%%%%%%%%%%%%%%%%%%%%%%%%%%%%%%%%%%%%%%%%%%%%%%%%%%%%%%

\subsection{RQ2: Overreliance on AI Responses}\label{sec:results.rq2}
\begin{figure*}
    %%%% nfc
    \begin{subfigure}[c]{0.47\textwidth}
       \centering
       \includegraphics[width=\textwidth]{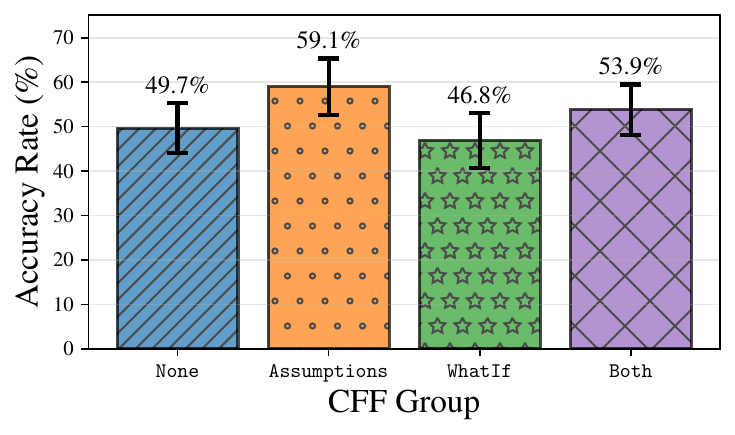}
        \caption{Accuracy per CFF Group}
        \label{fig:rq2_accuracy.alg}
    \end{subfigure}
    \hspace{2em}
    %%%% aot
    \begin{subfigure}[c]{0.47\textwidth}
       \centering
       \includegraphics[width=\textwidth]{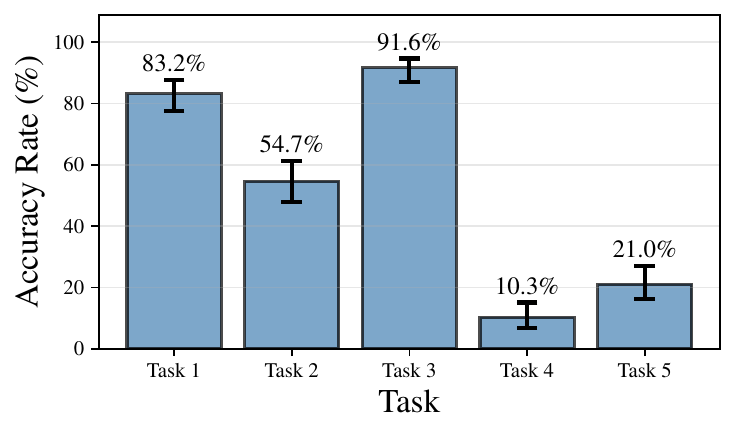}
        \caption{Accuracy per Project}
        \label{fig:rq2_accuracy.task}
    \end{subfigure}
    %%%%%%%%
    \\
    \begin{subfigure}[c]{0.47\textwidth}
       \centering
       \includegraphics[width=\textwidth]{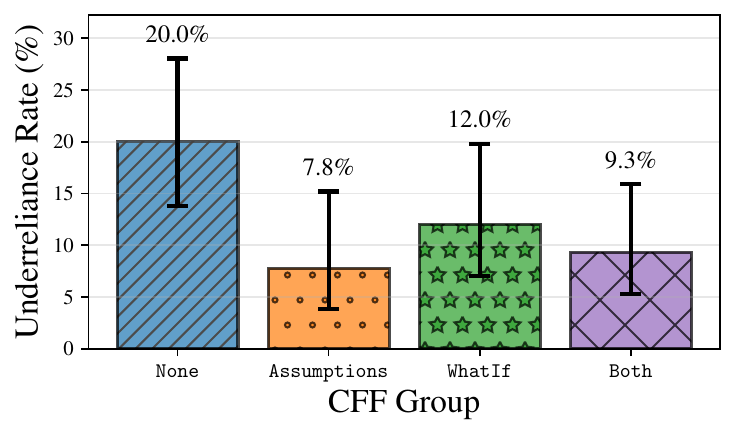}
        \caption{Underreliance per CFF Group}
        \label{fig:rq2_underreliance}
    \end{subfigure}
    \hspace{2em}
    %%%% aot
    \begin{subfigure}[c]{0.47\textwidth}
       \centering
       \includegraphics[width=\textwidth]{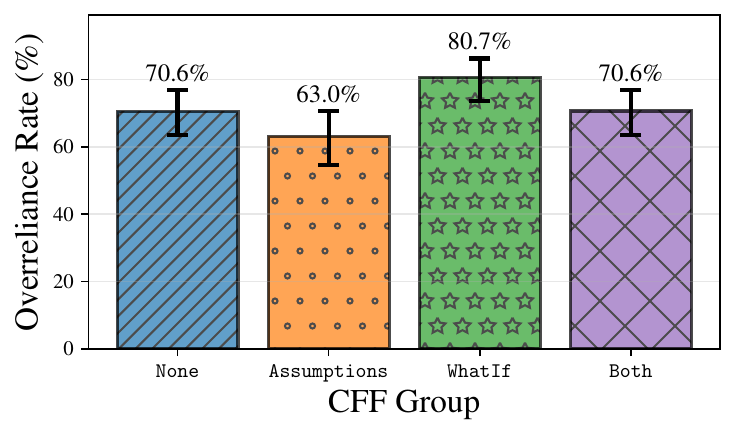}
        \caption{Overreliance per CFF Group}
        \label{fig:rq2_overreliance}
    \end{subfigure}
   %%%%%%%%%%%%%%%%%%
    \caption{\looseness-1 Accuracy of participants’ feedback on AI responses. Feedback was labeled as $1$ if it correctly identified bugs in the AI plan and draft (as described in Section~\ref{sec:setup.tasks}) or correctly noted the absence of issues, and $0$ otherwise. Higher accuracy is better. (a) presents the accuracy per CFF group and (b) presents the accuracy per knowledge work project. Underreliance and overreliance were computed from participants’ feedback on each project. For error-free drafts (Projects 1 and 3), feedback was labeled as underreliant if participants incorrectly rejected the draft, and non-underreliant otherwise. For projects with errors (Projects 2, 4, and 5), feedback was labeled as overreliant if participants incorrectly accepted the draft, and non-overreliant otherwise. Higher values indicate poorer calibration. (c) presents the underreliance rate per CFF group and (d) presents the overreliance rate per CFF group. Error bars represent standard error. }
    \Description{}
    \label{fig:rq2_accuracy}
\end{figure*}

To address the second research question on how different CFFs influenced users' overreliance on AI-generated responses, we analyzed participants' feedback accuracy, underreliance rate, and overreliance rate, computed following the measures described in Section~\ref{sec:setup.analysis}. Feedback accuracy captures whether participants correctly identified the presence or absence of errors in the AI-generated plan and draft. Underreliance and overreliance, in contrast, capture miscalibrated judgments: underreliance reflects unnecessary correction or doubt in error-free outputs, while overreliance reflects failure to detect errors in erroneous outputs. Figure~\ref{fig:rq2_accuracy.alg} presents feedback accuracy by CFF group (averaged across all tasks), and Figure~\ref{fig:rq2_accuracy.task} presents accuracy by task (averaged across all CFF groups). Figure~\ref{fig:rq2_underreliance} and Figure~\ref{fig:rq2_overreliance} present underreliance and overreliance rates by CFF group, respectively.

Across all tasks, \assumptioncff{} led to the highest feedback accuracy, followed by \bothcff{}, \nocff{}, and \whatifcff{} conditions. Compared to the \whatifcff{} condition, \assumptioncff{} achieved significantly higher accuracy ($\chi^2 = 7.20$, $p_{\text{BH}} = 0.04$, $V = 0.12$). No other pairwise differences in accuracy between CFF conditions were statistically significant after correction. When assessing underreliance rates, we found that \nocff{} condition had the highest rate while \assumptioncff{} had the lowest rate. \bothcff{} had the second lowest rates followed by the \whatifcff{} condition. However, no pairwise differences in underreliance rates between CFF conditions reached statistical significance after controlling for multiple comparisons. However, when assessing overreliance rates, we found that \whatifcff{} group had the highest rate and \assumptioncff{} had the lowest rate and this difference was significant ($\chi^2 = 10.25$, $p_{\text{BH}} = 0.008$, $V = 0.19$). No other pairwise differences in overreliance rates were statistically significant after correction.

When evaluated per task, feedback accuracy was generally lower for tasks containing synthetic errors (Tasks~2,~4, and~5) than for error-free tasks (Tasks~1 and~3). Furthermore, all pairwise comparisons between tasks yielded statistically significant differences in accuracy after controlling for multiple comparisons (all $p_{BH} < 0.05$).

%%%%%%%%%%%%%%%%%%%%%%%%%%%%%%%%%%%%%%%%%%%%%%%%%%%%%%%%
% !TEX root =  main.tex
%%%%%%%%%%%%%%%%%%%%%%%%%%%%%%%%%%%%%%%%%%%%%%%%%%%%%%%%%%
%%%%%%%%%%%%%%%%%%%%%%%%%%%%%%%%%%%%%%%%%%%%%%%%%%%%%%%%%%

%%%%%%%%%%%%%%%%%%%%%%%%%%%%%%%%%%%%%%%%%%%%%%%%%%%%%%%%

\paragraph{Robustness analysis with mixed-effects modeling.}
To assess whether the observed differences in feedback accuracy across CFF conditions persisted after accounting for repeated measures, task effects, and individual differences, we conducted a generalized linear mixed-effects analysis (GLMM) with accuracy as a binary outcome and participant-level random intercepts (Appendix C). Predicted accuracy rates were computed from the selected model after adjusting for presurvey covariates and task effects. Reported p-values are based on model-level tests of fixed effects, reflect planned contrasts between the CFF groups, and are not corrected for multiple comparisons. Consistent with the descriptive ordering reported above, the \assumptioncff{} condition exhibited the lowest model-adjusted accuracy rate, followed by the \bothcff{}, \nocff{} baseline, and \whatifcff{} conditions. Pairwise comparisons indicated that \assumptioncff{} differed significantly from \whatifcff{} ($p < 0.01$) but not w.r.t. the other CFF groups.

Assessment of fixed effects further showed that, after accounting for participant-level covariates, the accuracy advantages observed for \assumptioncff{} were not driven by a uniform main effect. In particular, exposure to \assumptioncff{} was associated with lower odds of providing accurate feedback relative to the \nocff{} condition ($OR = 0.45, p = 0.002$). \bothcff{} condition showed a smaller, marginal reduction in accuracy ($OR = 0.67, p = 0.09$), while \whatifcff{} condition did not differ significantly from \nocff{} ($OR = 1.24, p = 0.38$). Among individual differences, higher AOT scores were strongly associated with reduced accuracy across conditions ($OR = 0.72, p < 0.001$), whereas Need for Cognition (NFC) scores showed a marginal positive association. Task effects were substantial, with tasks containing erroneous AI outputs (Tasks 2, 4, and 5) showing higher baseline probabilities of accurate responses. Detailed GLMM analysis of accuracy is presented in Appendix C.

% !TEX root =  main.tex
%%%%%%%%%%%%%%%%%%%%%%%%%%%%%%%%%%%%%%%%%%%%%%%%%%%%%%%%%%
%%%%%%%%%%%%%%%%%%%%%%%%%%%%%%%%%%%%%%%%%%%%%%%%%%%%%%%%%%

\subsection{RQ3: Cognitive Load of Forcing Functions}\label{sec:results.rq3}
\begin{figure*}
    %%%% nfc
    \begin{subfigure}[c]{0.44\textwidth}
       \centering
       \includegraphics[width=\textwidth]{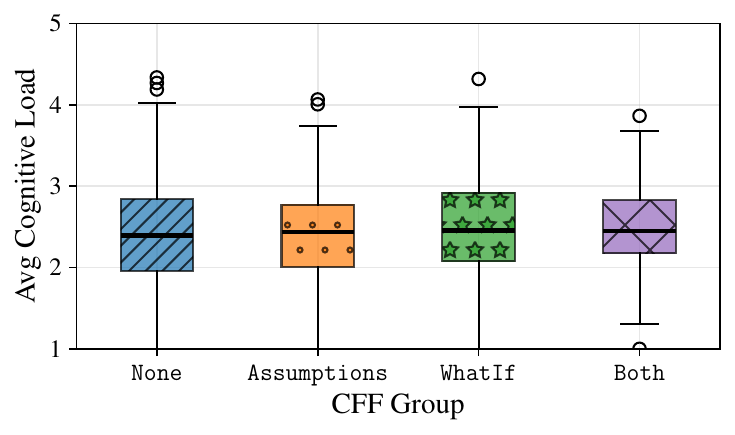}
        \caption{Mean TLX Load per CFF Group}
        \label{fig:rq3_cognitive_load_project.alg.box}
    \end{subfigure}
    %%%%%%%%%%%%%%
    \hspace{1em}
     %%%% opinion change and correctness
    \begin{subfigure}[c]{0.44\textwidth}
       \centering
       \includegraphics[width=\textwidth]{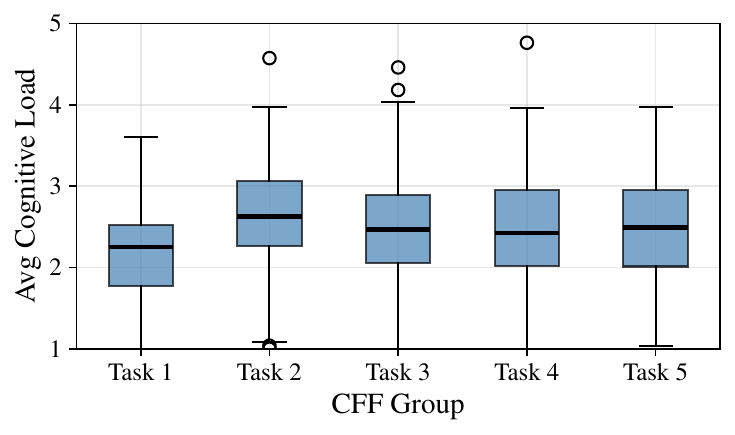}
        \caption{Mean TLX Load per Task}
        \label{fig:rq3_cognitive_load_project.task.box}
    \end{subfigure}
    % %%%%%%%%%%%%%%
    \caption{\looseness-1 Cognitive load of knowledge work projects. (a) and (b) show the distribution of average cognitive load measured over dimensions of mental load, temporal demand, performance, effort, and frustration taken from the NASA TLX survey (all dimensions except physical load). Each dimension was measured on a $5$-point Likert scale (5= A lot, 1=Not at all). 
    }
    \Description{}
    \label{fig:rq3_cognitive_load_project}
\end{figure*}

\begin{figure*}
    %%%% nfc
    \begin{subfigure}[c]{0.44\textwidth}
       \centering
       \includegraphics[width=\textwidth]{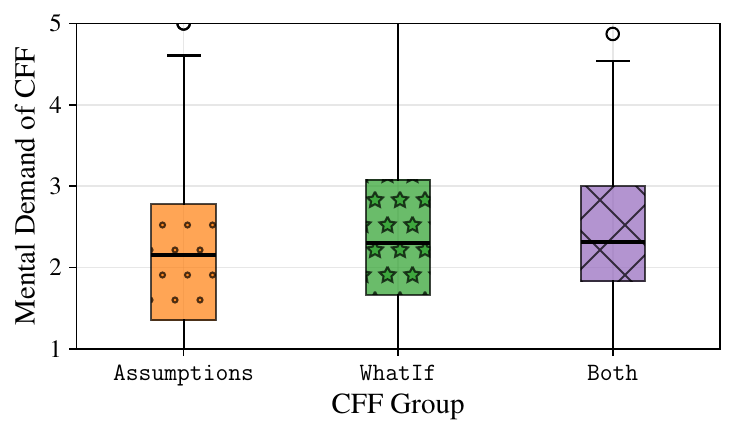}
        \caption{Mental Load per CFF Group}
        \label{fig:rq3_cognitive_load_cff.load}
    \end{subfigure}
    % \hspace{1em}
    % %%%% aot
    % \begin{subfigure}[c]{0.22\textwidth}
    %    \centering
    %    \includegraphics[width=\textwidth]{figs/5_rq3_results/cognitive_load_spider_algorithms.pdf}
    %     \caption{TLX Dimension per Group}
    %     \label{fig:rq3_cognitive_load_project.alg.spider}
    % \end{subfigure}
    %%%%%%%%%%%%%%
    \hspace{1em}
     %%%% opinion change and correctness
    \begin{subfigure}[c]{0.44\textwidth}
       \centering
       \includegraphics[width=\textwidth]{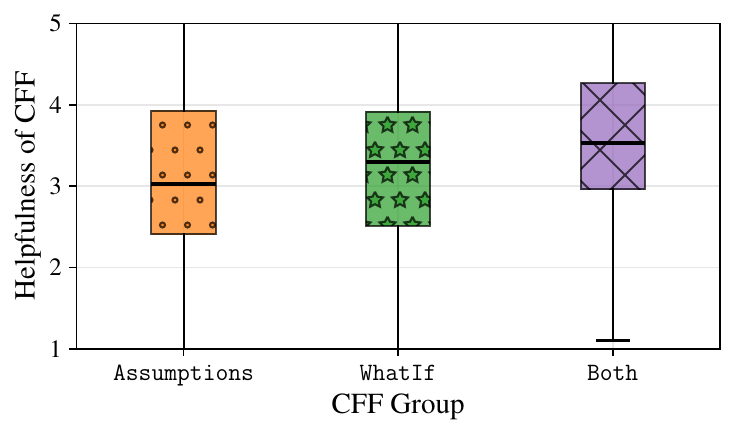}
        \caption{Helpfulness for feedback per CFF group}
        \label{fig:rq3_cognitive_load_cff.helpfulness}
    \end{subfigure}
    \caption{\looseness-1 Mental demand and perceived helpfulness of cognitive forcing functions. (a) shows participants' reported mental demand of the CFF method and (b) shows the perceived helpfulness of the CFF method in providing feedback on the AI response for the project. Both questions were measured on a $5$-point Likert scale (5= A lot, 1=Not at all).
    }
    \Description{}
    \label{fig:rq3_cognitive_load_cff}
\end{figure*}

To address the third research question regarding the perceived cognitive load associated with different CFFs during AI-assisted knowledge work tasks, we analyzed participants' reported cognitive load for each condition. Figures~\ref{fig:rq3_cognitive_load_project.alg.box} and~\ref{fig:rq3_cognitive_load_project.task.box} show the distribution of mean cognitive load per CFF group (averaged across all tasks) and per task (averaged across all CFF groups), respectively. Figure~\ref{fig:rq3_cognitive_load_cff} presents the distribution of participants' reported mental load for the CFFs, rated on a five-point Likert scale, where lower values indicate lower perceived load. The \nocff{} group is excluded from this plot, as participants in this condition were not presented with an explicit CFF intervention and therefore did not provide corresponding mental load ratings.

%%%%%%%%%%%%%%%%%%%%%%%%%%%%%%%%%%%%%%%%%%%%%%%%%%%%%%%%
% !TEX root =  main.tex
%%%%%%%%%%%%%%%%%%%%%%%%%%%%%%%%%%%%%%%%%%%%%%%%%%%%%%%%%%
%%%%%%%%%%%%%%%%%%%%%%%%%%%%%%%%%%%%%%%%%%%%%%%%%%%%%%%%%%

% !TEX root =  main.tex
%%%%%%%%%%%%%%%%%%%%%%%%%%%%%%%%%%%%%%%%%%%%%%%%%%%%%%%%%%
%%%%%%%%%%%%%%%%%%%%%%%%%%%%%%%%%%%%%%%%%%%%%%%%%%%%%%%%%%

%%%%%%%%%%%%%%%%%%%%%%%%%%%%%%%%%%%%%%%%%%%%%%%%%%%%%%%%

There was no significant difference in mean cognitive load across CFFs ($H = 2.99$, $p = 0.39$). However, when comparing mean cognitive load for specific tasks, there was a significant difference ($H = 73.21$, $p < .001$). In particular, Task~1 was associated with a significantly lower cognitive load compared to all other tasks (all $p_\text{BH} < 0.05$). Furthermore, Task~2 had significantly higher cognitive load compared to Task~5 ($U = 26827.5, p_\text{BH} = 0.002, r = 0.15$). No other pairwise differences in mean cognitive load were statistically significant after correction.

When evaluating the mental load attributed specifically to the CFFs, \assumptioncff{} was rated as having the lowest perceived mental load, while \whatifcff{} and \bothcff{} exhibited similar levels of mental demand. Moreover, the perceived mental load of \assumptioncff{} was significantly lower than that of \bothcff{}, ($U = 28{,}576.5$, $p_\text{BH} = 0.01$, $r = 0.12$). No other pairwise differences between CFF groups w.r.t. mental load were statistically significant after correction.
% !TEX root =  main.tex
%%%%%%%%%%%%%%%%%%%%%%%%%%%%%%%%%%%%%%%%%%%%%%%%%%%%%%%%%%
%%%%%%%%%%%%%%%%%%%%%%%%%%%%%%%%%%%%%%%%%%%%%%%%%%%%%%%%%%

\subsection{RQ4: Helpfulness of Forcing Functions}\label{sec:results.rq4}
To address the fourth research question on the perceived helpfulness of the CFFs in providing feedback as users' completed AI-assisted knowledge work tasks, we analyzed participants' reported helpfulness score in providing feedback on a 5-point Likert scale as shown in Figure~\ref{fig:rq3_cognitive_load_cff.helpfulness}. This plot omits the \nocff{} group as participants in this condition were not presented with an explicit CFF intervention and therefore did not provide corresponding mental load ratings. The results show that the perceived helpfulness was significantly higher for \bothcff{} compared to \assumptioncff{} and \whatifcff{} ($U = 26534.0, p_\text{BH} = 0.0001, r = 0.17$ w.r.t. \assumptioncff{}; $U = 32353.0,  p_\text{BH} = 0.01, r = 0.10$ w.r.t. \whatifcff{}). The difference between \assumptioncff{} and \whatifcff{} was not significant.

\begin{figure*}
    %%%% CT rating
    \begin{subfigure}[c]{0.48\textwidth}
       \centering
       \includegraphics[width=\textwidth]{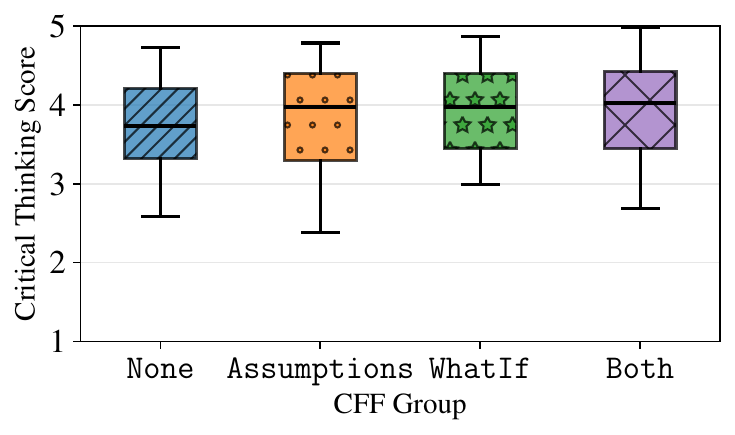}
        \caption{Critical Thinking per CFF Group}
        \label{fig:rq4_critical_thinking_cff.ct}
    \end{subfigure}
    %%%%%%%%%%%%%%
     %%%% CT skills
    \begin{subfigure}[c]{0.48\textwidth}
       \centering
       \includegraphics[width=\textwidth]{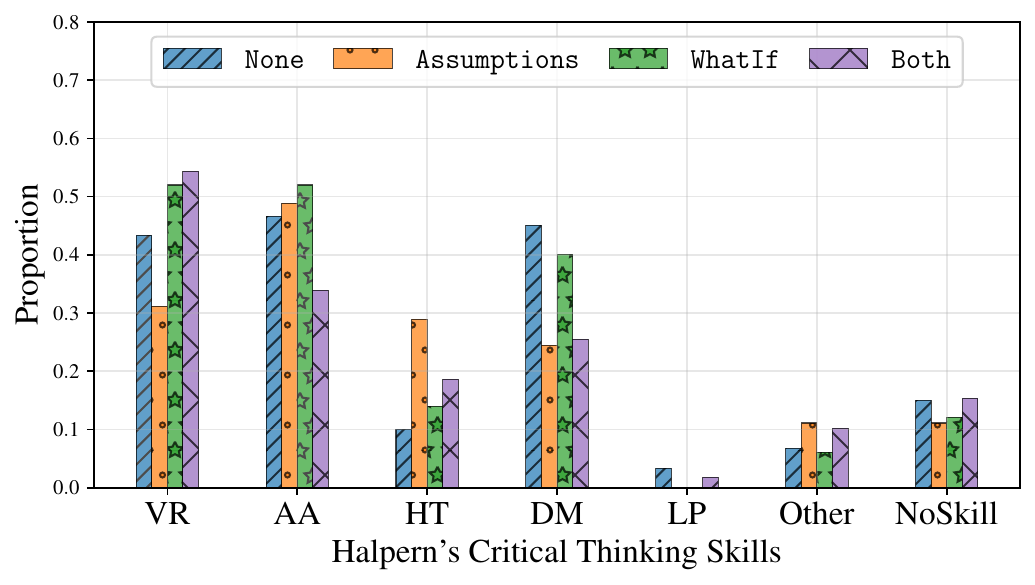}
        \caption{Halpern's CT skills per CFF group}
        \label{fig:rq4_critical_thinking.skills}
    \end{subfigure}
    %%%%%%%%%%%%
    \caption{\looseness-1 Perceived critical thinking (CT) skills induced by cognitive forcing functions (CFFs). (a) shows participants' self-reported CT skills prompted by the CFFs, rated on a five-point Likert scale, where higher scores indicate stronger perceived CT engagement. (b) presents the distribution of CT skill categories, based on Halpern’s CT framework, as identified in participants' open-ended responses describing how each CFF influenced their critical thinking. The CT skills are abbreviated as follows: verbal reasoning (VR), argument analysis (AA), hypothesis testing (HT), decision making and problem-solving (DM), and likelihood and uncertainty (LP). ``Other'' indicates presence of skills outside of Halpern's framework and ``NoSkill'' indicates the absence of evidence of any skill in the free-text response.
    }
    \Description{}
    \label{fig:rq4_critical_thinking}
\end{figure*}

We further analyzed the perceived helpfulness of the CFFs in terms of how effectively they supported participants' critical thinking about the AI-generated responses. Figure~\ref{fig:rq4_critical_thinking_cff.ct} shows participants' self-reported critical thinking (CT) scores induced by the different CFFs, measured on a 5-point Likert scale.

%%%%%%%%%%%%%%%%%%%%%%%%%%%%%%%%%%%%%%%%%%%%%%%%%%%%%%%%
% !TEX root =  main.tex
%%%%%%%%%%%%%%%%%%%%%%%%%%%%%%%%%%%%%%%%%%%%%%%%%%%%%%%%%%
%%%%%%%%%%%%%%%%%%%%%%%%%%%%%%%%%%%%%%%%%%%%%%%%%%%%%%%%%%

%%%%%%%%%%%%%%%%%%%%%%%%%%%%%%%%%%%%%%%%%%%%%%%%%%%%%%%%

The free-text responses were annotated according to Halpern's five core CT skills and supplemented with two additional labels—``Other'' and ``NoSkill''—for cases that could not be classified under any of the five categories. Particular attention was given to identifying instances of Analyzing Arguments (AA) and Hypothesis Testing (HT), as these two skills were expected to differentiate the CFF conditions. The remaining skills—Verbal Reasoning (VR), Decision Making and Problem-Solving (DM), and Likelihood and Uncertainty (LU)—were implicitly present across all tasks. This is because of the nature of the AI-generated responses, which presented a structured plan followed by a draft, and from the evaluation tasks that explicitly required participants to judge the quality and accuracy of the AI's responses through readiness and feedback ratings.

To ensure consistency in annotation, two authors independently coded $40$ of the $214$ responses ($10$ randomly selected from each of the four CFF groups) to refine the coding rubric. Inter-rater reliability, assessed using Cohen's kappa~\citep{cohen1960coefficient}, was $0.85$ for AA and $0.63$ for HT, indicating strong and moderate agreement, respectively. After reaching consensus on the coding scheme, one author completed the annotation of the remaining responses. The aggregated results of this analysis are shown in Figure~\ref{fig:rq4_critical_thinking.skills}.

The results show that the \nocff{} group had the lowest reported effect on participants' perceived critical thinking (CT) skills, while the other CFF conditions produced comparable effects. However, these differences were not statistically significant ($H = 1.82$, $p = 0.61$). Interestingly, participants in the \assumptioncff{} condition most frequently demonstrated evidence of hypothesis testing skills, whereas those in the \whatifcff{} condition most often exhibited argument analysis skills. This difference, however, was not statistically significant when comparing the two groups.

Across all conditions, the CT skills of verbal reasoning, decision making and problem-solving, and likelihood and uncertainty appeared consistently in participants' responses, likely due to the inherent demands of the knowledge work tasks and the structured nature of the AI-generated responses as a plan and draft. The main distinctions between the CFFs therefore emerged in the relative emphasis on argument analysis and hypothesis testing: \assumptioncff{} showed evidence of both skills, whereas the \emph{perception} of \whatifcff{} was predominantly that of encouraging argument analysis. Notably, this perception diverged from the intended design of the \whatifcff{}, which was grounded in eliciting hypothesis testing; this discrepancy reflects participants' subjective interpretations of the intervention rather than an objective assessment of the underlying cognitive processes engaged.

% !TEX root =  main.tex
%%%%%%%%%%%%%%%%%%%%%%%%%%%%%%%%%%%%%%%%%%%%%%%%%%%%%%%%%%
%%%%%%%%%%%%%%%%%%%%%%%%%%%%%%%%%%%%%%%%%%%%%%%%%%%%%%%%%%

\section{Qualitative Findings and Engagement with the Forcing Functions}\label{sec:qualitative_analysis}
To complement the quantitative findings, we qualitatively examined how participants interacted with the CFFs \assumptioncff{} and \whatifcff{}, and the no-CFF control (\nocff{}) during moderated interviews. The analysis provides insights on how participants engaged with AI-generated plans and drafts, how they interacted with the CFFs, and how these interactions shaped their reasoning during AI-assisted knowledge work. Henceforth we refer to the participants in the interview study as P01--P12.

First we characterize participants' observed interaction behaviors when reviewing AI-generated plans, drafts, and task artifacts (Section~\ref{sec:qualitative_analysis.precheck}). Next, we report engagement metrics that capture how participants engaged with CFFs, providing evidence that the interactions were meaningfully used (Section~\ref{sec:qualitative_analysis.cffengage}). We then analyze the reasoning processes elicited by the CFFs, based on the think-aloud protocols and participants' explanations (Section~\ref{sec:qualitative_analysis.reasoning}). Finally,we present participants' comparative reflections across the CFF conditions, based on structured interview questions presented after they experienced all three CFF conditions (Section~\ref{sec:qualitative_analysis.comparison}).

%%%%%%%%%%%%%%%%%%%%%%%%%%%%%%%%%%%%%%%%%%%%%%%%%%%%%%%%
% !TEX root =  main.tex
%%%%%%%%%%%%%%%%%%%%%%%%%%%%%%%%%%%%%%%%%%%%%%%%%%%%%%%%%%
%%%%%%%%%%%%%%%%%%%%%%%%%%%%%%%%%%%%%%%%%%%%%%%%%%%%%%%%%%

\begin{figure*}
    %%%% assumptions quizzes attempted
    \begin{subfigure}[c]{0.44\textwidth}
       \centering
       \includegraphics[width=\textwidth]{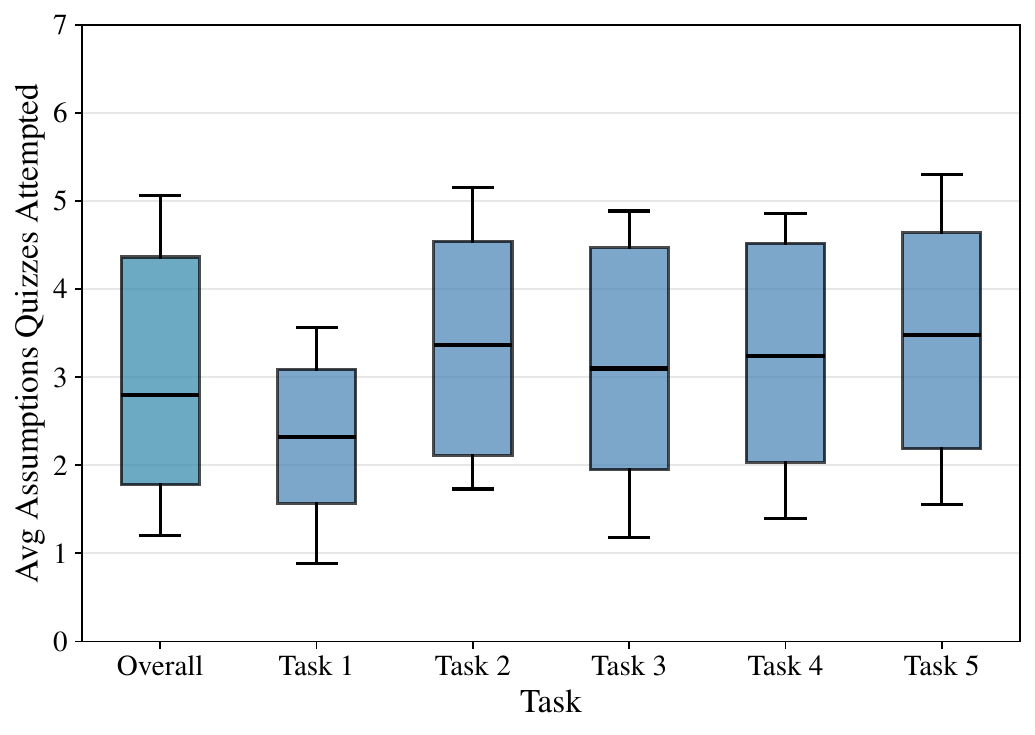}
        \caption{Mean Assumption Quizzes Attempted}
        \label{fig:cff_metrics.assumptions}
    \end{subfigure}
    \hspace{3em}
    %%%% aot
    \begin{subfigure}[c]{0.44\textwidth}
       \centering
       \includegraphics[width=\textwidth]{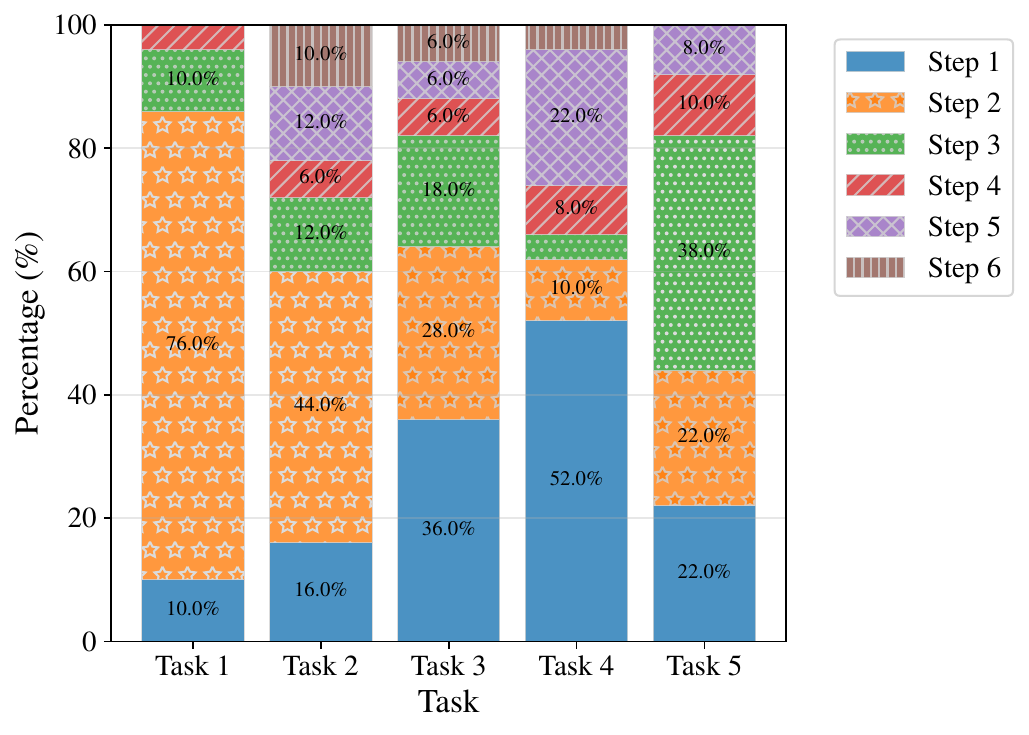}
        \caption{Distribution of Critical Plan Step Selection}
        \label{fig:cff_metrics.whatif}
    \end{subfigure}
    %%%%%%%%%%%%%%
    \caption{We present additional insights on the performance of the two CFF methods from our comparative online experiment. (a) presents the mean number of assumptions quizzes attempted per task. Note that, Task 1 had four plan steps and all the other tasks had six plan steps. (b) presents the distribution of critical plan steps as perceived by the participants for each knowledge work task.}
    \Description{}
    \label{fig:cff_metrics}
\end{figure*}
%%%%%%%%%%%%%%%%%%%%%%%%%%%%%%%%%%%%%%%%%%%%%%%%%%%%%%%%

%%%%%%%%%%%%%%%% interaction behaviors
\subsection{Interaction Behaviors During Plan, Draft, and Artifact Review}\label{sec:qualitative_analysis.precheck}
We first describe participants' interaction behaviors during the think-aloud sessions, focusing on how they navigated between the AI-generated execution plan, the resulting draft, and the task artifact \emph{prior} to the stage where they were presented with the CFF. The analysis shows whether and how participants engaged with the plan as an evaluative object and provides necessary context for interpreting later findings on reasoning and perceived usefulness of the CFFs.

Across the $12$ interview participants, engagement with the AI-generated plan was universal, as the study interface required plan generation prior to draft presentation. However, participants differed in the order, depth, and purpose of their engagement with the plan and draft. Some participants ($n = 4$) adopted a \emph{plan-first strategy}, reviewing the execution steps immediately after clicking ``AI Assist Me'' button and using the plan to orient their evaluation of the draft. Others ($n = 5$) adopted a \emph{draft-first strategy}, focusing primarily on the AI-generated text. A smaller subset ($n = 3$) treated the plan primarily as a reference artifact, returning to it selectively when needed instead of using it as the main focus of evaluation.

Participants also varied in their engagement with the task artifact.About half the participants ($n = 5$) reviewed the task artifact before invoking the AI, often describing this as a way to establish task context. Others ($ n = 7$) engaged with the artifact only after the AI-generated plan and draft were presented.

%%%%%%%%%%%%%% enagagement with the CFFs
\subsection{Engagement with CFFs}\label{sec:qualitative_analysis.cffengage}
In addition to the qualitative analyses from the think-aloud interviews, we examined behavioral engagement metrics to assess whether participants meaningfully interacted with the CFFs as designed. The engagement metrics reported in this subsection are derived from interaction logs collected during the comparative online experiment, and are included here to complement the qualitative findings from the interview study.

For the \assumptioncff{}, participants were required to engage with at least one plan step before proceeding. Figure~\ref{fig:cff_metrics.assumptions} shows the mean number of assumption prompts attempted per task. Across tasks, participants engaged with multiple plan steps on average, despite no requirement to do so. This pattern indicates that participants frequently explored assumptions associated with more than a single step of the AI-generated plan, suggesting active engagement with the prompt rather than treating it as a procedural hurdle.

For the \whatifcff{}, we analyzed participants' selection of the plan step they identified as most critical, as well as their subsequent reasoning about potential changes or failures. Figure~\ref{fig:cff_metrics.whatif} presents the distribution of plan steps selected as critical for each knowledge work task. Participants' selections were task-dependent and concentrated around semantically meaningful steps within each plan, indicating that step selection was not random. Instead, participants appeared to identify steps they perceived as central to task success, providing evidence of deliberate engagement with the hypothesis-testing prompt.

%%%%%%%%%%%%%% reasoning processed specific to each CFF
\subsection{Reasoning Processes Elicited by CFFs}\label{sec:qualitative_analysis.reasoning}
Building on the interaction patterns described in Section~\ref{sec:qualitative_analysis.precheck}, we examine how participants reasoned about the AI-generated plans and drafts under different CFFs. This analysis is based on participants' verbal explanations during task completion, focusing on the types of reasoning participants articulated rather than on task outcomes.

%%%% effect of plan centric CFFs -- connected to behaviors in Section 6.1
\paragraph{Effect of plan centric CFF design.} 
Across both the \assumptioncff{} and \whatifcff{} conditions, the forcing functions visibly altered participants' interaction patterns compared to their baseline behaviors described in Section~\ref{sec:qualitative_analysis.precheck}. More specifically,participants who initially focused primarily on the draft or skipped the plan altogether were required to explicitly engage with the plan as part of completing the CFF prompts. This requirement often led participants to inspect plan steps they had not previously examined, making plan engagement more salient even for those who did not naturally adopt a plan-first review strategy. In addition, across both CFFs, participants frequently returned to the task artifact while responding to the prompts, even when they had shown minimal or delayed artifact engagement earlier. In this way, plan-centered CFFs not only redirected attention toward the AI-generated plan, but also prompted participants to re-engage with the broader task context, for example by revisiting source materials they had previously overlooked. These shifts in attention are primarily driven by the plan-centric approach to the design of the CFFs.

%%%%%% assumption perceptions
\paragraph{Reasoning under \assumptioncff{}.} 
Participants interacting with \assumptioncff{} frequently articulated reasoning aligned with argument analysis, explicitly identifying implicit premises, constraints, and unstated expectations embedded in the AI-generated plan. The prompt encouraged participants to slow down and interrogate what the plan was taking for granted, rather than immediately evaluating surface-level quality.

Several participants described the assumptions prompt as a mechanism that ``brought a bit more reasoning'' into the review process (P01) and helped them ``make more sense of the plan'' (P02). For others, identifying assumptions served as a way to better understand the relationship between the plan and the resulting draft. As P04 explained, reflecting on assumptions ``helped me understand the draft more deeply and make better suggestions,'' while P09 noted that it clarified whether ``each part led logically from one to another.'' Participants also framed the \assumptioncff{} as a way to understand the AI's internal logic. P05 remarked that it ``helped me understand what the AI's motivation was,'' and P11 described the prompt as ``better, especially for understanding how the AI reached its plan.'' Similarly, P12 commented that ``looking at the assumptions gave insight into how the AI thinks,'' indicating that participants used assumptions as an interpretive lens for the system's behavior rather than solely as a correctness check. Importantly, the \assumptioncff{} provided structure to what participants often described as an otherwise unstructured or overwhelming evaluation task. P03 reflected that it ``guided me on what path to take [and] made it easier to streamline my focus,'' while P07 explained that it ``kept my focus on prioritizing how useful the plan and draft were, rather than just how logical or well written they were.'' These accounts suggest that the \assumptioncff{} functioned as a scaffold that organized participants' reasoning around the plan's underlying premises, supporting step-wise evaluation without disrupting task flow.

%%%%% whatif perceptions
\paragraph{Reasoning under \whatifcff{}.}
Participants interacting with the \whatifcff{} engaged in reflective reasoning that was often described as deeper but more cognitively demanding. The prompt encouraged participants to consider the consequences of critical plan steps failing or changing, which in principle aligns with hypothesis testing and counterfactual reasoning.

Several participants explicitly described this form of engagement. P02 noted that the \whatifcff{} ``helped me think more clearly about how the plan could change,'' while P03 said ``it helped to see the importance of each plan step.'' P04 similarly reported that the questions ``made me think more in depth about how each part connected,'' and P12 characterized the prompt as one that ``helped more for critical thinking.'' However, participants' verbalizations often revealed that they interpreted the \whatifcff{} less as an exercise in simulating alternative outcomes and more as a way to scrutinize the soundness of the plan itself. For example, P09 explained that the prompt ``made me rethink the AI plan and consider alternatives,'' but their reasoning primarily focused on whether the plan's logic was reasonable rather than on systematically tracing downstream consequences of hypothetical changes. 

This interpretation was accompanied by reports of increased cognitive effort. Some participants found the \whatifcff{} mentally taxing or intrusive. P07 described it as ``more mental work than I expected,'' and P05 stated that ``it interrupted my thought process.'' These qualitative accounts align with findings from the comparative online experiment reported in Section~\ref{sec:results.rq3}, where the \whatifcff{} was associated with higher self-reported mental load relative to other conditions. Importantly, participants' perceptions of the \whatifcff{} diverged from its intended design goal. Although the prompt was designed to elicit hypothesis testing, many participants described it in terms associated with argument analysis or justification, such as ``double-checking'' or ``re-evaluating'' the plan's logic.

% !TEX root =  main.tex
%%%%%%%%%%%%%%%%%%%%%%%%%%%%%%%%%%%%%%%%%%%%%%%%%%%%%%%%%%
%%%%%%%%%%%%%%%%%%%%%%%%%%%%%%%%%%%%%%%%%%%%%%%%%%%%%%%%%%

\subsection{Comparative Reflections Across CFFs}\label{sec:qualitative_analysis.comparison}
After completing tasks under all three CFF conditions, participants in the interview study reflected comparatively on the \nocff{}, \assumptioncff{}, and \whatifcff{} conditions along four dimensions: overall preference, support for critical evaluation, perceived limitations, and suitability for integration into everyday AI tools. These reflections were collected through structured comparison questions (see Appendix G) and provide insight into how participants experienced the practical impact of the CFFs across tasks.

\paragraph{Overall preference.} 
Most participants ($n = 6$) favored \assumptioncff{}, describing it as compatible with how they naturally approached reviewing AI outputs. This preference often emerged from sceanrios in which identifying assumptions provided a logical pathway for evaluation. For example, during Task~2, P01 initially skimmed the AI draft and felt unsure how to begin checking correctness. When prompted to identify assumptions, they returned to a specific plan step and articulated that the AI appeared to assume access to information not provided in the task artifact. Reflecting on this moment, P01 remarked that ``assumptions definitely made my evaluation more structured,'' explaining how the prompt helped them move from a vague impression to a concrete critique.

A smaller group of participants ($n = 4$) preferred \whatifcff{}, typically citing moments where imagining changes or failures revealed weaknesses in the plan's structure. For instance, P02 described selecting a plan step they believed was central to the task and reasoning through what would happen if it failed. This process, they explained, ``made me test the AI's reasoning,'' particularly by exposing how sensitive the overall plan was to a single step.

Only one participant (P05) expressed a preference for the \nocff{} condition, noting that it allowed them to ``concentrate on the task without extra steps.'' This participant described relying on their own internal review strategy and expressed confidence in evaluating the AI output without structured prompts.

%%%%%%
\paragraph{Support for critical thinking.} 
When asked which CFF best supported critical thinking, responses were more evenly divided. Participants who selected \assumptioncff{} ($n = 7$) emphasized how it surfaced hidden premises shaping the AI's approach. For example, P04 described an episode where identifying assumptions revealed that the AI's plan implicitly prioritized one stakeholder perspective over another, a choice that was not explicit in the task description. As they explained, the prompt helped them ``see what the AI was relying on and why,'' making the plan's logic more transparent.

Participants who identified \whatifcff{} as most supportive of critical evaluation ($n = 5$) pointed to moments of counterfactual reasoning that altered their confidence in the AI output. They also noted how the prompt had the ability to externalize evaluative reasoning. For example, P06 explained that it ``helped me check the plan against the draft,'' while P07 noted that the prompt ``put it in words for me,'' suggesting that \whatifcff{} prompts articulated a line of reasoning they were already attempting internally.

No participant identified the \nocff{} condition as supportive of critical thinking.

%%%%%%
\paragraph{Perceived limitations and least helpful CFF}
Participants' critiques of the CFFs centered on cognitive effort and perceived redundancy. The \nocff{} condition was most frequently described as least useful ($n = 7$), particularly in tasks involving complex artifacts or subtle errors. Participants described it as offering ``nothing there'' (P01) or lacking ``deeper interaction'' (P04), leaving them unsure how to structure their evaluation.

Criticism of \assumptioncff{} was less common ($n = 3$) and typically arose when participants felt the AI's reasoning was already clear. In such cases, the prompt was described as redundant or annoying. P06 characterized it as ``more annoying than helpful,'' while P07 felt it ``didn't give me any additional benefit.''

Only one participant identified \whatifcff{} as least useful, describing it as interruptive. P05 noted that the hypothetical reasoning ``interrupted [their] thought process,'' articulating concerns about increased cognitive load reported similar to the our findings described in Section~\ref{sec:results.rq3}.

%%%%%
\paragraph{Preferred integration into everyday AI tools.}
When imagining how CFFs could be embedded into everyday AI tools, most participants ($n = 5$) favored \assumptioncff{}. They claimed that it was the more practical option—P08 explained that it ``fits how I already check AI drafts,'' and P11 said it was ``better for understanding how the AI reached its plan.'. P12 favored assumptions because they were ``more detailed and adds clarity,'' while P09 valued their structured guidance.

An smaller number of participants selected \whatifcff{} ($n = 3$), describing it as useful for gaining alternative perspectives or stress-testing reasoning. Two participants (P03, P10) explicitly advocated keeping both forcing functions, suggesting that \assumptioncff{} could surface reasoning while \whatifcff{} could test robustness. 

Two participants (P01, P05) preferred \nocff{} condition by default, explaining that the other CFFs interfered with their analysis strategies and that they already considered themselves critical of AI-generated content. Although, P01—despite preferring \nocff{} overall—acknowledged that assumptioncff{} could be helpful in situations involving higher uncertainty.

% !TEX root =  main.tex
%%%%%%%%%%%%%%%%%%%%%%%%%%%%%%%%%%%%%%%%%%%%%%%%%%%%%%%%%%
%%%%%%%%%%%%%%%%%%%%%%%%%%%%%%%%%%%%%%%%%%%%%%%%%%%%%%%%%%

\section{Discussion}\label{sec:discussion}
In this section, we interpret our findings in relation to prior work on CFFs and human–AI collaboration, discuss their implications for the design of plan-centered CFFs in AI-assisted knowledge work, and discuss limitations and directions for future research.

\subsection{Plan-Centered CFFs Shape How Users Evaluate AI Output}\label{sec:discussion.plan}
Our findings build on and extend prior work on CFFs in human–AI interaction, while also highlighting important differences that arise when CFFs are applied to the evaluation of AI-generated execution plans rather than to single predictions or recommendations. Across both the large-scale comparative online experiment and the interview study, we observed that plan-centered CFFs systematically shaped how participants evaluated AI output, and that different forms of critical thinking elicited distinct patterns of engagement, trust recalibration, and cognitive effort.

%%%% relation to prior work: similarities
\looseness-1Prior work has shown that CFFs can effectively reduce overreliance by interrupting heuristic acceptance of AI recommendations and prompting users to reflect more carefully on AI outputs. For example, Buçinca et al.~\citep{DBLP:journals/pacmhci/BucincaMG21} demonstrated that requiring users to consider why an AI recommendation might be wrong reduced automation bias, but at the cost of increased perceived difficulty and frustration. Complementing this line of work, approaches grounded in cognitive science and education have used reflection prompts and process-oriented scaffolds to support reasoning, such as Socratic questioning frameworks for machine-assisted decision making~\citep{fischer2025taxonomy} and step-by-step scaffolding that elicits learners' reasoning in AI-assisted programming contexts~\citep{kazemitabaar2025exploring}. Our findings align with this perspective by showing that CFFs grounded in established critical thinking and metacognitive frameworks—when embedded in the structure of AI-generated plans—can meaningfully shape users' evaluation behavior without necessarily imposing excessive cognitive burden.

%%%%% how our findings are different
At the same time, our findings diverge from prior work in an important way. Much of the existing literature studies CFFs in settings where users evaluate a single outcome, label, or recommendation. In contrast, the present study focuses on AI-generated execution plans and long-form drafts, where the task is not simply to accept or reject an output, but to assess whether the AI has adopted an appropriate approach in the first place. In this context, we find that not all forms of cognitive forcing are equally effective. Specifically, the \assumptioncff{} CFF consistently produced lower underreliance and overreliance rates, without a corresponding increase in cognitive load. By contrast, the \whatifcff{} CFF, while often perceived as being more helpful, showed weaker behavioral effects and higher reported mental effort. Furthermore, sequential CFFs (\bothcff{}) do not necessarily offer additional benefits compared to single CFFs. Although participants found \bothcff{} to be more helpful, this perceived helpfulness did not translate into a corresponding reduction of overreliance. One possible explanation is that stacking multiple reflective prompts increases cognitive demands without proportionately improving users' ability to identify foundational issues in the AI's approach. Rather than deepening evaluation, sequential CFFs may diffuse users' attention across multiple forms of reflection, leading to diminished marginal returns. This suggests that, in plan-based evaluation contexts, selectively targeting a small number of critical reasoning processes may be more effective than encouraging broader but less focused reflection.

%%%% findings in the context of Halpern's framework
This pattern suggests that the effectiveness of a CFF depends not only on its strength, but on how well the targeted cognitive process aligns with the structure of the AI output being evaluated. In plan evaluation, identifying implicit assumptions appears to offer a low-friction entry point for metacognitive monitoring: assumptions are naturally embedded in plan steps, dependencies between the steps, and task constraints, and can be analyzed without requiring users to simulate counterfactual outcomes. This contrasts with hypothesis testing, which requires imagining alternative outcomes and propagating consequences across the plan, a cognitively demanding process that may be poorly suited to time-constrained review tasks. 

%%%% importance of CFFs based on AI plans
By centering CFFs on AI-generated plans rather than on final outputs or explanations, our results highlight a complementary pathway for supporting calibrated trust in AI-assisted knowledge work. Rather than blocking progress or challenging correctness directly, plan-centered CFFs restructure the evaluation task itself, guiding users' attention toward the AI's reasoning process and its underlying premises. In doing so, they preserve the core insight of prior CFF research: reflection can mitigate overreliance. However, they additionally demonstrate that the form of reflection matters, particularly in complex, open-ended knowledge work settings.

%%%%%%%%%%%%%%%%%%%%%%%%
\subsection{Transfer Beyond Writing-Centered Knowledge Work}\label{sec:discussion.transfer}
Although our study focused on AI-assisted writing tasks, the mechanisms revealed by our results are not inherently tied to writing alone. Instead, they stem from more general properties of AI-generated execution plans and critical thinking mechanisms required to evaluate them. At the same time, certain aspects of writing as a task domain may have shaped which CFFs were most effective. More specifically, writing tasks are characterized by open-ended goals, multi-level correctness criteria, and substantial reliance on judgment rather than formal verification. In such settings, evaluating AI output often requires assessment of the system's appropriate framing of the task, its requirements, and argumentative structure, along with checking for discrete factual or logical errors. This could be the reason why argument analysis based CFFs (such as \assumptioncff{}) were particularly effective in our study: identifying implicit assumptions aligns closely with the evaluative demands of writing, where omissions, framing choices, and implicit premises can significantly affect quality even when the output appears fluent and coherent, as was the case with all the writing tasks in our study.

At the same time, the plan-centered nature of our CFFs suggests clear avenues for transfer to other forms of knowledge work where AI systems increasingly generate structured procedures before producing artifacts. For example, in programming, or data analysis, AI-generated plans similarly encode assumptions about data availability, ordering of operations, and constraints. In these domains, argument analysis-focused prompts could support early detection of mismatches between the AI's inferred task constraints and the user's intent, while hypothesis-testing prompts may become more useful at later stages when verifying robustness, edge cases, or failure modes. In the work of Kazemitabaar et al.~\citep{kazemitabaar2025exploring} for example, the ``Lead and Reveal'' technique—which scaffolds step-by-step reasoning about intermediate structure before revealing the full output—was found to be the most effective engagement method, closely mirroring the role of argument analysis-focused prompts in our plan-based setting. 

More broadly, this suggests that plan-centered CFFs offer a flexible design pattern for AI-assisted knowledge work, but that their effectiveness depends on selecting prompts that align with how users naturally reason about quality, risk, and appropriateness in a given domain. Future work should therefore examine how plan-based CFFs perform across different task types and stages of work, and whether adaptive or staged forcing functions can better support critical evaluation as tasks evolve.

%%%%%%%%%%%%
\subsection{Perception, Adaptation, and Evaluation of CFFs}\label{sec:discussion.perception}
Another important implication of our findings is the mismatch between users' perceived helpfulness of CFFs and their actual effectiveness in reducing underreliance and overreliance on AI output. Although participants described \whatifcff{} as being more helpful in providing feedback to AI, \assumptioncff{} more consistently supported beneficial outcomes such as reduced overreliance and appropriate trust calibration, without increasing cognitive load. Similar perception–performance mismatches have been documented in prior human–AI interaction research, where interventions that most effectively reduced overreliance were not always those users preferred or found most intuitive~\citep{DBLP:journals/pacmhci/BucincaMG21,vaccaro2024combinations,bansal2021does}. 

Our results also highlight the importance of accounting for individual cognitive dispositions when designing CFFs. Mixed-effects analyses revealed that traits such as Actively Open-Minded Thinking significantly influenced how participants evaluated AI outputs, in some cases showing effects comparable to or stronger than the CFF condition itself (especially when considering readiness-revision). This suggests that a uniform approach to forcing functions may be suboptimal. Adaptive systems could instead tailor CFFs to users' cognitive profiles—for example, offering more structured, guidance-oriented prompts to users less inclined to revise initial judgments, while presenting lighter-touch scaffolding to users who already exhibit strong reflective tendencies. Such adaptation may help balance cognitive effort with performance benefits across a diverse user population.

\looseness-1Finally, our findings indicate that opinion revision alone is an insufficient marker of CFF effectiveness. While changes in readiness ratings capture whether users reconsider their judgments appropriately, revision behavior does not necessarily imply improved evaluation quality. In our study, higher rates of opinion change were not always accompanied by higher accuracy or lower overreliance, highlighting the need to interpret readiness-revision metrics together with overreliance measures. Future evaluations of CFFs should therefore triangulate across revision behavior, overreliance, and cognitive effort to more reliably assess how CFFs support calibrated trust and effective human–AI collaboration.

\begin{shaded}
\noindent
    \textbf{Design insight.} \assumptioncff{} reduced overreliance on AI without increasing cognitive load, making it the most effective intervention. This suggests that cognitive forcing functions centered on argument analysis better support metacognitive monitoring than those based on hypothesis testing alone or on a combination of argument analysis and hypothesis testing. While participants' cognitive disposition and prior GenAI familiarity primarily shaped their willingness to revise judgments, CFF design determined how critically they engaged with AI outputs.
\end{shaded}

% !TEX root =  main.tex
%%%%%%%%%%%%%%%%%%%%%%%%%%%%%%%%%%%%%%%%%%%%%%%%%%%%%%%%%%
%%%%%%%%%%%%%%%%%%%%%%%%%%%%%%%%%%%%%%%%%%%%%%%%%%%%%%%%%%

\subsection{Limitations and Future Work}\label{sec:discussion.limitations}
Next, we discuss limitations of our study that may have influenced our findings and propose directions for future work to mitigate or address them.

First, while our participant sample consisted of professional and semi-professional knowledge workers familiar with GenAI tools, it does not fully capture the diversity of expertise, organizational constraints, or time pressures present in real-world settings. In practice, such factors may shape how much cognitive effort users are willing to invest in reflective evaluation, potentially reducing or amplifying the effects of CFFs. In addition, our tasks were designed within simulated writing-focused knowledge work scenarios with predefined deliverables. Although this choice reflects a broad and heterogeneous user population, it limits direct generalization to more technical domains such as programming, data analysis, or spreadsheet modeling, where different error structures and evaluative strategies may apply. Future work should examine plan-centered CFFs in field deployments and across a wider range of knowledge work domains.

Second, our results may depend on the specific tasks and synthetic error types used in the study; although these errors were designed to reflect common plan-level errors in GenAI systems, we did not systematically vary error type difficulty. Outcomes may also be sensitive to the quality of the underlying LLM-generated plans and drafts, which are likely to evolve as models improve, potentially altering both baseline trust and the utility of forcing functions. To mitigate this limitation, future work should evaluate plan-centered CFFs across a broader range of tasks, error profiles, and interaction styles, ideally embedded within real-world AI tools and workflows.

Finally, designing effective plan-centered CFF prompts, particularly high-quality \assumptioncff{} and \whatifcff{} prompts, remains a non-trivial challenge that may influence effectiveness in real deployments. Future work should investigate methods for adaptive and context-aware generation of CFF prompts, including human-in-the-loop generation, model-assisted prompt validation, and dynamic tailoring of prompts to task complexity, plan structure, and user expertise, in order to ensure that forcing functions remain both meaningful and lightweight at scale.

% !TEX root =  main.tex
%%%%%%%%%%%%%%%%%%%%%%%%%%%%%%%%%%%%%%%%%%%%%%%%%%%%%%%%%%
%%%%%%%%%%%%%%%%%%%%%%%%%%%%%%%%%%%%%%%%%%%%%%%%%%%%%%%%%%

\section{Conclusion}\label{sec:conclusion}
In this paper, we investigate how cognitive forcing functions (CFFs) can support critical evaluation of AI-generated execution plans in AI-assisted writing. As GenAI systems increasingly present users with multi-step plans followed by long-form outputs, the evaluation task shifts from judging a single result to assessing whether the AI has adopted an appropriate approach. Despite the growing prevalence of such plan-based workflows, prior work has not examined the design of CFFs around on AI-generated plans.

Across a large-scale comparative online experiment and a complementary think-aloud interview study, we compared three plan-centered CFFs—\assumptioncff{}, targeting argument analysis, and \whatifcff{}, targeting hypothesis testing—their combination, along with a no-CFF control (\nocff{}). Our results show that \assumptioncff{}, which prompts users to analyze the assumptions and premises underlying AI plans, most effectively reduced overreliance on AI plans and drafts without increasing cognitive load. In contrast, \whatifcff{}, which prompts users to engage in counterfactual reasoning and test alternative hypothesis, was perceived as more helpful, but did not yield corresponding improvements in behavioral outcomes (such as underreliance and overreliance rates) and was associated with higher cognitive load.

These findings show that not all critical thinking processes are equally well suited to plan evaluation contexts. Specifically, in the context of AI-assisted writing, CFFs grounded in argument analysis, which make implicit assumptions and premises explicit, provide a low-friction and behaviorally effective mechanism for supporting metacognitive monitoring and trust calibration. In contrast, CFFs grounded in hypothesis testing require users to simulate alternative scenarios and propagate consequences across a plan, a cognitively demanding process that may misalign with the constraints of real-world ``review'' tasks. Finally, the observed gap between perceived helpfulness and actual performance highlights the importance of evaluating plan-centered CFFs using behavioral measures of calibration and overreliance, rather than relying on subjective impressions alone. Our results highlight the value of exploring plan-centered CFF designs that are explicitly grounded in established cognitive and educational theory, moving beyond ad hoc interaction heuristics toward more systematic support for critical evaluation.

% !TEX root =  main.tex
%%%%%%%%%%%%%%%%%%%%%%%%%%%%%%%%%%%%%%%%%%%%%%%%%%%%%%%%%%
%%%%%%%%%%%%%%%%%%%%%%%%%%%%%%%%%%%%%%%%%%%%%%%%%%%%%%%%%%

\section*{Acknowledgments}\label{sec:ack}
We would like to thank Lev Tankelevitch for insightful discussions on critical thinking and metacognitive frameworks that informed the design of the cognitive forcing functions, as well as for valuable feedback during the pilot study of the user study platform. We also thank Emily Doherty for her helpful insights during the pilot study of the platform.

\bibliographystyle{ACM-Reference-Format}
\bibliography{main}

\clearpage
\appendix
% !TEX root =  main.tex
%%%%%%%%%%%%%%%%%%%%%%%%%%%%%%%%%%%%%%%%%%%%%%%%%%%%%%%%%%
%%%%%%%%%%%%%%%%%%%%%%%%%%%%%%%%%%%%%%%%%%%%%%%%%%%%%%%%%%

\section{Appendix}\label{sec:appendix}
This appendix provides supplementary materials and analyses that support the main findings of the paper. It is organized as follows:
\begin{itemize}
    \item Appendix~\ref{appendix.rq1.results} presents the generalized linear mixed-effects modeling of the readiness-revision metric analyzed for RQ1 (Section 5.1).
    %%%%
    \item  Appendix~\ref{appendix.rq2.results} presents the generalized linear mixed-effects modeling of the accuracy metric analyzed for RQ2 (Section 5.2).
    %%%%
    \item Appendix~\ref{appendix.presurvey} presents the detailed questionnaire used in the presurvey phase of both the online experiment and the interview study (Section 4).
    %%%%%
    \item Appendix~\ref{appendix.projects} presents full details of the knowledge work tasks, including task descriptions, task artifacts, corresponding AI-generated execution plans and drafts, and the post-task survey questions used to assess cognitive load (Section 4.1).
    %%%%
    \item Appendix~\ref{appendix.postsurvey} presents the questionnaire used in the postsurvey phase of the online experiment (Section 4).
    %%%%
    \item Appendix~\ref{appendix.interview} presents the structured comparison questions used in the interview study to obtain participants' reflections on the three CFF conditions (Section 4).
    %%%%
\end{itemize}

%%%%%%%%%%%%%%%%%%%%%%%
% !TEX root =  main.tex
%%%%%%%%%%%%%%%%%%%%%%%%%%%%%%%%%%%%%%%%%%%%%%%%%%%%%%%%%%
%%%%%%%%%%%%%%%%%%%%%%%%%%%%%%%%%%%%%%%%%%%%%%%%%%%%%%%%%%

\section{Generalized Linear Mixed Model Analysis of Changes in AI Response Readiness}\label{appendix.rq1}
In this section, we describe the generalized linear mixed model (GLMM) analyses used to investigate changes in participants' AI response readiness. We first outline the model specification, then compare model variants, and finally present interpretations of the selected model.

\subsection{Model Specification and Variants}\label{appendix.rq1.models}
To investigate how participants' cognitive dispositions, need for cognition, and general familiarity with GenAI relate to their likelihood of revising judgments about an AI's response readiness, we modeled the outcome variable, ``Change in AI response readiness,'' as a binary logistic mixed-effects model. Each observation represents whether a participant revised their initial assessment of an AI response after completing a task, coded as a \emph{Bernoulli response}:

\begin{align*}  
y_{it} &=  
\begin{cases}  
1, & \text{if participant $i$ changed opinion on task $t$},\\  
0, & \text{if participant $i$ did not change opinion on task $t$},  
\end{cases} \\  
y_{it} &\sim \mathrm{Bernoulli}(\pi_{it}), \text{with} \; \mathrm{logit}(\pi_{it}) = \log\!\left(\frac{\pi_{it}}{1 - \pi_{it}}\right).  
\end{align*}  

The model uses the \emph{logit} link to estimate the log-odds of change in opinion, which are then transformed to probabilities. Repeated measures are accounted for through participant-level random intercepts, as each participant completed all five tasks. Participant-specific presurvey responses provide continuous/numeric covariates: need for cognition (NFC), actively open-minded thinking (AOT), and self-reported familiarity, usage, and trust in GenAI tools rated on a 5-point Likert scale. The primary model includes the CFF group as a fixed factor, with presurvey covariates and task identifiers as additional predictors as shown in Equation~\ref{eq:logit_model_alg}.

%%%%%%%%%%%%% Opinion change model: CFF group as fixed factor
\begin{equation}
\label{eq:logit_model_alg}
\mathrm{logit}(\pi_{it}) 
= \beta_0 
+ \boldsymbol{\beta}_{\text{alg}}^{\top} \mathbf{1}\{g_i\}
+ \boldsymbol{\beta}_{x}^{\top} \mathbf{x}_i
+ \boldsymbol{\gamma}^{\top} \mathbf{z}_t
+ b_i,
\tag{Model 1: Readiness outcome with CFF group factor.}
\end{equation}

\noindent
where
\[
y_{it} \sim \mathrm{Bernoulli}(\pi_{it}), 
\qquad
\mathrm{logit}(\pi_{it}) = 
\log\!\left(\frac{\pi_{it}}{1 - \pi_{it}}\right),
\qquad
b_i \sim \mathcal{N}(0, \sigma_b^2),
\]
and
\begin{itemize}
    \item $y_{it} \in \{0,1\}$ indicates whether participant $i$ changed opinion on task $t$;
    \item $\mathbf{1}\{g_i\}$ encodes the CFF condition for participant $i$ (a 4-level factor);
    \item $\mathbf{x}_i$ contains continuous/numeric presurvey covariates (AOT, NFC, familiarity, usage, trust);
    \item $\mathbf{z}_t$ encodes fixed proejct effects (since every participant completed all $5$ tasks);
    \item $b_i$ is a participant-level random intercept capturing repeated measures.
\end{itemize}
%%%%%%%%%%%%%%%%%%%%%%%%%%%%%%%%%%%

Two additional model variants were explored to assess the contribution of the CFF factor and the value of including its interaction with presurvey covariates as shown in Equations~\ref{eq:logit_model_noalg} and \ref{eq:logit_model_interaction}.

%%%%%%%%%%%%%%%%%%% No-alg variant %%%%%%%%%%
\begin{equation}
\label{eq:logit_model_noalg}
\mathrm{logit}(\pi_{it}) 
= \beta_0 
+ \boldsymbol{\beta}_{x}^{\top} \mathbf{x}_i
+ \boldsymbol{\gamma}^{\top} \mathbf{z}_t
+ b_i,
\tag{Model 2: Readiness outcome without CFF group factor.}
\end{equation}
%%%%%%%%%%%%%%%%%%%%%%%%%%%%

%%%%%%%%%%%%%%%%%% With interaction term between alg and presurvey
\begin{equation}
\label{eq:logit_model_interaction}
\mathrm{logit}(\pi_{it}) 
= \beta_0 
+ \boldsymbol{\beta}_{\text{alg}}^{\top} \mathbf{1}\{g_i\}
+ \boldsymbol{\beta}_{x}^{\top} \mathbf{x}_i
+ \boldsymbol{\beta}_{\text{int}}^{\top}\big(\mathbf{1}\{g_i\} \otimes \mathbf{x}_i\big)
+ \boldsymbol{\gamma}^{\top} \mathbf{z}_t
+ b_i,
\tag{Model 3: Readiness outcome with CFF $\times$ presurvey interactions.}
\end{equation}
%%%%%%%%%%%%%%%%%%%%%%%
\noindent
where, relative to the base model (Eq.~\ref{eq:logit_model_alg}),
\begin{itemize}
    \item Eq.~\eqref{eq:logit_model_noalg} excludes the CFF condition term 
          $\boldsymbol{\beta}_{\text{alg}}^{\top} \mathbf{1}\{g_i\}$ to test 
          whether the CFF condition contributes to model fit;
    \item Eq.~\eqref{eq:logit_model_interaction} extends the base model by adding 
          the interaction term 
          $\boldsymbol{\beta}_{\text{int}}^{\top}\big(\mathbf{1}\{g_i\} \otimes \mathbf{x}_i\big)$, 
          which allows the effects of the presurvey covariates $\mathbf{x}_i$ to vary 
          across CFF conditions.
\end{itemize}
%%%%%%%%%%%%%%%%%%%%%%%%%%%

%%%%%%%%%%%%%%%%%%%%%%%%%%%%%%%%%%%%%%%%
\subsection{Model Comparison and Selection}\label{appendix.rq1.modelcomparison}
To compare between these three models, we conducted a likelihood-ratio test (LRT) between the original model and the two variants to determine the effect of the CFF group on model fit, as well as inclusion of the more complex model containing the CFF interaction term with the presurvey covariates. We report significance (p-values) based on $\chi^2$ statistics for the comparison between the models, and their fit statistics. The results are presented in Table~\ref{tab:rq1.model_comparison_three}. We find that, removing the CFF group to the baseline model (with the CFF group) did not significantly improve model fit ($\chi^2 = 4.05, p = 0.26$). Furthermore, introducing the additional CFF~$\times$~presurvey interaction terms led to modest improvements in model likelihood, but this gain was not statistically significant ($\chi^2 = 23.21, p = 0.08$) and increased AIC/BIC values, suggesting overfitting. Although the inclusion of the CFF group did not significantly improve model fit statistically, it was retained in the final model due to its experimental relevance and interpretive importance in estimating adjusted opinion-change probabilities across conditions.

\begin{shaded}
\noindent    
    \textbf{Model selection.} The main-effects model including the CFF group (\ref{eq:logit_model_alg}) was retained as the most parsimonious and interpretable representation of changes in participants' judgments of AI response readiness (see highlighted row in Table~\ref{tab:rq1.model_comparison_three}).
\end{shaded}

%%%%%
\begin{table}[t]
\centering
\caption{Opinion-change model fit statistics and likelihood-ratio test (LRT) comparisons between nested GLMMs.}
\label{tab:rq1.model_comparison_three}
\scalebox{0.85}{
    \begin{tabular}{lccccccc}
    \hline
    \textbf{Model for readiness} & \textbf{npar} & \textbf{AIC} & \textbf{BIC} & \textbf{logLik} & \textbf{Compared to} & \textbf{$\Delta \chi^2$ (df)} & \textbf{$p$-value} \\
    %%%%
    \hline
    %%%%%
    Without CFF group & 11 & 926.81 & 981.54 & $-452.41$ & -- & -- & -- \\
    %%%%%
    \rowcolor{green!20}With CFF group & 14 & 928.76 & 998.41 & $-450.38$ & Without CFF group & 4.05 (3) & 0.256 \\
    %%%%%
    With CFF group $\times$ presurvey interactions & 29 & 935.54 & 1079.83 & $-438.77$ & With CFF group & 23.21 (15) & 0.080 \\
    %%%%%
    \hline
    \end{tabular}
}
\end{table}
%%%%%

%%%%%%%%%%%%%%%%%%%%%%%%%%%%%%%%%%%%%%%%%%%%%%%%%%%%%%%%%%%%
\subsection{Interpretation of the Selected Model}\label{appendix.rq1.results}
The selected model (Model~\ref{eq:logit_model_alg}) includes CFF condition as a between-subjects fixed factor, task as a within-subjects fixed factor, and participant as a random intercept. This formulation acknowledges the experimental manipulation of CFF conditions while maintaining parsimony, as adding CFF–presurvey interactions did not significantly enhance explanatory power.

\begin{figure*}[!t]
    \centering
    %%%%%%
    \begin{subfigure}[c]{0.55\textwidth}
        \centering
       \includegraphics[width=\textwidth]{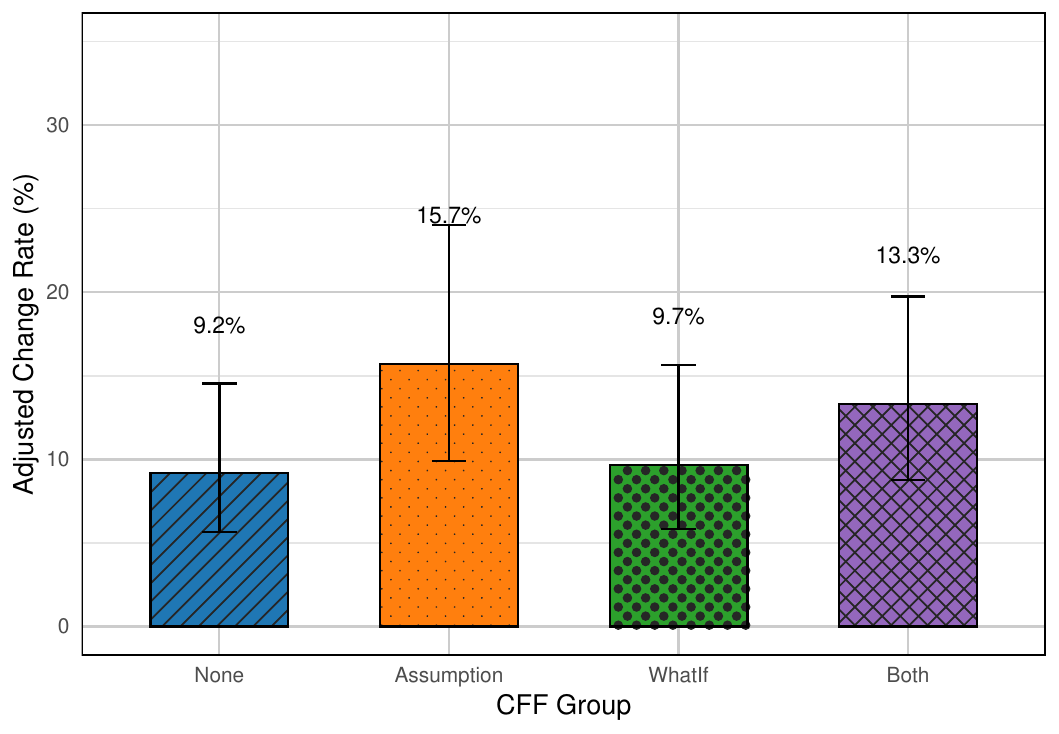}
        \caption{Adjusted Change Rate per CFF group}
        \label{fig:rq1.changerate.cff.adjusted}
    \end{subfigure}
    %%%%%%%
    \\
    \begin{subfigure}[c]{0.67\textwidth}
        \centering
       \includegraphics[width=\textwidth]{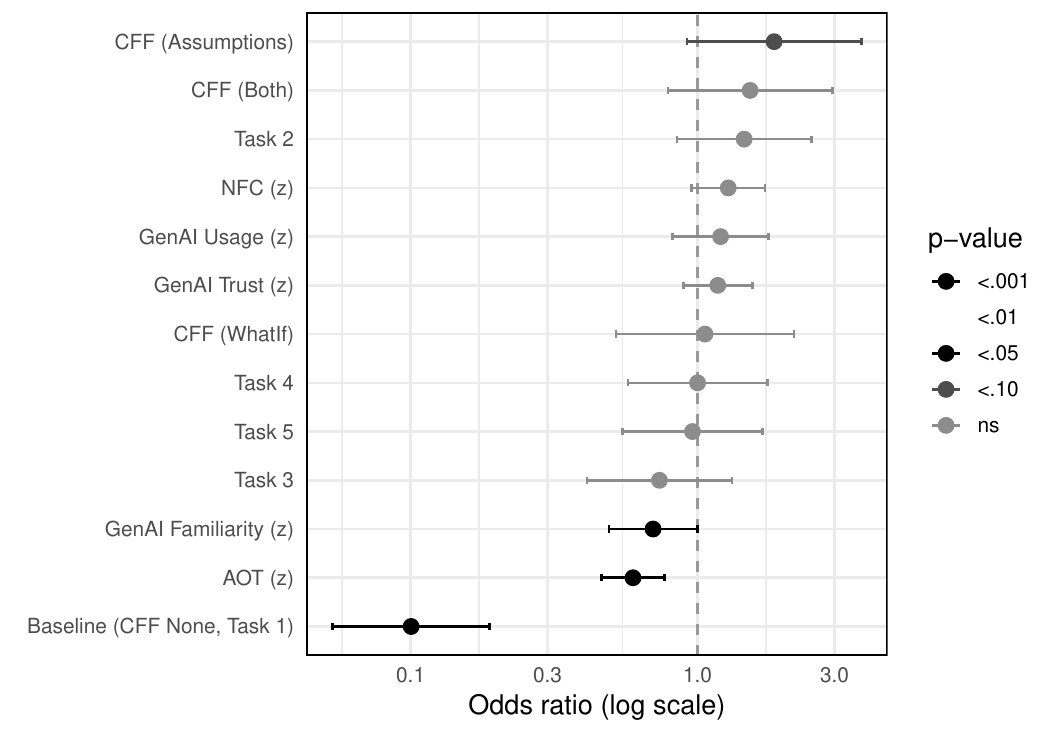}
        \caption{Odds Ratio per Factor}
        \label{fig:rq1.changerate.oddsratio}
    \end{subfigure}
    %%%%%%%%
    \caption{Results of the GLMM on change in AI response readiness: (a) model-adjusted predicted change rates for each CFF group, and 
(b) odds ratios with 95\% confidence intervals for all model factors.
}   
    \Description{}
    \label{fig:appendix.rq1.modelparams}
\end{figure*}

Predicted change rates (in AI response readiness) were computed from the model after adjusting for presurvey covariates and task effects.  
Figure~\ref{fig:rq1.changerate.cff.adjusted} shows these adjusted probabilities. Although \assumptioncff{} showed the highest estimated change rate, the differences between groups were not statistically significant. \nocff{} condition exhibited the lowest change rate overall. These model-adjusted results are consistent with the descriptive (raw) change-rate patterns observed in Section 5.1 —\assumptioncff{} continues to show the numerically highest rate—however, the previously significant difference between the \assumptioncff{} and \nocff{} groups is no longer statistically reliable once we account for individual covariates and repeated-measure effects.
       
To interpret the influence of individual predictors, Figure~\ref{fig:rq1.changerate.oddsratio} presents odds ratios for all fixed effects. Results show that participants' \textbf{AOT} and \textbf{GenAI Familiarity} scores had the strongest effects: higher AOT was associated with significantly lower odds of opinion change ($OR = 0.60$, 95\% CI [0.46, 0.77], $p < 0.001$), and higher GenAI Familiarity showed a weaker but still negative effect ($OR = 0.70$, 95\% CI [0.49, 1.00], $p = 0.048$). No other CFF conditions, presurvey measures, or task effects were statistically significant.

\begin{shaded}
\noindent
   \textbf{Key finding.} Participants' cognitive disposition and familiarity with GenAI tools influence their opinion rigidity regarding AI response readiness more strongly than the specific CFF condition to which they were exposed.
\end{shaded}

%%%%%%%%%%%%%%%%%%%%%%%%% plots for opinion change
%%%%%%%%%%%%%%%%%%%%%%%%%%%%%%%%%%%%%%%%%%%%%%
%%%%%%%%%%%%%%%%%%%%%

%%%%%%%%%%%%%%%%%%%%%%%
% !TEX root =  main.tex
%%%%%%%%%%%%%%%%%%%%%%%%%%%%%%%%%%%%%%%%%%%%%%%%%%%%%%%%%%
%%%%%%%%%%%%%%%%%%%%%%%%%%%%%%%%%%%%%%%%%%%%%%%%%%%%%%%%%%

\section{Generalized Linear Mixed Model Analysis of Accuracy Rate of AI Response}\label{appendix.rq2}
In this section, we describe the GLMM analyses used to investigate participants' accuracy in assessing AI responses. We first outline the model specification, then compare model variants, and finally present interpretations of the selected model.

\subsection{Model Specification and Variants}\label{appendix.rq2.models}
To investigate how participants' cognitive dispositions, need for cognition, general familiarity with GenAI, and the tendency to change their opinion on AI response readiness, relate to their accuracy rate on the AI response, we modeled the outcome variable, ``Accuracy Rate,'' as a binary logistic mixed-effects model (GLMM). Each observation corresponds to a participant's correctness in identifying the presence or absence of an error in an AI response following task completion, modeled as a Bernoulli outcome:

\begin{align*}  
y_{it} &=  
    \begin{cases}
      1, & \text{if participant $i$ correctly identified an error for task $t$}, \\
      0, & \text{otherwise},
    \end{cases} \\  
y_{it} &\sim \mathrm{Bernoulli}(\pi_{it}), \text{with} \; \mathrm{logit}(\pi_{it}) = \log\!\left(\frac{\pi_{it}}{1 - \pi_{it}}\right).  
\end{align*} 

The model uses the \emph{logit} link to estimate the log-odds of accuracy, which are then transformed to probabilities.  Repeated measures are accounted for through participant-level random intercepts, as each participant completed all five tasks. Participant-specific presurvey responses provide continuous/numeric covariates: need for cognition (NFC), actively open-minded thinking (AOT), and self-reported familiarity, usage, and trust in GenAI tools rated on a 5-point Likert scale. The primary model includes the CFF group as a fixed factor, with presurvey covariates and task identifiers as additional predictors as shown in Equation~\ref{eq:logit_model_alg_overrel}.

%%%%%%%%%%%%% Overrel model: CFF group as fixed factor
\begin{equation}
\label{eq:logit_model_alg_overrel}
\mathrm{logit}(\pi_{it}) 
= \beta_0 
+ \boldsymbol{\beta}_{\text{alg}}^{\top} \mathbf{1}\{g_i\}
+ \boldsymbol{\beta}_{x}^{\top} \mathbf{x}_i
+ \boldsymbol{\gamma}^{\top} \mathbf{z}_t
+ b_i,
\tag{Model 1: Accuracy outcome with CFF group factor.}
\end{equation}

\noindent
where
\[
y_{it} \sim \mathrm{Bernoulli}(\pi_{it}), 
\qquad
\mathrm{logit}(\pi_{it}) = 
\log\!\left(\frac{\pi_{it}}{1 - \pi_{it}}\right),
\qquad
b_i \sim \mathcal{N}(0, \sigma_b^2),
\]
and
\begin{itemize}
    \item $y_{it} \in \{0,1\}$ indicates whether participant $i$ was accurate on task $t$;
    \item $\mathbf{1}\{g_i\}$ encodes the CFF condition for participant $i$ (a 4-level factor);
    \item $\mathbf{x}_i$ contains continuous/numeric presurvey covariates (AOT, NFC, familiarity, usage, trust);
    \item $\mathbf{z}_t$ encodes fixed task effects (since every participant completed all $5$ tasks);
    \item $b_i$ is a participant-level random intercept capturing repeated measures.
\end{itemize}
%%%%%%%%%%%%%%%%%%%%%%%%%%%%%%%%%%%

Three additional model variants were explored to assess the contribution of the CFF factor, change in opinion about AI response readiness factor, and the value of including CFF groups' interaction with presurvey covariates as shown in Equations~\ref{eq:logit_model_noalg_overrel}, \ref{eq:logit_model_change_overrel} and \ref{eq:logit_model_interaction_overrel}.

%%%%%%%%%%%%%%%%%%% No-alg variant %%%%%%%%%%
\begin{equation}
\label{eq:logit_model_noalg_overrel}
\mathrm{logit}(\pi_{it}) 
= \beta_0 
+ \boldsymbol{\beta}_{x}^{\top} \mathbf{x}_i
+ \boldsymbol{\gamma}^{\top} \mathbf{z}_t
+ b_i,
\tag{Model 2: Accuracy outcome without CFF group factor.}
\end{equation}
%%%%%%%%%%%%%%%%%%%%%%%%%%%%

%%%%%%%%%%%%%%%%%% With change as a factor and alg
\begin{equation}
\label{eq:logit_model_change_overrel}
\mathrm{logit}(\pi_{it}) 
= \beta_0 
+ \boldsymbol{\beta}_{\text{alg}}^{\top} \mathbf{1}\{g_i\}
+ \boldsymbol{\beta}_{x}^{\top} \mathbf{x}_i
+ \beta_{c}\, c_{it}
+ \boldsymbol{\gamma}^{\top} \mathbf{z}_t
+ b_i,
\tag{Model 3: Accuracy outcome with opinion change as a factor.}
\end{equation}
%%%%%%%%%%%%%%%%%%%%%%%

%%%%%%%%%%%%%%%%%% With interaction term between alg and presurvey
\begin{equation}
\label{eq:logit_model_interaction_overrel}
\mathrm{logit}(\pi_{it}) 
= \beta_0 
+ \boldsymbol{\beta}_{\text{alg}}^{\top} \mathbf{1}\{g_i\}
+ \boldsymbol{\beta}_{x}^{\top} \mathbf{x}_i
+ \boldsymbol{\beta}_{\text{int}}^{\top}\big(\mathbf{1}\{g_i\} \otimes \mathbf{x}_i\big)
+ \boldsymbol{\gamma}^{\top} \mathbf{z}_t
+ b_i,
\tag{Model 4: Accuracy outcome with CFF $\times$ presurvey interactions.}
\end{equation}
%%%%%%%%%%%%%%%%%%%%%%%
\noindent
where, relative to the base model (Eq.~\ref{eq:logit_model_alg_overrel}),
\begin{itemize}
    \item Eq.~\eqref{eq:logit_model_noalg_overrel} excludes the CFF condition term 
          $\boldsymbol{\beta}_{\text{alg}}^{\top} \mathbf{1}\{g_i\}$ to test 
          whether the CFF condition contributes to model fit;
    \item Eq.~\eqref{eq:logit_model_change_overrel} includes the per–task opinion-change indicator
          $c_{it}$ with coefficient $\beta_c$, capturing whether revising AI readiness relates to accuracy;
    \item Eq.~\eqref{eq:logit_model_interaction_overrel} extends the base model by adding 
          the interaction term 
          $\boldsymbol{\beta}_{\text{int}}^{\top}\big(\mathbf{1}\{g_i\} \otimes \mathbf{x}_i\big)$, 
          which allows the effects of the presurvey covariates $\mathbf{x}_i$ to vary 
          across CFF conditions.
\end{itemize}
%%%%%%%%%%%%%%%%%%%%%%%%%%%

%%%%%%%%%%%%%%%%%%%%%%%%%%%%%%%%%%%%%%%%
\subsection{Model Comparison and Selection}\label{appendix.rq2.modelcomparison}
To compare between these four models, we conducted a likelihood-ratio test (LRT) between the original model and the three variants to determine the effect of the CFF group on model fit, opinion-change factor, as well as inclusion of the more complex model containing the CFF interaction term with the presurvey covariates. We report significance (p-values) based on $\chi^2$ statistics for the comparision between the models, and their fit statistics. The results are presented in Table~\ref{tab:rq2.model_comparison_four}. We find that including the CFF group significantly improved model fit relative to the baseline model without it ($\chi^2 = 17.23$, $p = 0.0006$), indicating that participants' likelihood of accuracy on AI responses varied meaningfully across CFF conditions. Adding the opinion-change factor showed a marginal improvement in model fit ($\chi^2 = 3.55$, $p = 0.06$), suggesting a weak trend toward reduced accuracy among participants who revised their AI response readiness ratings. In contrast, the inclusion of the more complex model with CFF~$\times$~presurvey interaction terms did not improve fit ($\chi^2 = 11.04$, $p = 0.27$) and resulted in higher AIC and BIC values, consistent with model overfitting. Although including the opinion-change factor slightly improved fit statistics, the model with only the CFF group was retained as the most parsimonious and theoretically meaningful specification, as the additional variable did not reach statistical significance and the primary focus remained on estimating CFF-related differences in accuracy. Nevertheless, we also examined the odds ratios from the model including the opinion-change factor to assess the direction and magnitude of this effect, which we discuss in Appendix~\ref{appendix.rq2.results}.

\begin{shaded}
\noindent
    \textbf{Model selection.} The main-effects model including the CFF group (\ref{eq:logit_model_alg_overrel}) was retained as the most parsimonious and interpretable representation of changes in participants' accuracy w.r.t. AI responses (see highlighted row in Table~\ref{tab:rq2.model_comparison_four}).
\end{shaded}

%%%%%
\begin{table}[t]
\centering
\caption{Accuracy model fit statistics and likelihood-ratio test (LRT) comparisons between nested GLMMs.}
\label{tab:rq2.model_comparison_four}
\scalebox{0.85}{
    \begin{tabular}{lccccccl}
    \hline
    \textbf{Model for accuracy} & \textbf{npar} & \textbf{AIC} & \textbf{BIC} & \textbf{logLik} & \textbf{Compared to} & \textbf{$\Delta \chi^2$ (df)} & \textbf{$p$-value} \\
    %%%%
    \hline
    %%%%%
    Without CFF group & 11 & 980.59 & 1035.3 & $-479.29$ & -- & -- & -- \\
    %%%%%
    \rowcolor{green!20}With CFF group & 14 & 969.35 & 1039.0 & $-470.68$ & Without CFF group & 17.23 (3) & 0.0006 \\
    %%%%%
    %%%%%
    With opinion change & 15 & 967.81 & 1042.4 & $-468.90$ & With CFF group & 3.55 (1) & 0.06 \\
    %%%%%
    With CFF group $\times$ presurvey interactions & 23 & 976.32 & 1090.8 & $-465.16$ & With CFF group & 11.04 (9) & 0.27 \\
    %%%%%
    \hline
    \end{tabular}
}
\end{table}
%%%%%

%%%%%%%%%%%%%%%%%%%%%%%%%%%%%%%%%%%%%%%%%%%%%%%%%%%%%%%%%%%%
\subsection{Interpretation of the Selected Model}\label{appendix.rq2.results}
The selected model (\ref{eq:logit_model_alg_overrel}) includes CFF condition as a between-subjects fixed factor, task as a within-subjects fixed factor, and participant as a random intercept. This formulation acknowledges the experimental manipulation of CFF conditions while maintaining parsimony, as adding CFF–presurvey interactions did not significantly enhance explanatory power.

\begin{figure*}[!t]
    \centering
    %%%%%%
    \begin{subfigure}[c]{0.55\textwidth}
        \centering
       \includegraphics[width=\textwidth]{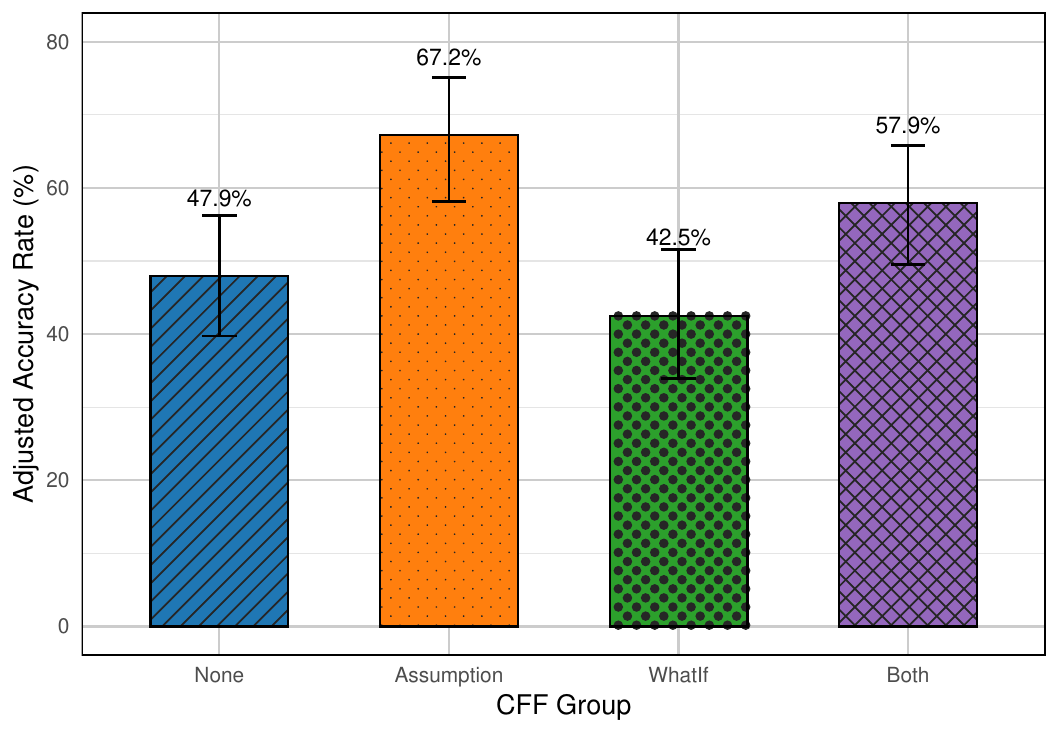}
        \caption{Adjusted Accuracy Rate per CFF group}
        \label{fig:rq2.overrel.cff.adjusted}
    \end{subfigure}
    %%%%%%%
    \\
    \begin{subfigure}[c]{0.67\textwidth}
        \centering
       \includegraphics[width=\textwidth]{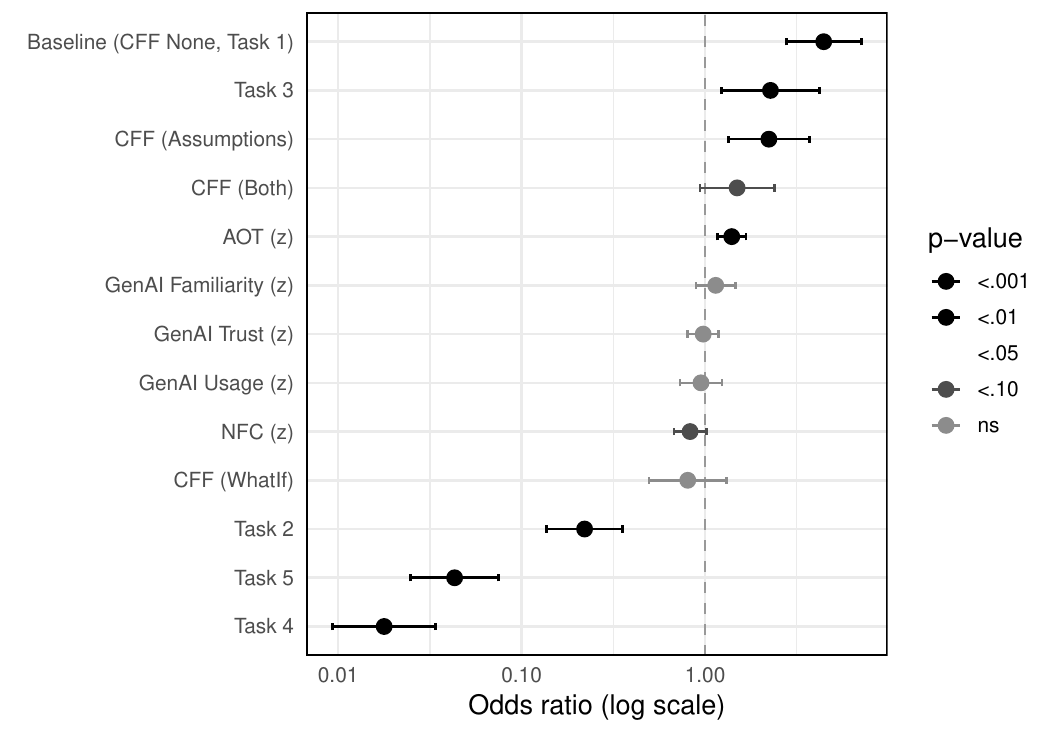}
        \caption{Odds Ratio per Factor}
        \label{fig:rq2.overrel.oddsratio}
    \end{subfigure}
    %%%%%%%%
    \caption{Results of the GLMM on accuracy in detecting errors in AI response: (a) model-adjusted predicted accuracy rates for each CFF group, and 
(b) odds ratios with 95\% confidence intervals for all model factors.
}
    \Description{}
    \label{fig:appendix.rq2.modelparams}
\end{figure*}

Predicted accuracy rate ($\pi_{it} \times 100$) were computed from the selected model after adjusting for presurvey covariates and task effects. Figure~\ref{fig:rq2.overrel.cff.adjusted} shows the adjusted accuracy rates for each CFF condition. Participants in the \assumptioncff{} condition exhibited the lowest model-adjusted accuracy rate ($33\%$), followed by the \bothcff{} condition ($42\%$), the \nocff{} baseline condition ($52\%$), and the \whatifcff{} condition, which showed the highest estimated rate ($58\%$). Pairwise comparisons (Tukey-adjusted) indicated that the \assumptioncff{} condition differed significantly from \whatifcff{} ($p = 0.01$) and \bothcff{} ($p = 0.06$), but not from the \nocff{} baseline ($p = 0.33$). These results suggest that exposing participants to the \assumptioncff{} CFF reduced their likelihood of incorrectly detecting the absence/presence of errors in AI outputs compared with other CFF conditions. These model-adjusted results are consistent with the descriptive (raw) change-rate patterns observed in Section 5.2.

To interpret the influence of individual predictors, Figure~\ref{fig:rq2.overrel.oddsratio} presents odds ratios for all fixed effects. Consistent with the pairwise contrasts, the \assumptioncff{} condition was associated with significantly lower odds of accurate responses relative to the control ($OR = 0.45$, 95\%~CI [0.27, 0.75], $p = .002$). The \bothcff{} condition showed a smaller, marginal reduction ($OR = 0.67$, 95\%~CI [0.42, 1.07], $p = .09$), while the \whatifcff{} condition did not differ from the control ($OR = 1.24$, $p = .38$). Among individual difference measures, higher \textbf{AOT} was strongly associated with reduced accuracy rates ($OR = 0.72$, 95\%~CI [0.60, 0.86], $p < .001$), whereas higher \textbf{NFC} showed a marginal positive association ($OR = 1.21$, 95\%~CI [0.98, 1.48], $p = .07$). Presurvey GenAI familiarity, usage, and trust did not significantly predict accuracy rates once CFF condition and cognitive dispositions were accounted for. Task effects were substantial, with certain tasks (especially Tasks~2,4 and 5) showing higher higher probabilities of accurate responses, indicating strong task-level influences.

\begin{shaded}
\noindent
    \textbf{Key finding.} Participants exposed to the \assumptioncff{} CFF group showed the lowest adjusted accuracy rates, even after controlling for presurvey and task effects. In addition, higher actively open-minded thinking (AOT) scores predicted reduced accuracy across conditions, suggesting that both CFF condition and dispositional (cognitive style) factors jointly shape how participants evaluate AI response.
\end{shaded}

%%%%%%%%%%%%%%%%%%%%%%%%% plots for opinion change
%%%%%%%%%%%%%%%%%%%%%%%%%%%%%%%%%%%%%%%%%%%%%%

\paragraph{\textbf{Corollary.}} Although the addition of the opinion-change factor (\textit{Change in AI response readiness}) did not significantly improve overall model fit, we investigated the extended model (\ref{eq:logit_model_change_overrel}) to explore the potential influence of this variable, given its relevance to reflective evaluation. We found that participants who revised their initial judgments of AI responses were less likely to incorrectly identify issues or lack thereof in AI responses ($OR = 0.66$, 95\%~CI [0.42, 1.02], $p = .06$). This marginal trend suggests that engaging in evaluative reflection via revising one's assessment of the AI's response, may slightly reduce the likelihood of incorrectly identifying errors in AI response, even though the effect does not reach statistical significance.

%%%%%%%%%%%%%%%%%%%%%

%%%%%%%%%%%%%%%%%%%%%%%
% !TEX root =  main.tex
%%%%%%%%%%%%%%%%%%%%%%%%%%%%%%%%%%%%%%%%%%%%%%%%%%%%%%%%%%
%%%%%%%%%%%%%%%%%%%%%%%%%%%%%%%%%%%%%%%%%%%%%%%%%%%%%%%%%%

\section{Details of the Presurvey}\label{appendix.presurvey}
In this section, we present the detailed questionnaire that the participants had to answer as part of the presurvey. This was presented before they began completing the knowledge work tasks assisted by AI, and was hosted on the same web-application.

%%%%%%%%%%%%%%%%%%%%%%%%%%%%%%%%%%
\subsection{Background Information}\label{appendix.presurvey.background}
\emph{Please tell us about your background}

\begin{enumerate}
    %%%%%%%
    \item What is your age-range?
        \begin{enumerate}[label=(\alph*)]
            \item 18--24
            \item 25--34
            \item 35--44
            \item 45--54
            \item 55+
            \item Prefer not to say
        \end{enumerate}
    %%%%%%%
    \item What is your gender?
        \begin{enumerate}[label=(\alph*)]
            \item Man
            \item Woman
            \item Non-binary/gender diverse
            \item Self-described: [Free response]
            \item Prefer not to say
        \end{enumerate}
    %%%%%%%
    \item What is your primary country of residence? [Free response]
    %%%%%%%
    \item What is your job-title? [Free response]
    %%%%%%%
    \item Which industry do you primarily work in?
        \begin{enumerate}[label=(\alph*)]
            \item Technology
            \item Healthcare
            \item Education
            \item Finance
            \item Manufacturing
            \item Retail
            \item Hospitality
            \item Other: [Free response]
        \end{enumerate}
    %%%%%%%
    \item How would you describe your level of programming experience?
        \begin{enumerate}[label=(\alph*)]
            \item (Score 1) None -- I have never programmed before. 
            \item (Score 2) Minimal -- I have not programmed but I can read and understand code. 
            \item (Score 3) Moderate -- I have done some programming but not regularly.
            \item (Score 4) Proficient -- I have programmed quite a bit but would not call myself an expert.
            \item (Score 5) Expert -- I regularly write code and consider myself an expert.
        \end{enumerate}
     %%%%%%%
     \item How would you describe your level of experience with structured writing, such as academic papers, technical reports, or creative writing tasks?
        \begin{enumerate}[label=(\alph*)]
            \item (Score 1) None -- I have little to no experience with structured writing.
            \item (Score 2) Minimal -- I have done some writing but not extensively.
            \item (Score 3) Moderate -- I write occasionally and feel comfortable doing so.
            \item (Score 4) Proficient -- I write frequently and with confidence.
            \item (Score 5) Expert -- I write regularly and consider myself highly skilled.
        \end{enumerate}
    %%%%%%%
\end{enumerate}

%%%%%%%%%%%%%%%%%%%%%%%%%%%%%%%%%%

\subsection{Experience of GenAI Tools'}\label{appendix.presurvey.genaiexp}
\emph{Please tell us about your experience with using GenAI tools.}

\begin{enumerate}
    %%%%%%%
    \item Are you familiar with GenAI tools such as ChatGPT, Copilot and others?
        \begin{enumerate}[label=(\alph*)]
            \item (Score 1) Not at all -- I have never heard of these tools before.
            \item (Score 2) Minimal –- I have heard of these tools but never used them.
            \item (Score 3) Moderate -- I have heard of these tools and have used them occasionally.
            \item (Score 4) Proficient -- I use these tools frequently and feel confident using them.
            \item (Score 5) Expert -- I use these tools regularly and consider myself highly skilled.
        \end{enumerate}
    %%%%%%%
    \item How often do you use GenAI tools at work?
        \begin{enumerate}[label=(\alph*)]
            \item (Score 1) Never -- I do not use GenAI tools in my work.
            \item (Score 2) Rarely -- I use them infrequently, only when prompted or required.
            \item (Score 3) Sometimes -- I use them occasionally for specific tasks.
            \item (Score 4) Often -- I use them regularly for a variety of work-related tasks.
            \item (Score 5) Very Frequently -- I rely on them heavily and use them in my daily workflow.
        \end{enumerate}
    %%%%%%%
   \item What GenAI tools do you typically use at work? [Free reponse]
    %%%%%%%
   \item What is your level of trust on GenAI tools?
        \begin{enumerate}[label=(\alph*)]
            \item (Score 1) Not at all -- I do not trust these tools at all.
            \item (Score 2) Minimal -- I have limited trust in these tools.
            \item (Score 3) Moderate -- I trust these tools in some situations.
            \item (Score 4) Quite a bit -- I generally trust these tools and find them reliable.
            \item (Score 5) A lot -- I have high trust in these tools and confidently rely on their outputs.
        \end{enumerate}
    %%%%%%%
\end{enumerate}

%%%%%%%%%%%%%%%%%%%%%%%%%%%%%%%%%%
\subsection{General Thinking Preference}
\label{appendix.presurvey.nfc}

\emph{This is a short questionnaire\footnote{These questions are adopted from the validated shorter version of the \emph{Need for Cognition} scale~\citep{lins2020very}. It consists of 6 items.} designed to understand your thinking preferences. There are no right or wrong responses. Please rate how much you agree with each statement on a scale of 1 (not at all) to 5 (fully agree).}

\begin{enumerate}
    %%%%%%%
    \item I would prefer complex to simple problems.
        \begin{enumerate}[label=(\alph*)]
            \item (Score 1) Not at all
            \item (Score 2) Slightly
            \item (Score 3) Moderately
            \item (Score 4) Quite a bit
            \item (Score 5) Fully agree
        \end{enumerate}
    %%%%%%%
    \item I would rather do something that requires little thought than something that is sure to challenge my thinking abilities.\textsuperscript{$\dagger$}
        \begin{enumerate}[label=(\alph*)]
            \item (Score 1) Not at all
            \item (Score 2) Slightly
            \item (Score 3) Moderately
            \item (Score 4) Quite a bit
            \item (Score 5) Fully agree
        \end{enumerate}
    %%%%%%%
    \item I really enjoy a task that involves coming up with new solutions to problems.
        \begin{enumerate}[label=(\alph*)]
            \item (Score 1) Not at all
            \item (Score 2) Slightly
            \item (Score 3) Moderately
            \item (Score 4) Quite a bit
            \item (Score 5) Fully agree
        \end{enumerate}
    %%%%%%%
     \item I would prefer a task that is intellectual, difficult, and important to one that is somewhat important but does not require much thought.
        \begin{enumerate}[label=(\alph*)]
            \item (Score 1) Not at all
            \item (Score 2) Slightly
            \item (Score 3) Moderately
            \item (Score 4) Quite a bit
            \item (Score 5) Fully agree
        \end{enumerate}
    %%%%%%%
    \item Thinking is not my idea of fun.\textsuperscript{$\dagger$}
        \begin{enumerate}[label=(\alph*)]
            \item (Score 1) Not at all
            \item (Score 2) Slightly
            \item (Score 3) Moderately
            \item (Score 4) Quite a bit
            \item (Score 5) Fully agree
        \end{enumerate}
    %%%%%%%
     \item I like to have the responsibility of handling a situation that requires a lot of thinking.
        \begin{enumerate}[label=(\alph*)]
            \item (Score 1) Not at all
            \item (Score 2) Slightly
            \item (Score 3) Moderately
            \item (Score 4) Quite a bit
            \item (Score 5) Fully agree
        \end{enumerate}
    %%%%%%%
\end{enumerate}
\textsuperscript{$\dagger$}\,Reverse-scored item.

%%%%%%%%%%%%%%%%%%%%%%%%%%%%%%%%%%
\subsection{General Thinking Disposition}
\label{appendix.presurvey.aot}

\emph{This is a short questionnaire\footnote{These questions are adopted from the validated \emph{Actively Open-Minded Thinking scale}~\citep{stanovich2023actively}. It consists of 13 items.} designed to understand how you approach new information. There are no right or wrong responses. Please rate how much you agree with each statement on a scale of 1 (disagree strongly) to 6 (agree strongly).}

\begin{enumerate}
    %%%%%%%
    \item People should always take into consideration evidence that goes against their opinions.
        \begin{enumerate}[label=(\alph*)]
            \item (Score 1) Disagree strongly
            \item (Score 2) Disagree moderately
            \item (Score 3) Disagree slightly
            \item (Score 4) Agree slightly
            \item (Score 5) Agree moderately
            \item (Score 6) Agree strongly
        \end{enumerate}
    %%%%%%%
   \item Changing your mind is a sign of weakness.\textsuperscript{$\dagger$}
        \begin{enumerate}[label=(\alph*)]
            \item (Score 1) Disagree strongly
            \item (Score 2) Disagree moderately
            \item (Score 3) Disagree slightly
            \item (Score 4) Agree slightly
            \item (Score 5) Agree moderately
            \item (Score 6) Agree strongly
        \end{enumerate}
    %%%%%%%
    %%%%%%%
   \item I like to think that my actions are motivated by sound reasons.
        \begin{enumerate}[label=(\alph*)]
            \item (Score 1) Disagree strongly
            \item (Score 2) Disagree moderately
            \item (Score 3) Disagree slightly
            \item (Score 4) Agree slightly
            \item (Score 5) Agree moderately
            \item (Score 6) Agree strongly
        \end{enumerate}
    %%%%%%%
    %%%%%%%
    \item It is important to stick to your opinions even when evidence is brought to bear against them.\textsuperscript{$\dagger$}
        \begin{enumerate}[label=(\alph*)]
            \item (Score 1) Disagree strongly
            \item (Score 2) Disagree moderately
            \item (Score 3) Disagree slightly
            \item (Score 4) Agree slightly
            \item (Score 5) Agree moderately
            \item (Score 6) Agree strongly
        \end{enumerate}
    %%%%%%%
    %%%%%%%
    \item Intuition is the best guide in making decisions.\textsuperscript{$\dagger$}
        \begin{enumerate}[label=(\alph*)]
            \item (Score 1) Disagree strongly
            \item (Score 2) Disagree moderately
            \item (Score 3) Disagree slightly
            \item (Score 4) Agree slightly
            \item (Score 5) Agree moderately
            \item (Score 6) Agree strongly
        \end{enumerate}
    %%%%%%%
    %%%%%%%
   \item Considering too many different opinions often leads to muddled thinking.\textsuperscript{$\dagger$}
        \begin{enumerate}[label=(\alph*)]
            \item (Score 1) Disagree strongly
            \item (Score 2) Disagree moderately
            \item (Score 3) Disagree slightly
            \item (Score 4) Agree slightly
            \item (Score 5) Agree moderately
            \item (Score 6) Agree strongly
        \end{enumerate}
    %%%%%%%
    %%%%%%%
     \item One should disregard evidence that conflicts with your current opinions.\textsuperscript{$\dagger$}
        \begin{enumerate}[label=(\alph*)]
            \item (Score 1) Disagree strongly
            \item (Score 2) Disagree moderately
            \item (Score 3) Disagree slightly
            \item (Score 4) Agree slightly
            \item (Score 5) Agree moderately
            \item (Score 6) Agree strongly
        \end{enumerate}
    %%%%%%%
    %%%%%%%
     \item Coming to decisions quickly is a sign of wisdom.\textsuperscript{$\dagger$}
        \begin{enumerate}[label=(\alph*)]
            \item (Score 1) Disagree strongly
            \item (Score 2) Disagree moderately
            \item (Score 3) Disagree slightly
            \item (Score 4) Agree slightly
            \item (Score 5) Agree moderately
            \item (Score 6) Agree strongly
        \end{enumerate}
    %%%%%%%
    %%%%%%%
     \item Allowing oneself to be convinced by a solid opposing argument is a sign of good character.
        \begin{enumerate}[label=(\alph*)]
            \item (Score 1) Disagree strongly
            \item (Score 2) Disagree moderately
            \item (Score 3) Disagree slightly
            \item (Score 4) Agree slightly
            \item (Score 5) Agree moderately
            \item (Score 6) Agree strongly
        \end{enumerate}
    %%%%%%%
    %%%%%%%
     \item If something I think feels right then I am comfortable, whether or not it is true.\textsuperscript{$\dagger$}
        \begin{enumerate}[label=(\alph*)]
            \item (Score 1) Disagree strongly
            \item (Score 2) Disagree moderately
            \item (Score 3) Disagree slightly
            \item (Score 4) Agree slightly
            \item (Score 5) Agree moderately
            \item (Score 6) Agree strongly
        \end{enumerate}
    %%%%%%%
    %%%%%%%
     \item A person should always consider new information.
        \begin{enumerate}[label=(\alph*)]
            \item (Score 1) Disagree strongly
            \item (Score 2) Disagree moderately
            \item (Score 3) Disagree slightly
            \item (Score 4) Agree slightly
            \item (Score 5) Agree moderately
            \item (Score 6) Agree strongly
        \end{enumerate}
    %%%%%%%
    %%%%%%%
     \item People should revise their conclusions in response to relevant new information.
        \begin{enumerate}[label=(\alph*)]
            \item (Score 1) Disagree strongly
            \item (Score 2) Disagree moderately
            \item (Score 3) Disagree slightly
            \item (Score 4) Agree slightly
            \item (Score 5) Agree moderately
            \item (Score 6) Agree strongly
        \end{enumerate}
    %%%%%%%
    %%%%%%%
     \item Certain opinions are just too important to abandon no matter how good a case can be made against them.\textsuperscript{$\dagger$}
        \begin{enumerate}[label=(\alph*)]
            \item (Score 1) Disagree strongly
            \item (Score 2) Disagree moderately
            \item (Score 3) Disagree slightly
            \item (Score 4) Agree slightly
            \item (Score 5) Agree moderately
            \item (Score 6) Agree strongly
        \end{enumerate}
    %%%%%%%
    %%%%%%%
\end{enumerate}
\textsuperscript{$\dagger$}\,Reverse-scored item.

%%%%%%%%%%%%%%%%%%%%%%%%%%%%%%%%%%%%%%%%%%%%%%%%%%%%

%%%%%%%%%%%%%%%%%%%%%

%%%%%%%%%%%%%%%%%%%%%%%
% !TEX root =  main.tex
%%%%%%%%%%%%%%%%%%%%%%%%%%%%%%%%%%%%%%%%%%%%%%%%%%%%%%%%%%
%%%%%%%%%%%%%%%%%%%%%%%%%%%%%%%%%%%%%%%%%%%%%%%%%%%%%%%%%%

\section{Details of the Knowledge Work Tasks}\label{appendix.projects}
In this section, we present the details of each knowledge-work task, the common questions asked to elicit participants' trust, feedback to the AI responses, cognitive forcing function (CFF) perceptions, and the questions asked as part of the \assumptioncff{} and \whatifcff{}.

%%%%%%%%%%%%%%%%%%%%%%%%%%%%%%%%%%%%%%%%%%%%%%%%%%%%%%%%%%%%
\subsection{Initial Assessment of Participants' Understanding and Acceptance of AI Response}\label{appendix.projects.pre}
These questions were presented to every CFF group just after the AI response was given and before the CFF was presented. These questions were presented in the ``INTERACTION PANE'' of the knowledge-work task page, and participants compulsorily had to answer these questions.

\begin{enumerate}
    %%%%%%%
    \item How well do you understand the AI plan?
        \begin{enumerate}[label=(\alph*)]
            \item (Score 5) I understand the plan fully.
            \item (Score 4) I understand most parts of the plan.
            \item (Score 3) I understand some parts of the plan and not others.
            \item (Score 2) I understand only a few parts of the plan.
            \item (Score 1) I do not understand the plan at all.
        \end{enumerate}
    %%%%%%%
    \item How close to ready is the AI plan and draft for the task?
        \begin{enumerate}[label=(\alph*)]
            \item (Score 5) It is completely ready to use as is.
            \item (Score 4) It is mostly ready and only minor tweaks are needed.
            \item (Score 3) It is somewhat ready and moderate changes are needed.
            \item (Score 2) It is barely ready and major changes are needed.
            \item (Score 1) It is not ready at all and entirely incorrect.
            \item (Score N/A) I don't know.
        \end{enumerate}
    %%%%%%%
\end{enumerate}

%%%%%%%%%%%%%%%%%%%%%%%%%%%%%%%%%%%%%%%%%%%%%%%%%%%%%%%%%%%%
\subsection{Final Assessment of Participants' Understanding and Acceptance of AI Response}\label{appendix.projects.post}
These questions were presented to every CFF group after the participants completed interacting with the CFF. These questions were presented in the ``INTERACTION PANE'' of the knowledge-work task page, and participants compulsorily had to answer these questions.

\begin{enumerate}
    %%%%%%%
    \item How close to ready is the AI plan and draft for the task?
        \begin{enumerate}[label=(\alph*)]
            \item (Score 5) It is completely ready to use as is.
            \item (Score 4) It is mostly ready and only minor tweaks are needed.
            \item (Score 3) It is somewhat ready and moderate changes are needed.
            \item (Score 2) It is barely ready and major changes are needed.
            \item (Score 1) It is not ready at all and entirely incorrect.
            \item (Score N/A) I don't know.
        \end{enumerate}
    %%%%%%%
    \item State your feedback needed to improve the plan/draft for the task. Feel free to be as detailed as you would like. If there is no feedback, specify ``No feedback''. Your performance will be reviewed based on the correctness and quality of the feedback: [Free response]
    %%%%%%%
\end{enumerate}

%%%%%%%%%%%%%%%%%%%%%%%%%%%%%%%%%%%%%%%%%%%%%%%%%%%%%%%%%%%%
\subsection{Post-task Survey Questionnaire}\label{appendix.projects.postsurvey}
These questions\footnote{The questions are adopted from the validated \emph{NASA TLX} scale~\citep{hart1988development} to assess cognitive load.} were presented to every CFF group after the participants completed each knowledge-work task. These questions were presented as a new dialog box within the knowledge-work task page, and participants compulsorily had to answer these questions before proceeding to the next task.

\begin{enumerate}
    %%%%%%%
    \item (Mental Demand) How mentally demanding was the task?
        \begin{enumerate}[label=(\alph*)]
            \item (Score 1) Not at all
            \item (Score 2) Slightly
            \item (Score 3) Moderately
            \item (Score 4) Very
            \item (Score 5) Extremely
        \end{enumerate}
    %%%%%%%
      \item (Temporal Demand) How hurried or rushed was the pace of the task?
        \begin{enumerate}[label=(\alph*)]
            \item (Score 1) Not at all
            \item (Score 2) Slightly
            \item (Score 3) Moderately
            \item (Score 4) Very
            \item (Score 5) Extremely
        \end{enumerate}
    %%%%%%%
     \item (Performance) How successful were you in providing feedback to improve the AI plan and draft for the task?
        \begin{enumerate}[label=(\alph*)]
            \item (Score 1) Not at all
            \item (Score 2) Slightly
            \item (Score 3) Moderately
            \item (Score 4) Very
            \item (Score 5) Extremely
        \end{enumerate}
    %%%%%%%
     \item (Effort) How hard did you have to work to accomplish your level of performance in providing the feedback?
        \begin{enumerate}[label=(\alph*)]
            \item (Score 1) Not at all
            \item (Score 2) Slightly
            \item (Score 3) Moderately
            \item (Score 4) Very
            \item (Score 5) Extremely
        \end{enumerate}
    %%%%%%%
     \item (Frustration) How insecure, discouraged, irritated, stressed and annoyed were you while completing the task?
       \begin{enumerate}[label=(\alph*)]
            \item (Score 1) Not at all
            \item (Score 2) Slightly
            \item (Score 3) Moderately
            \item (Score 4) Very
            \item (Score 5) Extremely
        \end{enumerate}
    %%%%%%%
     \item (Quiz Demand) How mentally demanding were the (assumptions/what-if) questions in the INTERACTION PANE?\textsuperscript{$\dagger$}
        \begin{enumerate}[label=(\alph*)]
            \item (Score 1) Not at all
            \item (Score 2) Slightly
            \item (Score 3) Moderately
            \item (Score 4) Very
            \item (Score 5) Extremely
        \end{enumerate}
    %%%%%%%
     \item (Quiz Helpfulness) How helpful were the (assumptions/what-if) questions in the INTERACTION PANE in providing feedback about the AI plan and draft?\textsuperscript{$\dagger$}
       \begin{enumerate}[label=(\alph*)]
            \item (Score 1) Not at all
            \item (Score 2) Slightly
            \item (Score 3) Moderately
            \item (Score 4) Very
            \item (Score 5) Extremely
        \end{enumerate}
    %%%%%%%
\end{enumerate}
\textsuperscript{$\dagger$}\,These questions were only presented to the groups which had quiz-based CFFs such as: \assumptioncff{}, \whatifcff{}, and \bothcff{}. They were not presented to the CFF group \nocff{}.

%%%%%%%%%%%%%%%%%%%%%%%%%%%%%%%%%%%%%%%%%%%%%%%%%%%%%%%%%
\subsection{Task Details}\label{appendix.projects.details}
We present the full materials for each task below, including the task description, task artifact, the AI-generated plan and draft, and the corresponding prompts used in the \assumptioncff{} and \whatifcff{} CFF conditions. For participants in the \bothcff{} group, the interface first displayed the \assumptioncff{} questions followed by the \whatifcff{} questions. All AI plans and drafts were generated using GPT-4o, including versions intentionally prompted to contain bugs. When presenting the AI plan and draft for each task, any intentionally erroneous plan step is highlighted in red. The accompanying AI draft is consistent with the (erroneous) plan by design, and therefore its content is not separately highlighted.

Figures 2c and 2d illustrate the user interface for the \assumptioncff{} and \whatifcff{}-based questions. The \assumptioncff{} prompts were customized to each individual step in the AI plan, and the answer options were generated with support from GPT-4o. In contrast, the \whatifcff{} prompts were applied at the level of the full plan and followed a consistent template across all tasks.

\clearpage
%%%%% Project 1
% !TEX root =  main.tex
%%%%%%%%%%%%%%%%%%%%%%%%%%%%%%%%%%%%%%%%%%%%%%%%%%%%%%%%%%
%%%%%%%%%%%%%%%%%%%%%%%%%%%%%%%%%%%%%%%%%%%%%%%%%%%%%%%%%%

\begin{AppendixDocumentBox}{Task 1: Library Notice}
You are part of the communications team at the local library. Recently three service upgrades have been approved by the team (see the \textbf{library\_upgrades.txt} artifact below). You must email a notice that tells patrons what is new. Write a short notice that highlights the three upgrades, so every member immediately understands the benefits.

\medskip

\begin{ArtifactBox}{library\_upgrades.txt}
Approved Library Upgrades:
\begin{enumerate}
    \item Weekend hours extend to 8pm
    \item New quiet-study wing opens 1st September
    \item Online room-booking is now available
\end{enumerate}
\end{ArtifactBox}

\end{AppendixDocumentBox}
%%%%%%%%%%%%%%%%%%%%%%%%%%%%%%%%%%%%%%%%%%%%%%%%%

\begin{AppendixDocumentBox}[yellow!20]{AI Response for Task 1}
\textbf{AI Plan}\\
\textcolor{red}{For this task, there are no synthetic errors introduced in the AI plan (as also described in Section 4.1, Table 3).}

\begin{WhiteContainer}
\setcounter{plancardctr}{0}

\begin{numberedPlanCard}
Start with an engaging sentence indicating service upgrades.
\end{numberedPlanCard}

\begin{numberedPlanCard}
List each of the three upgrades clearly and concisely.
\end{numberedPlanCard}

\begin{numberedPlanCard}
Highlight key benefits (longer hours, quiet space, online booking).
\end{numberedPlanCard}

\begin{numberedPlanCard}
Keep the tone informative, friendly, and short.
\end{numberedPlanCard}

\end{WhiteContainer}

\medskip

\textbf{AI Draft}

\begin{WhiteContainer}
Great news! We’ve made three exciting upgrades: weekend hours now extend to 8pm,
a new quiet-study wing opens on 1st September, and you can now book rooms online.
Enjoy more flexibility, focused study, and easy access, designed to make your
library experience better than ever!
\end{WhiteContainer}

\end{AppendixDocumentBox}
%%%%%%%%%%%%%%%%%%%%%%%%%%%%%%%%%%%%%%%%%%%%%%%%%

\begin{AppendixDocumentBox}[orange!20]{Assumption Quizzes for Task 1}
Click the ``Reflect Assumptions'' buttons in the plan cards of the AI generated plan to think about assumptions for each step in the AI plan. Feel free to reflect on as many plan steps as you like. \textbf{After you reflect on at least one plan step, the Next button will be enabled.}

\medskip
%%%%%%%%%%%%%%%%%%%%%%%%%%%%%%%%%%%%%%%%%%%%%%%%%

    \textbf{Assumption Quiz: AI Plan Step 1}
    
    \begin{quizCard}
    \textbf{Question:}
    What assumptions underlie plan step 1: \emph{Start with an engaging sentence indicating service upgrades}? (Select all that apply)
    
    \begin{enumerate}[label=(\alph*), leftmargin=1.2cm]
        \item Members are more likely to read the notice if it begins with a clear, appealing summary.
        \item Dramatic and overly promotional language is more engaging for library notices.
        \item Other: [Free response]
    \end{enumerate}
    \end{quizCard}

%%%%%%%%%%%%%%%%%%%%%%%%%%%%%%%%%%%%%
    \textbf{Assumption Quiz: AI Plan Step 2}
    
    \begin{quizCard}
    \textbf{Question:}  
    What assumptions underlie plan step 2: \emph{List each of the three upgrades clearly and concisely}? (Select all that apply)
    
    \begin{enumerate}[label=(\alph*), leftmargin=1.2cm]
        \item Listing the upgrades in technical detail will make the message more trustworthy.
        \item Clarity and brevity will help patrons quickly grasp the value of each upgrade.
        \item Other: [Free response]
    \end{enumerate}
    \end{quizCard}

    %%%%%%%%%%%%%%%%%%%%%%%%%%%%%%%%%%%%%

    \textbf{Assumption Quiz: AI Plan Step 3}
    
    \begin{quizCard}
    \textbf{Question:}  
    What assumptions underlie plan step 3: \emph{Highlight key benefits (longer hours, quiet space, online booking)}? (Select all that apply) 
    
    \begin{enumerate}[label=(\alph*), leftmargin=1.2cm]
        \item Emphasizing user-relevant benefits helps patrons understand why the upgrades matter.
        \item Stating the upgrade features without context is enough for readers to value them.
        \item Other: [Free response]
    \end{enumerate}
    \end{quizCard}

    %%%%%%%%%%%%%%%%%%%%%%%%%%%%%%%%%%%%%%%
    \textbf{Assumption Quiz: AI Plan Step 4}
    
    \begin{quizCard}
    \textbf{Question:}  
    What assumptions underlie plan step 4: \emph{Keep the tone informative, friendly, and short.}? (Select all that apply)    
    
    \begin{enumerate}[label=(\alph*), leftmargin=1.2cm]
        \item A concise, friendly tone is best for engaging a wide range of library patrons.
        \item Using formal or bureaucratic language adds authority and is preferred by readers.
        \item Other: [Free response]
    \end{enumerate}
    \end{quizCard}
\end{AppendixDocumentBox}
%%%%%%%%%%%%%%%%%%%%%%%%%%%%%%%%%%%%%%%%%%%%%%%%%

\begin{AppendixDocumentBox}[orange!20]{What-If Questions for Task 1}

    \textbf{What-If Question 1}
    
    \begin{quizCard}
    \textbf{Question:}  
    Which is the most critical step in the plan?
    \begin{enumerate}[label=(\alph*), leftmargin=1.2cm]
        \item Step 1: \emph{Start with an engaging sentence indicating service upgrades.}
        \item Step 2: \emph{List each of the three upgrades clearly and concisely.}
        \item Step 3: \emph{Highlight key benefits (longer hours, quiet space, online booking).}
        \item Step 4: \emph{Keep the tone informative, friendly, and short.}
    \end{enumerate}
    \end{quizCard}
    
    \textbf{What-If Question 2}
    
    \begin{quizCard}
    \textbf{Question:}
    What happens if this step fails or changes? Describe in a few words the impact on the plan/draft. If there is no impact, you can answer ``No impact''.
    [Free response]
    \end{quizCard}

\end{AppendixDocumentBox}
%%%%%%%%%%%%%%%%%%%%%%%%%%%%%%%%%%%%%%%%%%%%%%%%%

%%%%%%%%%%%%%%%%%%%%%%%%%%%
\clearpage
%%%%% Project 2
% !TEX root =  main.tex
%%%%%%%%%%%%%%%%%%%%%%%%%%%%%%%%%%%%%%%%%%%%%%%%%%%%%%%%%%
%%%%%%%%%%%%%%%%%%%%%%%%%%%%%%%%%%%%%%%%%%%%%%%%%%%%%%%%%%

\begin{AppendixDocumentBox}{Task 2: Analyzing Product Reviews}
    You are a UX researcher preparing a short internal report for the product design team. User reviews of a productivity app that were submitted within the last month are available in the text file **user\_reviews.txt** below. Your goal is to analyze the reviews and derive high level insights that could guide the product strategy. Support each insight with a few example phrases or quotes from the reviews. Your response must be presented as one coherent document.
    %%%%%%%
    \medskip

    % Replaces the tcolorbox (artifact file box)
    % Note: If you need smaller text inside, keep \small within the environment.
    \begin{ArtifactBox}{user\_reviews.txt}
    {\small
        - Best note-taking app I’ve used—simple, elegant, and efficient.\\
        - Love how quickly I can access everything from the dashboard.\\
        - The sync feature is so unreliable. I never know if my notes are updated across devices.\\
        - Dark mode looks great and saves my eyes during late nights.\\
        - Love the interface, but please fix the recuring reminders; they keep disappearing randomly.\\
        - Battery usage has gotten worse since the last update. Why is it always running in the background?\\
        - Great app overall, but the calendar integration is inconsistent with my Google calendar.\\
        - The desktop version is slick, but the mobile lags badly when opening shared documents.\\
        - I'm confused by the settings layout. Took me forever to find the export option.\\
        - Recuring reminders are completely broken on Android 13.\\
        - The new update looks nice, but it killed the quick note feature. That was the main reason I used this app.\\
        - Notes don’t always show up on my tablet until hours later.\\
        - The app looks clean, but the core features just aren’t reliable.\\
        - Notifications are well-timed and unobtrusive.\\
        - I constantly have to restart the app to get it to sync.\\
        - Recurring tasks vanished again after the latest patch.\\
        - I like the design, but why is the background sync so aggressive?\\
        - My reminders work fine on iOS, but not on Android.\\
        - Exporting files should be easier to find. The menu isn’t intuitive.\\
        - The UI is modern, but I wish it didn’t crash when loading shared files.\\
        - Syncing between work and personal devices is hit or miss.\\
        - Still can’t trust that my calendar entries are accurate.\\
        - Calendar integration worked perfectly with my Outlook setup.\\
        - I rely on this app daily and it's never let me down.\\
        - Super easy to onboard—great job with the tutorials and first-time use.\\
        - Recurring tasks work well on my iPhone—very reliable.\\
        - Fast sync and stable performance on both Android and desktop.\\\
        $\ldots$\\
        $\ldots$\\
        $\ldots$  \textbf{(57 more reviews)}
    }
    \end{ArtifactBox}

\end{AppendixDocumentBox}

%%%%%%%%% AI Plan and Draft
\begin{AppendixDocumentBox}[yellow!20]{AI Response for Task 2}
    \textbf{AI Plan}
    \\
    \textcolor{red}{For this task, the synthetic error is introduced in plan step 5. According to the error taxonomy in Section 4.1, Table 3, this step is classified as an incorrect plan step. The step is erroneous because it encourages highlighting only positive reviews rather than providing a balanced depiction, despite many user reviews being clearly negative.}

    % Replaces the tcolorbox (white container around plan cards)
    \begin{WhiteContainer}
        \setcounter{plancardctr}{0} % reset numbering for this group

        \begin{numberedPlanCard}
        Read and categorize user reviews based on common themes.
        \end{numberedPlanCard}
        
        \begin{numberedPlanCard}
        Identify frequently mentioned issues or praise points across reviews.
        \end{numberedPlanCard}
        
        \begin{numberedPlanCard}
        Group feedback into broader UX or technical categories (e.g., performance, design).
        \end{numberedPlanCard}
        
        \begin{numberedPlanCard}
        Select representative quotes to support each identified insight.
        \end{numberedPlanCard}

        \begin{numberedPlanCard}
        \textcolor{red}{Emphasize commonly cited positives and prioritize general usability improvements.}
        \end{numberedPlanCard}

        \begin{numberedPlanCard}
        Synthesize insights into a concise, coherent narrative.
        \end{numberedPlanCard}
    \end{WhiteContainer}

    \medskip

    \textbf{AI Draft}

    % Replaces the tcolorbox (white container around draft)
    \begin{WhiteContainer}
        {\small
        Recent user reviews of the productivity app highlight a largely positive user experience, with consistent appreciation for the app’s visual design, cross-platform availability, and ease of use. The clean, modern interface continues to resonate strongly with users, as reflected in comments like “The interface is clean and modern—really enjoy using it on desktop,” “Dark mode looks great and saves my eyes during late nights,” and “Love how simple the app feels visually. Easy on the eyes.” Users consistently describe the visual design as intuitive, minimal, and pleasing, reinforcing that the app is achieving high design standards. Many users also complimented the latest update for improving the look and feel of the app—“The latest update made things look more polished overall.”

        Performance on desktop platforms is frequently highlighted as a strength. Users describe the experience as “smooth and responsive,” with others noting that “The desktop app is fast and reliable for my daily tasks.” The app’s speed and responsiveness make it a compelling tool for managing notes and tasks, and it integrates seamlessly into desktop workflows. Several users also praised the export and search capabilities, noting they are “very pleased with the ability to search across all notes effortlessly” and that the export options “support my workflow perfectly.”
        }
    \end{WhiteContainer}
    \begin{WhiteContainer}
        {\small
         Cross-platform consistency and onboarding flow were also celebrated. Comments such as “Super easy to onboard—great job with the tutorials and first-time use” and “The cross-platform consistency really stands out—looks and works the same everywhere” demonstrate that users value the predictability and familiarity of the app across devices. Syncing functionality, while mentioned in a few reviews, was also cited as effective in many cases—“Sync has been working flawlessly for me between my laptop and phone,” and “When it works, the sync feature is a real time-saver.” These experiences suggest that the app delivers seamless data continuity for many users.
         
        Additionally, the app’s reminder and notification system receives praise for enhancing user productivity. Quotes such as “The reminder feature helps me stay organized every single day” and “Notifications are timely and subtle—doesn’t interrupt my focus” show how the app supports users in staying on track without becoming intrusive. Users also appreciate mobile responsiveness and fast performance on newer devices, saying “The mobile app is snappy and loads instantly on my Pixel 8.”

        Notably, the app has earned enthusiastic endorsements as a preferred daily tool. Statements like “Best note-taking app I’ve used—simple, elegant, and efficient” and “I rely on this app daily and it’s never let me down” underscore the app’s effectiveness and user satisfaction. While every product can continue to evolve, the current trajectory of updates and visual improvements is clearly resonating with a growing number of users.

        To build on this momentum, the product team should continue investing in design consistency, speed, and user-centric interactions across platforms. The strong praise for interface design, ease of onboarding, and reliable desktop performance reflects meaningful alignment with user expectations. Minor friction points observed in a small number of reviews can be addressed incrementally, but the overarching takeaway is that the app is well-received and delivering significant value in users’ daily workflows.}
    \end{WhiteContainer}

\end{AppendixDocumentBox}

%%%%%%%%%%%%%%%%%%%%%%%%%%%%%%%%%%

%%%%%%%%%%%%%%%%%%% Assumptions quizzes, whatif questions
\begin{AppendixDocumentBox}[orange!20]{Assumption Quizzes for Task 2}
    Click the ``Reflect Assumptions'' buttons in the plan cards of the AI generated plan to think about assumptions for each step in the AI plan. Feel free to reflect on as many plan steps as you like. \textbf{After you reflect on at least one plan step, the Next button will be enabled.}

    \medskip
    \textbf{Assumption Quiz: AI Plan Step 1}
    
    \begin{quizCard}
    \textbf{Question:}  
    What assumptions underlie plan step 1: \emph{Read and categorize user reviews based on common themes}? (Select all that apply)
    
    \begin{enumerate}[label=(\alph*), leftmargin=1.2cm]
        \item User reviews can be categorized into meaningful themes.
        \item Common themes can be identified through careful reading and pattern recognition.
        \item Other: [Free response]
    \end{enumerate}
    \end{quizCard}

    %%%%%%%%%%%%%%%%%%%%%%%%%%%%%%%%%%%%%
    \textbf{Assumption Quiz: AI Plan Step 2}
    
    \begin{quizCard}
    \textbf{Question:}  
    What assumptions underlie plan step 2: \emph{Identify frequently mentioned issues or praise points across reviews}? (Select all that apply)
    
    \begin{enumerate}[label=(\alph*), leftmargin=1.2cm]
        \item Frequency of mention indicates relative importance to users.
        \item Reviews can be classified into issues or praise points based on sentiment.
        \item Other: [Free response]
    \end{enumerate}
    \end{quizCard}

    %%%%%%%%%%%%%%%%%%%%%%%%%%%%%%%%%%%%%

    \textbf{Assumption Quiz: AI Plan Step 3}
    
    \begin{quizCard}
    \textbf{Question:}  
    What assumptions underlie plan step 3: \emph{Group feedback into broader UX or technical categories (e.g., performance, design)}? (Select all that apply) 
    
    \begin{enumerate}[label=(\alph*), leftmargin=1.2cm]
        \item UX and technical categories are meaningful distinctions for product teams.
        \item Issues can be clearly classified as either UX-related or technical.
        \item Other: [Free response]
    \end{enumerate}
    \end{quizCard}

    %%%%%%%%%%%%%%%%%%%%%%%%%%%%%%%%%%%%%%%
    \textbf{Assumption Quiz: AI Plan Step 4}
    
    \begin{quizCard}
    \textbf{Question:}  
    What assumptions underlie plan step 4: \emph{Select representative quotes to support each identified insight}? (Select all that apply)    
    
    \begin{enumerate}[label=(\alph*), leftmargin=1.2cm]
        \item Direct user quotes provide more credible evidence than paraphrasing.
        \item Selected quotes accurately represent the broader sentiment in that category.
        \item Other: [Free response]
    \end{enumerate}
    \end{quizCard}

    %%%%%%%%%%%%%%%%%%%%%%%%%%%%%%%%%%%%%%%
    \textbf{Assumption Quiz: AI Plan Step 5}
    
    \begin{quizCard}
    \textbf{Question:}  
    What assumptions underlie plan step 5: \emph{Emphasize commonly cited positives and prioritize general usability improvements}? (Select all that apply)    
    
    \begin{enumerate}[label=(\alph*), leftmargin=1.2cm]
        \item Highlighting positives maintains team confidence and that is most important.
        \item Presenting a balanced view of positives and negatives is more persuasive.
        \item Other: [Free response]
    \end{enumerate}
    \end{quizCard}

    %%%%%%%%%%%%%%%%%%%%%%%%%%%%%%%%%%%%%%%
    \textbf{Assumption Quiz: AI Plan Step 6}
    
    \begin{quizCard}
    \textbf{Question:}  
    What assumptions underlie plan step 6: \emph{Synthesize insights into a concise, coherent narrative}? (Select all that apply)    
    
    \begin{enumerate}[label=(\alph*), leftmargin=1.2cm]
        \item A coherent narrative is more actionable than a list of disconnected findings.
        \item The 300-word limit forces focus on the most important insights
        \item Other: [Free response]
    \end{enumerate}
    \end{quizCard}

\end{AppendixDocumentBox}

%%%%%%%%%%%%%%%%%%%%%%%%%%%%%%%%%%%%%%%%%%%%%%

\begin{AppendixDocumentBox}[orange!20]{What-If Questions for Task 2}

    \textbf{What-If Question 1}
    
    \begin{quizCard}
    \textbf{Question:}  
    Which is the most critical step in the plan?
    \begin{enumerate}[label=(\alph*), leftmargin=1.2cm]
        \item Step 1: \emph{Read and categorize user reviews based on common themes.}
        \item Step 2: \emph{Identify frequently mentioned issues or praise points across reviews.}
        \item Step 3: \emph{Group feedback into broader UX or technical categories (e.g., performance, design).}
        \item Step 4: \emph{Select representative quotes to support each identified insight.}
        \item Step 5: \emph{Emphasize commonly cited positives and prioritize general usability improvements.}
        \item Step 6: \emph{Synthesize insights into a concise, coherent narrative.}
    \end{enumerate}
    \end{quizCard}
    
    \textbf{What-If Question 2}
    
    \begin{quizCard}
    \textbf{Question:}
    What happens if this step fails or changes? Describe in a few words the impact on the plan/draft. If there is no impact, you can answer ``No impact''.
    [Free response]
    \end{quizCard}

\end{AppendixDocumentBox}

%%%%%%%%%%%%%%%%%%%%%%%%%%%
\clearpage
%%%%% Project 3
% !TEX root =  main.tex
%%%%%%%%%%%%%%%%%%%%%%%%%%%%%%%%%%%%%%%%%%%%%%%%%%%%%%%%%%
%%%%%%%%%%%%%%%%%%%%%%%%%%%%%%%%%%%%%%%%%%%%%%%%%%%%%%%%%%

\begin{AppendixDocumentBox}{Task 3: Persuasive Essay}
    Write a persuasive essay in support of the claim: *Prescription drug importation should be allowed to increase access and lower costs*. The comprehensive essay has to be presented in an executive meeting with the health ministry for an internal policy debate. You are also provided with a few relevant facts in the file, \textbf{drug\_importation\_facts.md}, below, that can help you make your claim.
    %%%%%%%
    \medskip

    % Replaces the tcolorbox for the facts file
    \begin{ArtifactBox}{drug\_importation\_facts.md}
    {\small
        \textbf{\#\# 30 Credible Facts on Prescription Drug Importation}\\

        1. Prescription drugs in Canada cost 25\%–80\% less than in the United States.\\
        2. Legalizing drug importation can help alleviate the burden of high medication costs in the U.S.\\
        3. Studies show a significant portion of Americans import drugs for personal use to save money.\\
        4. Importation increases patient access to necessary medications that might otherwise be unaffordable.\\
        5. Safety concerns about imported drugs can be managed through regulations and approvals by agencies like the FDA.\\
        6. Canada has expressed concerns about potential drug shortages if U.S. importation increases substantially.\\
        7. Personal drug importation behavior in the U.S. has increased as domestic drug prices rise.\\
        8. Expanding FDA personal use exemptions could legally boost access to affordable medication.\\
        9. Importation from countries with strong drug regulation, such as Canada, Australia, and New Zealand, can ensure safety.\\
        10. Brand name drugs are often cheaper in Canada and are a major focus for U.S. importation demand.\\
        $\ldots$\\
        $\ldots$\textbf{(20 more facts)}\\

        \medskip
        \textbf{\#\# Sources}\\
        
        - Congressional Research Service, "Prescription Drug Importation: Federal Legislation and Congressional Issues," 2022.\\
        - KFF, "How do prescription drug costs in the United States compare to other countries?" 2019.\\
        - Government of Canada, "How Canada Regulates Drug Prices," 2021.\\
        - U.S. Department of Health \& Human Services, "Trump Administration Finalizes Plan Allowing the Importation of Certain Prescription Drugs," 2020.\\
        - U.S. Food \& Drug Administration, "Importation of Prescription Drugs," 2023.\\
        - The Commonwealth Fund, "Prescription Drug Importation: A Window into U.S. Drug Pricing Policy," 2020.\\
        - Gallup, "Nearly 20\% of Americans Say They Cannot Afford Prescribed Medication," 2019.\\
        - Centers for Disease Control and Prevention, "Strategies Used by Adults Aged 18–64 to Reduce Their Prescription Drug Costs," 2022.\\
        $\ldots$\\
        $\ldots$\textbf{(30 more sources)}\\
    }
    \end{ArtifactBox}

\end{AppendixDocumentBox}

%%%%%%%%% AI Plan and Draft
\begin{AppendixDocumentBox}[yellow!20]{AI Response for Task 3}
    \textbf{AI Plan}
    \\
    \textcolor{red}{For this task, there are no synthetic errors introduced in the AI plan (as also described in Section 4.1, Table 3).}

    % Replaces the tcolorbox (white container around plan cards)
    \begin{WhiteContainer}
        \setcounter{plancardctr}{0} % reset numbering for this group

        \begin{numberedPlanCard}
        Analyze the drug\_importation\_facts.md file to extract relevant facts supporting the claim.
        \end{numberedPlanCard}
        
        \begin{numberedPlanCard}
        Define the thesis supporting drug importation as a means to lower costs and improve access.
        \end{numberedPlanCard}
        
        \begin{numberedPlanCard}
        Present economic benefits using quoted facts to highlight cost differences and savings.
        \end{numberedPlanCard}
        
        \begin{numberedPlanCard}
        Demonstrate access improvements with data to support the impact on vulnerable populations.
        \end{numberedPlanCard}

        \begin{numberedPlanCard}
        Address safety concerns using evidence from the file and regulatory examples.
        \end{numberedPlanCard}

        \begin{numberedPlanCard}
        Conclude with a call to action, reinforcing the data-backed benefits of the policy.
        \end{numberedPlanCard}
    \end{WhiteContainer}

    \medskip

    \textbf{AI Draft}

    % Replaces the tcolorbox (white container around draft)
    \begin{WhiteContainer}
    {\small
        Americans face some of the highest prescription drug costs in the world, a reality that forces many to make impossible choices between essential medication and basic necessities such as food, rent, or utilities. For families managing chronic conditions or elderly patients on fixed incomes, these costs can be devastating. Legalizing prescription drug importation is a practical, evidence-based solution that has the potential to dramatically reduce prices, expand access, and maintain safety standards. With proper regulation, drug importation can provide immediate and widespread benefits to consumers and public health systems alike.

        A 2021 study cited in the drug\_importation\_facts.md file reveals that U.S. patients pay an average of 2.56 times more for prescription drugs than patients in 32 other OECD countries. This staggering price gap is not due to superior quality or innovation but is largely a consequence of limited competition and the absence of meaningful price controls. While countries like Canada and those in the European Union negotiate directly with pharmaceutical companies to cap drug prices, the U.S. allows manufacturers to set prices with little constraint. Importing medications from nations with proven safety standards and transparent pricing—especially from close allies like Canada—can provide Americans with access to the same life-saving drugs at a fraction of the cost.
        
         The economic implications of legalizing drug importation are significant. According to a 2020 report by the Department of Health and Human Services, safe drug importation could save the U.S. healthcare system up to 6 billion dollars annually. These savings would benefit more than just individual consumers. Publicly funded programs such as Medicare and Medicaid would also see cost reductions, making these programs more sustainable in the long term and freeing up resources for other critical areas of care.
        }
    \end{WhiteContainer}
    \begin{WhiteContainer}
        {\small
       
        Equally pressing is the issue of access. High drug prices serve as a barrier to care for millions of Americans, particularly vulnerable populations such as seniors, low-income families, and those without comprehensive health insurance. Studies show that a significant portion of Americans skip doses, delay refills, or forgo medications entirely because they simply cannot afford them. Legalizing drug importation would reduce prices and increase availability, allowing patients to maintain consistent treatment and avoid costly medical complications from untreated conditions.
        
        Of course, one of the most frequently cited counterarguments is safety. Critics argue that imported drugs may not meet U.S. regulatory standards. However, this concern does not hold up under scrutiny. Canada—the most commonly proposed source for imported drugs—has pharmaceutical safety standards that are highly comparable to those of the U.S. Food and Drug Administration (FDA). In fact, many of the drugs sold in Canada are manufactured in the same facilities as those sold in the U.S., often by the same companies. Furthermore, several U.S. states are already taking steps toward safe importation. Pilot programs in Florida and Colorado have submitted detailed drug importation plans to the FDA, offering blueprints for how safety, traceability, and quality control can be managed under federal oversight.
        
        Opposition to drug importation is often driven more by lobbying pressure than genuine safety or logistical concerns. The pharmaceutical industry has a financial interest in maintaining the status quo, but the cost of inaction is borne by patients. The time has come for the U.S. government to prioritize public health and economic fairness over corporate profit margins.
        
        In conclusion, prescription drug importation is not a fringe idea or an untested experiment—it is a viable, well-supported policy solution with the potential to significantly lower drug costs, increase access to treatment, and maintain rigorous safety standards. The facts are clear: American patients are overpaying, alternative sources of medication are safe and affordable, and real-world pilot programs are already paving the way. Lawmakers should act swiftly to legalize and scale up prescription drug importation efforts. Doing so is not only fiscally responsible—it is a moral imperative to ensure that no one in this country has to choose between their health and their livelihood.
        }
    \end{WhiteContainer}

\end{AppendixDocumentBox}

%%%%%%%%%%%%%%%%%%%%%%%%%%%%%%%%%%

%%%%%%%%%%%%%%%%%%% Assumptions quizzes, whatif questions
\begin{AppendixDocumentBox}[orange!20]{Assumption Quizzes for Task 3}
    Click the ``Reflect Assumptions'' buttons in the plan cards of the AI generated plan to think about assumptions for each step in the AI plan. Feel free to reflect on as many plan steps as you like. \textbf{After you reflect on at least one plan step, the Next button will be enabled.}

    \medskip
    \textbf{Assumption Quiz: AI Plan Step 1}
    
    \begin{quizCard}
    \textbf{Question:}  
    What assumptions underlie plan step 1: \emph{Analyze the drug\_importation\_facts.md file to extract relevant facts supporting the claim}? (Select all that apply)
    
    \begin{enumerate}[label=(\alph*), leftmargin=1.2cm]
        \item The facts in the file are relevant, accurate, and supportive of the pro-importation stance.
        \item The essay must summarize the entire contents of the file regardless of relevance.
        \item Other: [Free response]
    \end{enumerate}
    \end{quizCard}

    %%%%%%%%%%%%%%%%%%%%%%%%%%%%%%%%%%%%%
    \textbf{Assumption Quiz: AI Plan Step 2}
    
    \begin{quizCard}
    \textbf{Question:}  
    What assumptions underlie plan step 2: \emph{Define the thesis supporting drug importation as a means to lower costs and improve access.}? (Select all that apply)
    
    \begin{enumerate}[label=(\alph*), leftmargin=1.2cm]
        \item The thesis should present both pros and cons of drug importation equally to maintain fairness.
        \item The reader expects a strong position early in the essay because of its persuasive nature.
        \item Other: [Free response]
    \end{enumerate}
    \end{quizCard}

    %%%%%%%%%%%%%%%%%%%%%%%%%%%%%%%%%%%%%

    \textbf{Assumption Quiz: AI Plan Step 3}
    
    \begin{quizCard}
    \textbf{Question:}  
    What assumptions underlie plan step 3: \emph{Present economic benefits using quoted facts to highlight cost differences and savings}? (Select all that apply) 
    
    \begin{enumerate}[label=(\alph*), leftmargin=1.2cm]
        \item Highlighting savings supports both individual and systemic benefits.
        \item Economic benefits is the strongest form of evidence used in persuasive writing.
        \item Other: [Free response]
    \end{enumerate}
    \end{quizCard}

    %%%%%%%%%%%%%%%%%%%%%%%%%%%%%%%%%%%%%%%
    \textbf{Assumption Quiz: AI Plan Step 4}
    
    \begin{quizCard}
    \textbf{Question:}  
    What assumptions underlie plan step 4: \emph{Demonstrate access improvements with data to support the impact on vulnerable populations}? (Select all that apply)    
    
    \begin{enumerate}[label=(\alph*), leftmargin=1.2cm]
        \item Access issues are mainly due to poor insurance literacy, not drug pricing.
        \item Showing human impact enhances the emotional and ethical appeal of the argument.
        \item Other: [Free response]
    \end{enumerate}
    \end{quizCard}

    %%%%%%%%%%%%%%%%%%%%%%%%%%%%%%%%%%%%%%%
    \textbf{Assumption Quiz: AI Plan Step 5}
    
    \begin{quizCard}
    \textbf{Question:}  
    What assumptions underlie plan step 5: \emph{Address safety concerns using evidence from the file and regulatory examples}? (Select all that apply)    
    
    \begin{enumerate}[label=(\alph*), leftmargin=1.2cm]
        \item Safety is a common and credible counterargument to drug importation.
        \item Safety concerns are no longer relevant because all imported drugs are already FDA-approved.
        \item Other: [Free response]
    \end{enumerate}
    \end{quizCard}

    %%%%%%%%%%%%%%%%%%%%%%%%%%%%%%%%%%%%%%%
    \textbf{Assumption Quiz: AI Plan Step 6}
    
    \begin{quizCard}
    \textbf{Question:}  
    What assumptions underlie plan step 6: \emph{Conclude with a call to action, reinforcing the data-backed benefits of the policy}? (Select all that apply)    
    
    \begin{enumerate}[label=(\alph*), leftmargin=1.2cm]
        \item The conclusion should restate the entire essay word-for-word in summary format.
        \item A call to action motivates support or engagement with the policy issue.
        \item Other: [Free response]
    \end{enumerate}
    \end{quizCard}

\end{AppendixDocumentBox}

%%%%%%%%%%%%%%%%%%%%%%%%%%%%%%%%%%%%%%%%%%%%%%

\begin{AppendixDocumentBox}[orange!20]{What-If Questions for Task 3}

    \textbf{What-If Question 1}
    
    \begin{quizCard}
    \textbf{Question:}  
    Which is the most critical step in the plan?
    \begin{enumerate}[label=(\alph*), leftmargin=1.2cm]
        \item Step 1: \emph{Analyze the drug\_importation\_facts.md file to extract relevant facts supporting the claim.}
        \item Step 2: \emph{Define the thesis supporting drug importation as a means to lower costs and improve access.}
        \item Step 3: \emph{Present economic benefits using quoted facts to highlight cost differences and savings.}
        \item Step 4: \emph{Demonstrate access improvements with data to support the impact on vulnerable populations.}
        \item Step 5: \emph{Address safety concerns using evidence from the file and regulatory examples.}
        \item Step 6: \emph{Conclude with a call to action, reinforcing the data-backed benefits of the policy.}
    \end{enumerate}
    \end{quizCard}
    
    \textbf{What-If Question 2}
    
    \begin{quizCard}
    \textbf{Question:}
    What happens if this step fails or changes? Describe in a few words the impact on the plan/draft. If there is no impact, you can answer ``No impact''.
    [Free response]
    \end{quizCard}

\end{AppendixDocumentBox}

%%%%%%%%%%%%%%%%%%%%%%%%%%%
\clearpage
%%%%% Project 4
% !TEX root =  main.tex
%%%%%%%%%%%%%%%%%%%%%%%%%%%%%%%%%%%%%%%%%%%%%%%%%%%%%%%%%%
%%%%%%%%%%%%%%%%%%%%%%%%%%%%%%%%%%%%%%%%%%%%%%%%%%%%%%%%%%

\begin{AppendixDocumentBox}{Task 4: Structuring Meeting Minutes}
As a member of the Long Beach City Council administrative staff, you are reviewing the official minutes from a council meeting. Please prepare a concise summary highlighting the key decisions, discussions, and action items. This summary will be presented to the Mayor in next week’s briefing and should clearly convey the most relevant points for executive awareness and follow-up. The meeting minutes are provided in the file \textbf{202504\_1753\_meeting\_minutes.txt}.
%%%%%%%
\medskip

\begin{ArtifactBox}{202504\_1753\_meeting\_minutes.txt}
{\small
Thank you very much. Congratulations. We have and again, because we had the budget hearing, everything is just taking longer than it normally would. We have one more hearing tonight and that's hearing for or the third hearing on the agenda, and then we'll go into the regular agenda. So this is hearing item number four, which is an early vacancy. So, madam, please read the item. Report from Public Works recommendation to receive supporting documentation into the record, conclude the public hearing. Find that the area to be vacated is not needed for present or prospective public use and adopt resolution ordering the vacation of the north south alley west of Long Beach Boulevard between East Waldo Road and 35th Street, and a portion of sidewalk right of way along Locust Avenue, District seven. Thank you, Mr. Modica. That report would be given by Craig Beck, our public works director. It's members of the council. I think we did a really good job of describing what we're proposing in the recommended action. The staff have been working closely with laser fish to help them expand their footprint here in the city. They're looking to add more office space for their operations. They own two parcels where an alley cuts those parcels. And we're looking to vacate that alley to allow for the development. To move forward. That concludes my report and I'm available for questions. Thank you. I appreciate that. Let me since we're this is a hearing, let me go ahead and just continue. I want let me go out and do public comment first for the hearing. So for the Ali vacation. Carelessly, Robert Fox, Mr. Avaya and Jeff Miller, please come forward now. Mr. Miller. No. Okay. So concludes public comment. Let me go ahead and go back to Councilman Ringo. Thank you. I think this is a great project. It's opportunity for Lady Fish to complete its its expansion that it's doing to maintain their business leader fish. It's a great corporate partner here for Long Beach, and this is a great project that will help them expand. And I would appreciate the support of my colleagues. Thank you, Councilmember Austin. I'm just going to I want to just add that this is a fantastic project. This Ali vacation will allow us to expand and support the incredible work of Lazar Fish, which is an early tech company within the city of Long Beach that took a chance on Long Beach. And I always want to recognize that they went in, took a chance. Now they are growing by hundreds and hundreds of jobs on their campus and creating a model facility in Bixby Knolls as a gateway. They're clearing oil fields are replacing fences, they're doing landscaping, they're fixing alleys. And they're the exact example of the type of corporate partner you would want in your community. And so I want to thank Mr. Wacker and his entire team. They're really doing a great job and thank the councilmember for his incredible support. And this project would not be where it's at if it wasn't for his support as well. And with that, there's in motion any second. So let me please ask you to cast their votes. Councilmember Richardson. That's why Richardson motion carries. Thank you. We're moving on now to public comment and consent. I made you consent calendar first. Can I get a motion any second on the consent calendar, please? I have carelessly Robert fox carlos over here on the consent calendar.
}
\end{ArtifactBox}

\end{AppendixDocumentBox}

%%%%%%%%%%
\clearpage %%% added to not split \textcolor{red}{....}
%%%%%%%%%%%

%%%%%%%%% AI Plan and Draft
\begin{AppendixDocumentBox}[yellow!20]{AI Response for Task 4}
\textbf{AI Plan}\\
\textcolor{red}{For this task, the synthetic error is that the plan is missing a critical step
(as also described in Section 4.1, Table 3).
The critical step that has been skipped is that the structured summary does not highlight
the follow-up action items despite it being an important component of the task description.}

\begin{WhiteContainer}
\setcounter{plancardctr}{0} % reset numbering for this group

\begin{numberedPlanCard}
Identify and summarize the key agenda items discussed in the meeting.
\end{numberedPlanCard}

\begin{numberedPlanCard}
Explain the background and purpose of the item.
\end{numberedPlanCard}

\begin{numberedPlanCard}
Note key stakeholders involved in the discussion and their contributions.
\end{numberedPlanCard}

\begin{numberedPlanCard}
Summarize key points made by councilmembers.
\end{numberedPlanCard}

\begin{numberedPlanCard}
Record the final key decision and voting outcome.
\end{numberedPlanCard}

\begin{numberedPlanCard}
Mention how the meeting concluded after the decision with the note of thanks.
\end{numberedPlanCard}

\end{WhiteContainer}

\medskip

\textbf{AI Draft}

\begin{WhiteContainer}
{\small
\textbf{Key Agenda Item}\\
The primary focus of yesterday’s council meeting was a public hearing on the proposed vacation of a north-south alley located west of Long Beach Boulevard, between East Waldo Road and 35th Street, as well as a portion of sidewalk right-of-way along Locust Avenue in District 7. This agenda item was a central component of the hearing, reflecting the City’s continued commitment to facilitating local development projects that support economic growth and urban improvement.

\textbf{Background and Purpose}\\
The alley vacation proposal was initiated to support Laserfiche, a well-established local technology company seeking to expand its campus in Long Beach. Currently, the alley separates two parcels owned by Laserfiche, creating a barrier to unified development. By vacating this alley, the company would be able to combine the parcels into a single, larger property, enabling the construction of additional office space to accommodate its growing operations. The City’s Public Works department conducted a thorough evaluation of the alley’s use and determined that it is no longer needed for present or future public purposes, making it eligible for vacation. This step aligns with the City’s strategy to encourage business growth and revitalization in the area.
}
\end{WhiteContainer}

\begin{WhiteContainer}
    {\small
    \textbf{Stakeholders and Contributions}\\
Craig Beck, the Public Works Director, provided the staff report to the Council. He detailed the technical rationale behind the recommendation and described the coordination efforts with Laserfiche representatives to ensure the proposal met both city planning and community standards. Beck emphasized the collaborative nature of the process and assured the Council that due diligence had been exercised. During the public comment portion, several individuals were invited to speak. However, no opposition or concerns were raised regarding the alley vacation, indicating general community support or acceptance of the project.

\textbf{Councilmembers’ Points}\\
Councilmember Ringo expressed strong enthusiasm for the proposal, noting it as an excellent opportunity for Laserfiche to strengthen its presence in Long Beach. He highlighted the company’s role as a significant local employer and its contributions to the city’s economic vitality. Councilmember Austin also voiced support, praising Laserfiche as a model corporate citizen that not only creates jobs but also contributes positively to environmental and community improvements. Both councilmembers underscored how the project aligns with broader city objectives to foster innovation, economic development, and sustainable urban growth.

\textbf{Decision and Voting Outcome}\\
After deliberation, the Council proceeded to vote on the resolution to vacate the alley and sidewalk right-of-way. The motion was introduced by Councilmember Richardson and passed unanimously, signaling full Council support. This resolution officially authorizes Laserfiche to move forward with their planned expansion, marking a key milestone in the company’s growth trajectory within the city limits and reaffirming the City’s commitment to supporting local businesses.

\textbf{Meeting Conclusion}\\
Following the approval, the meeting transitioned seamlessly to the next order of business, including public comments and consideration of the consent calendar. The Council expressed appreciation to all participants for their involvement, thanking them for their engagement and contributions during the hearing. This courteous closure provided a respectful and organized end to this segment of the meeting, maintaining the efficient flow of Council business.}
\end{WhiteContainer}

\end{AppendixDocumentBox}

%%%%%%%%%%%%%%%%%%%%%%%%%%%%%%%%%%

%%%%%%%%%%%%%%%%%%% Assumptions quizzes, whatif questions
\begin{AppendixDocumentBox}[orange!20]{Assumption Quizzes for Task 4}
Click the ``Reflect Assumptions'' buttons in the plan cards of the AI generated plan to think about assumptions for each step in the AI plan. Feel free to reflect on as many plan steps as you like. \textbf{After you reflect on at least one plan step, the Next button will be enabled.}

\medskip
\textbf{Assumption Quiz: AI Plan Step 1}

\begin{quizCard}
\textbf{Question:}
What assumptions underlie plan step 1: \emph{Identify and summarize the key agenda items discussed in the meeting}? (Select all that apply)

\begin{enumerate}[label=(\alph*), leftmargin=1.2cm]
    \item The most important agenda item(s) can be clearly identified from the minutes and form the foundation of the summary.
    \item All agenda items are equally important and should be given roughly the same weight in the summary.
    \item Other: [Free response]
\end{enumerate}
\end{quizCard}

%%%%%%%%%%%%%%%%%%%%%%%%%%%%%%%%%%%%%
\textbf{Assumption Quiz: AI Plan Step 2}

\begin{quizCard}
\textbf{Question:}
What assumptions underlie plan step 2: \emph{Explain the background and purpose of the item}? (Select all that apply)

\begin{enumerate}[label=(\alph*), leftmargin=1.2cm]
    \item Explaining the rationale behind key decisions provides necessary context for executive understanding.
    \item Only financial implications of decisions need to be emphasized for executive awareness.
    \item Other: [Free response]
\end{enumerate}
\end{quizCard}

%%%%%%%%%%%%%%%%%%%%%%%%%%%%%%%%%%%%%
\textbf{Assumption Quiz: AI Plan Step 3}

\begin{quizCard}
\textbf{Question:}
What assumptions underlie plan step 3: \emph{Explain the background and purpose of the item}? (Select all that apply)

\begin{enumerate}[label=(\alph*), leftmargin=1.2cm]
    \item Public comments from non-council members are irrelevant and should be omitted entirely.
    \item Including the voices and positions of influential stakeholders helps convey the significance of the decision.
    \item Other: [Free response]
\end{enumerate}
\end{quizCard}

%%%%%%%%%%%%%%%%%%%%%%%%%%%%%%%%%%%%%%
\textbf{Assumption Quiz: AI Plan Step 4}

\begin{quizCard}
\textbf{Question:}
What assumptions underlie plan step 4: \emph{Note key stakeholders involved in the discussion and their contributions.
}? (Select all that apply)

\begin{enumerate}[label=(\alph*), leftmargin=1.2cm]
    \item Council member endorsements and rationale provide critical political context for the decision’s importance.
    \item Personal opinions of council members are less important than procedural details of the meeting.
    \item Other: [Free response]
\end{enumerate}
\end{quizCard}

%%%%%%%%%%%%%%%%%%%%%%%%%%%%%%%%%%%%%%
\textbf{Assumption Quiz: AI Plan Step 5}

\begin{quizCard}
\textbf{Question:}
What assumptions underlie plan step 5: \emph{Record the final key decision and voting outcome}? (Select all that apply)

\begin{enumerate}[label=(\alph*), leftmargin=1.2cm]
    \item The tone and emotion of the vote (e.g., unanimous or contested) is irrelevant to executive briefing.
    \item The official resolution and vote conclude the decision-making process and are key for formal record-keeping.
    \item Other: [Free response]
\end{enumerate}
\end{quizCard}

%%%%%%%%%%%%%%%%%%%%%%%%%%%%%%%%%%%%%%
\textbf{Assumption Quiz: AI Plan Step 6}

\begin{quizCard}
\textbf{Question:}
What assumptions underlie plan step 6: \emph{Mention how the meeting concluded after the decision with the note of thanks}? (Select all that apply)

\begin{enumerate}[label=(\alph*), leftmargin=1.2cm]
    \item Action-items are not the focus of executive awareness.
    \item Including the conclusion and a thank you note provides a natural and polite closure to the summary, giving the executive a sense of the meeting’s flow and completion without needing additional details.
    \item Other: [Free response]
\end{enumerate}
\end{quizCard}

\end{AppendixDocumentBox}

%%%%%%%%%%%%%%%%%%%%%%%%%%%%%%%%%%%%%%%%%%%%%%

\begin{AppendixDocumentBox}[orange!20]{What-If Questions for Task 4}

\textbf{What-If Question 1}

\begin{quizCard}
\textbf{Question:}
Which is the most critical step in the plan?
\begin{enumerate}[label=(\alph*), leftmargin=1.2cm]
    \item Step 1: \emph{Identify and summarize the key agenda items discussed in the meeting.}
    \item Step 2: \emph{Explain the background and purpose of the item.}
    \item Step 3: \emph{Note key stakeholders involved in the discussion and their contributions.}
    \item Step 4: \emph{Summarize key points made by councilmembers.}
    \item Step 5: \emph{Record the final key decision and voting outcome.}
    \item Step 6: \emph{Mention how the meeting concluded after the decision with the note of thanks.}
\end{enumerate}
\end{quizCard}

\textbf{What-If Question 2}

\begin{quizCard}
\textbf{Question:}
What happens if this step fails or changes? Describe in a few words the impact on the plan/draft. If there is no impact, you can answer ``No impact''.
[Free response]
\end{quizCard}

\end{AppendixDocumentBox}

%%%%%%%%%%%%%%%%%%%%%%%%%%%
\clearpage
%%%%% Project 5
% !TEX root =  main.tex
%%%%%%%%%%%%%%%%%%%%%%%%%%%%%%%%%%%%%%%%%%%%%%%%%%%%%%%%%%
%%%%%%%%%%%%%%%%%%%%%%%%%%%%%%%%%%%%%%%%%%%%%%%%%%%%%%%%%%

\begin{AppendixDocumentBox}{Task 5: News Article Summary}
You are presented with a news article adapted from a popular news agency. Your task is to read the article provided in the \textbf{news\_article.txt} file carefully and generate a concise summary that captures the key points or main idea of the article focusing strictly on the factual findings. The summary must be presented to the Anne Frank House executives so that they may update the museum artifacts with the latest information.
%%%%%%%
\medskip

\begin{ArtifactBox}{news\_article.txt}
{\small
(CNN) Seventy years ago, Anne Frank died of typhus in a Nazi concentration camp at the age of 15. Just two weeks after her supposed death on March 31, 1945, the Bergen-Belsen concentration camp where she had been imprisoned was liberated -- timing that showed how close the Jewish diarist had been to surviving the Holocaust. But new research released by the Anne Frank House shows that Anne and her older sister, Margot Frank, died at least a month earlier than previously thought. Researchers re-examined archives of the Red Cross, the International Training Service and the Bergen-Belsen Memorial, along with testimonies of survivors. They concluded that Anne and Margot probably did not survive to March 1945 -- contradicting the date of death which had previously been determined by Dutch authorities. In 1944, Anne and seven others hiding in the Amsterdam secret annex were arrested and sent to the Auschwitz-Birkenau concentration camp. Anne Frank's final entry. That same year, Anne and Margot were separated from their mother and sent away to work as slave labor at the Bergen-Belsen camp in Germany. Days at the camp were filled with terror and dread, witnesses said. The sisters stayed in a section of the overcrowded camp with no lighting, little water and no latrine. They slept on lice-ridden straw and violent storms shredded the tents, according to the researchers. Like the other prisoners, the sisters endured long hours at roll call. Her classmate, Nannette Blitz, recalled seeing Anne there in December 1944: "She was no more than a skeleton by then. She was wrapped in a blanket; she couldn't bear to wear her clothes anymore because they were crawling with lice." Listen to Anne Frank's friends describe her concentration camp experience. As the Russians advanced further, the Bergen-Belsen concentration camp became even more crowded, bringing more disease. A deadly typhus outbreak caused thousands to die each day. Typhus is an infectious disease caused by lice that breaks out in places with poor hygiene. The disease causes high fever, chills and skin eruptions. "Because of the lice infesting the bedstraw and her clothes, Anne was exposed to the main carrier of epidemic typhus for an extended period," museum researchers wrote. They concluded that it's unlikely the sisters survived until March, because witnesses at the camp said the sisters both had symptoms before February 7. "Most deaths caused by typhus occur around twelve days after the first symptoms appear," wrote authors Erika Prins and Gertjan Broek. The exact dates of death for Anne and Margot remain unclear. Margot died before Anne. "Anne never gave up hope," said Blitz, her friend. "She was absolutely convinced she would survive." Her diary endures as one of the world's most popular books. Read more about Anne Frank's cousin, a keeper of her legacy.
}
\end{ArtifactBox}

\end{AppendixDocumentBox}

%%%%%%%%%
\clearpage %%% added to not split \textcolor{red}{....}
%%%%%%%%%%

%%%%%%%%% AI Plan and Draft
\begin{AppendixDocumentBox}[yellow!20]{AI Response for Task 5}
\textbf{AI Plan}\\
\textcolor{red}{For this task, the synthetic error arises from the inclusion of unnecessary, hallucinated plan steps (as also described in Section 4.1, Table 3). The additional steps are plan steps 2 and 6. Because the task description specifies that the short summary is intended for the Anne Frank House management and should focus on new findings, there is no justification for dividing the article into thematic categories or emphasizing Anne Frank’s emotional resilience and the symbolic significance of her diary. These points are already well established. The summary should mainly concentrate on the newly uncovered information without dilution.}

\begin{WhiteContainer}
\setcounter{plancardctr}{0} % reset numbering for this group

\begin{numberedPlanCard}
Identify the main subject of the article — Anne Frank and the circumstances of her death.
\end{numberedPlanCard}

\begin{numberedPlanCard}
\textcolor{red}{Break the article into thematic categories: historical context, emotional testimony, medical explanation, and personal legacy.}
\end{numberedPlanCard}

\begin{numberedPlanCard}
Note the new discovery — revised death date based on recent research.
\end{numberedPlanCard}

\begin{numberedPlanCard}
Analyze the sources of evidence used — Red Cross, survivor accounts, and Bergen-Belsen records.
\end{numberedPlanCard}

\begin{numberedPlanCard}
Summarize the harsh conditions of the camp and how typhus played a role.
\end{numberedPlanCard}

\begin{numberedPlanCard}
\textcolor{red}{Highlight Anne Frank’s emotional resilience and emphasize the symbolic importance of her diary to modern audiences and summarize all the themes.}
\end{numberedPlanCard}

\end{WhiteContainer}

\medskip

\textbf{AI Draft}

\begin{WhiteContainer}
{\small
Anne Frank, the young Jewish diarist whose legacy continues to move millions, endured harrowing conditions during her imprisonment at Bergen-Belsen. The camp was overcrowded, lacked basic sanitation, and was infested with lice, which led to a deadly outbreak of typhus. Based on a detailed re-examination of Red Cross archives, survivor testimonies, and Bergen-Belsen Memorial records, researchers now believe that Anne and her sister Margot likely died at least a month earlier than the previously assumed date of March 1945. Eyewitness accounts describe Anne as severely weakened, wrapped in a blanket, unable to wear lice-ridden clothes. Despite her suffering, Anne never gave up hope — a spirit immortalized in her famous diary, which remains a symbol of resilience and humanity
}
\end{WhiteContainer}

\end{AppendixDocumentBox}

%%%%%%%%%%%%%%%%%%%%%%%%%%%%%%%%%%

%%%%%%%%%%%%%%%%%%% Assumptions quizzes, whatif questions
\begin{AppendixDocumentBox}[orange!20]{Assumption Quizzes for Task 5}
Click the ``Reflect Assumptions'' buttons in the plan cards of the AI generated plan to think about assumptions for each step in the AI plan. Feel free to reflect on as many plan steps as you like. \textbf{After you reflect on at least one plan step, the Next button will be enabled.}

\medskip
\textbf{Assumption Quiz: AI Plan Step 1}

\begin{quizCard}
\textbf{Question:}
What assumptions underlie plan step 1: \emph{Identify the main subject of the article — Anne Frank and the circumstances of her death}? (Select all that apply)

\begin{enumerate}[label=(\alph*), leftmargin=1.2cm]
    \item A summary should start by broadly identifying the central figure and setting the context of the event.
    \item You must explain the historical significance of the main subject before addressing the article’s specific update.
    \item Other: [Free response]
\end{enumerate}
\end{quizCard}

%%%%%%%%%%%%%%%%%%%%%%%%%%%%%%%%%%%%%
\textbf{Assumption Quiz: AI Plan Step 2}

\begin{quizCard}
\textbf{Question:}
What assumptions underlie plan step 2: \emph{Identify the main subject of the article — Anne Frank and the circumstances of her death}? (Select all that apply)

\begin{enumerate}[label=(\alph*), leftmargin=1.2cm]
    \item All relevant themes in an article must be represented equally in the final summary.
    \item Organizing content thematically helps distill key points, important in short summaries especially for this audience.
    \item Other: [Free response]
\end{enumerate}
\end{quizCard}

%%%%%%%%%%%%%%%%%%%%%%%%%%%%%%%%%%%%%
\textbf{Assumption Quiz: AI Plan Step 3}

\begin{quizCard}
\textbf{Question:}
What assumptions underlie plan step 3: \emph{Note the new discovery — revised death date based on recent research}? (Select all that apply)

\begin{enumerate}[label=(\alph*), leftmargin=1.2cm]
    \item The discovery is central but fits within a broader narrative framework.
    \item New findings should be framed alongside past assumptions and presented with detailed contrast.
    \item Other: [Free response]
\end{enumerate}
\end{quizCard}

%%%%%%%%%%%%%%%%%%%%%%%%%%%%%%%%%%%%%%
\textbf{Assumption Quiz: AI Plan Step 4}

\begin{quizCard}
\textbf{Question:}
What assumptions underlie plan step 4: \emph{Analyze the sources of evidence used — Red Cross, survivor accounts, and Bergen-Belsen records}? (Select all that apply)

\begin{enumerate}[label=(\alph*), leftmargin=1.2cm]
    \item Source credibility adds depth and supports the accuracy of the summary.
    \item Credible summaries must explicitly mention all major sources of information.
    \item Other: [Free response]
\end{enumerate}
\end{quizCard}

%%%%%%%%%%%%%%%%%%%%%%%%%%%%%%%%%%%%%%
\textbf{Assumption Quiz: AI Plan Step 5}

\begin{quizCard}
\textbf{Question:}
What assumptions underlie plan step 5: \emph{Summarize the harsh conditions of the camp and how typhus played a role}? (Select all that apply)

\begin{enumerate}[label=(\alph*), leftmargin=1.2cm]
    \item Including minimal context on camp conditions helps explain the cause of death.
    \item The summary should describe the full physical environment and disease transmission details to be complete.
    \item Other: [Free response]
\end{enumerate}
\end{quizCard}

%%%%%%%%%%%%%%%%%%%%%%%%%%%%%%%%%%%%%%
\textbf{Assumption Quiz: AI Plan Step 6}

\begin{quizCard}
\textbf{Question:}
What assumptions underlie plan step 6: \emph{Highlight Anne Frank’s emotional resilience and emphasize the symbolic importance of her diary to modern audiences and summarize all the themes}? (Select all that apply)

\begin{enumerate}[label=(\alph*), leftmargin=1.2cm]
    \item Anne Frank’s emotional strength and legacy are emotionally resonant and is required for this audience.
    \item No summary of Anne Frank is complete without reference to her diary’s global impact.
    \item Other: [Free response]
\end{enumerate}
\end{quizCard}

\end{AppendixDocumentBox}

%%%%%%%%%%%%%%%%%%%%%%%%%%%%%%%%%%%%%%%%%%%%%%

\begin{AppendixDocumentBox}[orange!20]{What-If Questions for Task 5}

\textbf{What-If Question 1}

\begin{quizCard}
\textbf{Question:}
Which is the most critical step in the plan?
\begin{enumerate}[label=(\alph*), leftmargin=1.2cm]
    \item Step 1: \emph{Identify the main subject of the article — Anne Frank and the circumstances of her death.}
    \item Step 2: \emph{Break the article into thematic categories: historical context, emotional testimony, medical explanation, and personal legacy.}
    \item Step 3: \emph{Note the new discovery — revised death date based on recent research.}
    \item Step 4: \emph{Analyze the sources of evidence used — Red Cross, survivor accounts, and Bergen-Belsen records.}
    \item Step 5: \emph{Summarize the harsh conditions of the camp and how typhus played a role.}
    \item Step 6: \emph{Highlight Anne Frank’s emotional resilience and emphasize the symbolic importance of her diary to modern audiences and summarize all the themes.}
\end{enumerate}
\end{quizCard}

\textbf{What-If Question 2}

\begin{quizCard}
\textbf{Question:}
What happens if this step fails or changes? Describe in a few words the impact on the plan/draft. If there is no impact, you can answer ``No impact''.
[Free response]
\end{quizCard}

\end{AppendixDocumentBox}

%%%%%%%%%%%%%%%%%%%%%%%%%%
\clearpage
%%%%%%%%%%%%%%%%%%%%%

%%%%%%%%%%%%%%%%%%%%%%%
% !TEX root =  main.tex
%%%%%%%%%%%%%%%%%%%%%%%%%%%%%%%%%%%%%%%%%%%%%%%%%%%%%%%%%%
%%%%%%%%%%%%%%%%%%%%%%%%%%%%%%%%%%%%%%%%%%%%%%%%%%%%%%%%%%

\section{Details of the Postsurvey}\label{appendix.postsurvey}
In this section, we present the detailed questionnaire that the participants had to answer as part of the postsurvey. This was presented after they completed the knowledge work tasks assisted by AI, and was hosted on the same web-application.

%%%%%%%%%%%%%%%%%%%%%%%%%%%%%%%%%%
\subsection{Task Experience}\label{appendix.postsurvey.projectexp}

\emph{Please rate your agreement with the following statements on a scale of 1 (not at all) to 5 (very much so). Note: The phrase ``questions in the INTERACTION PANE'' in the statements below refer to the questions shown before the feedback stage.}

\begin{enumerate}
   \item I enjoyed doing the tasks on the platform.
        \begin{enumerate}[label=(\alph*)]
            \item (Score 1) Not at all
            \item (Score 2) Slightly
            \item (Score 3) Moderately
            \item (Score 4) Quite a bit
            \item (Score 5) Very much so
        \end{enumerate}
    %%%%%%%
    \item The questions in the INTERACTION PANE encouraged me to think critically about the AI plan and draft. Critical thinking is careful, deliberate judgment that involves interpreting, analyzing, and evaluating information, and explaining the evidence, concepts, or context behind your conclusions.
        \begin{enumerate}[label=(\alph*)]
            \item (Score 1) Not at all
            \item (Score 2) Slightly
            \item (Score 3) Moderately
            \item (Score 4) Quite a bit
            \item (Score 5) Very much so
            \item If you felt the questions in the INTERACTION PANE encouraged critical thinking, can you describe how they influenced your thought process? [Free response]
        \end{enumerate}
    %%%%%%%
     \item The questions in the INTERACTION PANE helped me engage more deeply with the AI plan and draft.
        \begin{enumerate}[label=(\alph*)]
            \item (Score 1) Not at all
            \item (Score 2) Slightly
            \item (Score 3) Moderately
            \item (Score 4) Quite a bit
            \item (Score 5) Very much so
        \end{enumerate}
    %%%%%%%
      \item The questions in the INTERACTION PANE helped me better understand the AI plan and draft and improved my sense of agency in completing the task.
        \begin{enumerate}[label=(\alph*)]
            \item (Score 1) Not at all
            \item (Score 2) Slightly
            \item (Score 3) Moderately
            \item (Score 4) Quite a bit
            \item (Score 5) Very much so
        \end{enumerate}
    %%%%%%%
      \item The questions in the INTERACTION PANE interfered with my process of understanding the AI’s plan and draft.
        \begin{enumerate}[label=(\alph*)]
            \item (Score 1) Not at all
            \item (Score 2) Slightly
            \item (Score 3) Moderately
            \item (Score 4) Quite a bit
            \item (Score 5) Very much so
        \end{enumerate}
    %%%%%%%
     \item If AI tools were integrated into my workflow, I would prefer having questions similar to those in the INTERACTION PANE alongside the AI plan and draft.
        \begin{enumerate}[label=(\alph*)]
            \item (Score 1) Not at all
            \item (Score 2) Slightly
            \item (Score 3) Moderately
            \item (Score 4) Quite a bit
            \item (Score 5) Very much so
        \end{enumerate}
    %%%%%%%
\end{enumerate}
%%%%%%%%%%%%%%%%%%%%%

%%%%%%%%%%%%%%%%%%%%%%%
% !TEX root =  main.tex
%%%%%%%%%%%%%%%%%%%%%%%%%%%%%%%%%%%%%%%%%%%%%%%%%%%%%%%%%%
%%%%%%%%%%%%%%%%%%%%%%%%%%%%%%%%%%%%%%%%%%%%%%%%%%%%%%%%%%

\section{Details of the Interview Questions}\label{appendix.interview}
In this section, we present the comparative questions that the participants in the interview study had to answer after they completed experiencing all the CFF conditions. 

\begin{enumerate}
    \item Which of the three CFFs did you like the most and why?
    %%%%%
    \item Which of the three CFFs helped you most critically review AI response and why?
    %%%%%
    \item Which CFF felt least useful and why?
    %%%%%
    \item If you had to choose one of these CFFs to keep in your day-to-day work with AI tools, which would it be?
\end{enumerate}

%%%%%

%%%%%%%%%%%%%%%%%%%%%

\end{document}